%% file: ANA-HMBS-2024-07-PAPER.tex
\pdfoutput=1
\documentclass[PAPER, atlasdraft=false, UKenglish,cernpreprint, texmf, orcidlogo]{atlasdoc}
\usepackage{atlaspackage}
\usepackage{atlasbiblatex}

\usepackage[BSM=true]{atlasphysics}
\graphicspath{{logos/}{figures/}}

\usepackage{ANA-HMBS-2024-07-PAPER-defs}
\usepackage{adjustbox}
\usepackage{multirow}
\usepackage{makecell}
\AtlasTitle{Search for compressed electroweak SUSY production in final states with two $\tau$-leptons in $\sqrt{s} = 13$ and $13.6$ TeV \textit{pp} collisions with the ATLAS detector}

\AtlasAbstract{%
Searches for the production of $\tau$-sleptons ($\tilde{\tau}$), charginos ($\tilde{\chi}^{\pm}_1$), and neutralinos ($\tilde{\chi}^0_2$) in events with at least one hadronically decaying $\tau$-lepton are presented, targeting compressed mass spectra scenarios. For charginos and neutralinos, the cascade decays producing either a $\tau$-slepton or a $\tau$-sneutrino are considered. Two final selections are studied, requiring either one hadronically decaying $\tau$-lepton and one other charged lepton, or two or more hadronically decaying $\tau$-leptons. To enhance sensitivity to the compressed mass spectra, production modes with initial-state radiation (ISR) jets are considered. The searches use 140~$\mathrm{fb}^{-1}$ and 53~$\mathrm{fb}^{-1}$ of proton--proton collision data recorded by the ATLAS detector at the Large Hadron Collider during 2015--2018 at a centre-of-mass energy of $\sqrt{s} = 13$ TeV and 2022--2023 at a centre-of-mass energy of $\sqrt{s} = 13.6$ TeV, respectively.
The searches provide sensitivity to the compressed mass spectra scenarios with light $\tau$-sleptons. The results are generally consistent with the Standard Model, although several deviations with local significances of $2.1-2.4 \sigma$ are observed. These deviations weaken the observed exclusion limits.
For direct $\tau$-slepton production, in the ISR channel alone the limits are set for $\Delta m(\tilde{\tau}, \tilde{\chi}^0_1) = 30$ GeV for $\tau$-slepton masses up to $115$ GeV at 95\% confidence level. Given the observed data, no exclusion limits are set for direct $\tau$-sleptons production when statistically combining all considered channels. For electroweakino production, limits are set for $\Delta m(\tilde{\chi}^0_2, \tilde{\chi}^0_1) = 30$ GeV for $\tilde{\chi}^0_2$ masses up to $200$ GeV. For pure chargino pair-production, limits are set for chargino masses up to $175$ GeV for the same mass splitting.
}
\AtlasRefCode{HMBS-2024-07}

\PreprintIdNumber{CERN-EP-2026-272}

\AtlasJournal{JHEP}
\hypersetup{pdftitle={ATLAS document},pdfauthor={The ATLAS Collaboration}}

\begin{document}

\maketitle

\tableofcontents

\section{Introduction}
\label{sec:intro}


Supersymmetry (SUSY)~\cite{Golfand:1971iw,Volkov:1973ix,Wess:1974tw,Wess:1974jb,Ferrara:1974pu,Salam:1974ig,Martin:1997ns}
postulates a bosonic (fermionic) superpartner for every fermionic (bosonic) particle in the Standard Model (SM), with a spin difference of one-half unit relative to the corresponding SM particle.
In SUSY models with $R$-parity conservation~\cite{Farrar:1978xj}, the SUSY particles must be produced in pairs and the lightest supersymmetric particle (LSP) is stable and weakly interacting, thus providing an excellent dark matter candidate.

In the simplified SUSY models considered in this \paper, all strongly interacting SUSY particles are assumed to be decoupled, leaving only the electroweak‐only SUSY sector which includes charginos ($\susy{\chi}^{\pm}_{i}$, $i$ = 1, 2 in order of increasing mass), neutralinos
($\susy{\chi}^{0}_{j}$, $j$ = 1, 2, 3, 4 in order of increasing mass), charged sleptons (\slepton), and sneutrinos ($\susy{\nu}$).
Charginos and neutralinos are mass eigenstates obtained as linear combinations of the superpartners of the Higgs bosons and electroweak gauge bosons.
Charged sleptons are the superpartners of the charged leptons and, following the convention for their SM counterparts, are denoted \sleptonL and \sleptonR, respectively.
The slepton mass eigenstates are a mixture of \sleptonL and \sleptonR, and are labelled as $\susy{\ell}_{k}$ ($k$ = 1, 2 in order of increasing mass).
In this \paper, the scalar superpartners of the left-handed $\tau$-lepton ($\stau_\mathrm{L}$) and right-handed $\tau$-lepton ($\stau_\mathrm{R}$) are assumed to be mass degenerate and are referred to as \enquote{$\tau$-sleptons} (\stau), or generically, \enquote{staus.}

In SUSY models with small mass difference between the LSP and the parent SUSY particles, also referred to as \enquote{compressed mass spectra}, co‑annihilation processes involving nearly mass-degenerate SUSY particles can naturally explain the observed dark matter relic density~\cite{GDuan}. Final states containing $\tau$-leptons are particularly promising: light $\tau$-slepton decays into neutralino LSPs are well‑motivated by dark matter phenomenology~\cite{Vasquez:2011}. Compressed-spectrum \stau scenarios satisfy Planck relic-density constraints~\cite{Chakraborti}.
In many SUSY models $\tau$-sleptons are the lightest slepton generation, escaping the bounds set by searches for selectrons and smuons. Compressed-mass electroweak SUSY signals with $\tau$-leptons remain largely unexplored~\cite{SUSY-2023-03}, due to the challenges of soft $\tau$-lepton reconstruction, limited trigger bandwidth, and large SM backgrounds. The primary focus of the presented analysis is on final states containing events with an initial-state radiation (ISR) jet that provides a Lorentz boost to the SUSY system, thus enhancing the experimental sensitivity to the signal. In addition, events containing one leptonically decaying $\tau$-lepton are considered, which permits the use of single-lepton triggers with lower transverse momentum (\pt) thresholds than the $\tau$-lepton triggers used in previous studies~\cite{SUSY-2023-03}.

\begin{figure}[!htb]
\centering
\subfloat[][]{\includegraphics[width=.3\textwidth]{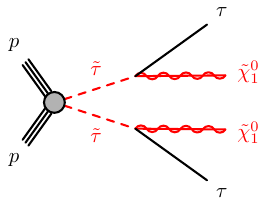}\label{fig:feynmanslepton_DSNoISR}}\hspace{0.5cm}
\subfloat[][]{\includegraphics[width=.3\textwidth]{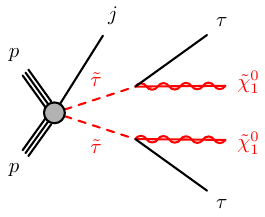}\label{fig:feynmanslepton_DSISR}}\\
\subfloat[][]{\includegraphics[width=.3\textwidth]{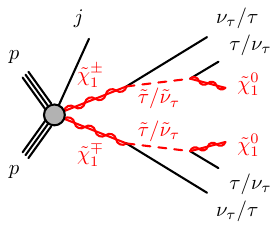}\label{fig:feynmanslepton_C1C1}}\hspace{0.5cm}
\subfloat[][]{\includegraphics[width=.3\textwidth]{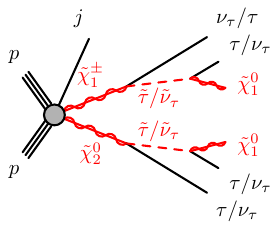}\label{fig:feynmanslepton_C1N2}}
\caption{Representative diagrams of SUSY scenarios that are being searched for in this \paper. Direct \stau production (\staunoisr) is shown in (a), and direct \stau associated with ISR production (\stauisr) is shown in (b). The processes for the intermediate stau channel is shown in (c) and (d).  In all cases, the subsequent decays contain two $\tau$-leptons in the final state. }
\label{fig:feynmanslepton_summary}
\end{figure}

Four search channels are presented in this \paper, all including at least one hadronically decaying $\tau$-lepton in the final state in addition to either low jet activity or one ISR jet candidate.
The first scenario considers the direct production of $\tau$-slepton pairs with mass-degenerate $\stau_\mathrm{L,R}$, which are assumed to decay exclusively into a $\tau$-lepton and the LSP, referred to as the \enquote{\staunoisr channel}. The second scenario makes the same assumptions about $\tau$-sleptons, but focuses on the production mode with an associated ISR jet and missing transverse momentum, $\pTmiss$, from the neutralinos and neutrinos. This is referred to as the \enquote{\stauisr channel}.
The third and fourth scenarios involve the production of a chargino pair (\ConepConemFancy) or chargino-neutralino (\ConepmNtwoFancy) in association with an ISR jet, respectively. These states decay into the lightest neutralino exclusively via intermediate $\tau$-sleptons and $\tau$-sneutrinos (\tausneut) with equal branching fractions, defining the so‐called \enquote{Intermediate stau} channels.
Representative diagrams of the targeted signal processes are shown in Figure~\ref{fig:feynmanslepton_summary}. In all scenarios the $\tau$-sleptons are assumed to decay promptly into an LSP and a $\tau$-lepton.

Previous analyses from the ATLAS Collaboration have set exclusion limits at 95\% confidence level (CL) on these SUSY models with the \RunOne and \RunTwo data~\cite{SUSY-2013-14,SUSY-2014-05,SUSY-2016-23,SUSY-2018-04,SUSY-2023-03} as part of the SUSY searches programme~\cite{SUSY-2023-10_fix,SUSY-2014-06}. The CMS Collaboration has also set similar exclusion limits~\cite{CMS-SUS-21-001,CMS-SUS-19-012,CMS-SUS-19-002}.
This \paper extends the ATLAS experiment sensitivity to smaller mass splittings between $\ninotwo$/$\chinoonepm/\stau$ and $\ninoone$, using the full data sample collected in \RunTwo and the \RunThr data sample collected from 2022 to 2023.
Furthermore, it is the first ATLAS experiment search with sensitivity to the compressed prompt $\tau$-sleptons and electroweak SUSY scenarios with $\tau$-lepton final states, exploring the region of mass splittings \masssplit or \dmcn  of less than $50$~\GeV. The increased sensitivity is achieved by adjusting the trigger strategy, improved analysis techniques, and dedicated calibration of soft $\tau$-leptons.

%

%


%
\FloatBarrier
\section{ATLAS detector}
\label{sec:detector}


The ATLAS detector~\cite{PERF-2007-01,GENR-2019-02} at the LHC covers nearly the entire solid angle around the collision point.\footnote{ATLAS uses a right-handed coordinate system with its origin at the nominal interaction point (IP)
in the centre of the detector and the \(z\)-axis along the beam pipe.
The \(x\)-axis points from the IP to the centre of the LHC ring,
and the \(y\)-axis points upwards.
Polar coordinates \((r,\phi)\) are used in the transverse plane,
\(\phi\) being the azimuthal angle around the \(z\)-axis.
The pseudorapidity is defined in terms of the polar angle \(\theta\) as \(\eta = -\ln \tan(\theta/2)\) and is equal to the rapidity
$ y = \frac{1}{2} \ln \left( \frac{E + p_z}{E - p_z} \right) $ in the relativistic limit.
Angular distance is defined as \(\Delta R \equiv \sqrt{(\Delta y)^{2} + (\Delta\phi)^{2}}\).}
It consists of an inner tracking detector surrounded by a thin superconducting solenoid, electromagnetic and hadronic calorimeters,
and a muon spectrometer incorporating three large superconducting air-core toroidal magnets.

The inner-detector (ID) system is immersed in a \qty{2}{\tesla} axial magnetic field
and provides charged-particle tracking in the range of \(|\eta| < 2.5\).
The high-granularity silicon pixel detector covers the vertex region and typically provides four measurements per track,
the first hit generally being in the insertable B-layer (IBL).
It is followed by the SemiConductor Tracker (SCT), which usually provides eight measurements per track.
These silicon detectors are complemented by the transition radiation tracker (TRT),
which enables radially extended track reconstruction up to \(|\eta| = 2.0\).
The TRT also provides electron identification information
based on the fraction of hits (typically 30 in total) above a higher energy-deposit threshold corresponding to transition radiation.

The calorimeter system covers the pseudorapidity range \(|\eta| < 4.9\).
Within the region \(|\eta|< 3.2\), electromagnetic calorimetry is provided by barrel and
endcap high-granularity lead/liquid-argon (LAr) calorimeters,
with an additional thin LAr presampler covering \(|\eta| < 1.8\)
to correct for energy loss in material upstream of the calorimeters.
Hadronic calorimetry is provided by the steel/scintillator-tile calorimeter,
segmented into three barrel structures within \(|\eta| < 1.7\), and two copper/LAr hadronic endcap calorimeters.
The solid angle coverage is completed with forward copper/LAr and tungsten/LAr calorimeter modules
optimised for electromagnetic and hadronic energy measurements, respectively.

The muon spectrometer (MS) comprises separate trigger and
high-precision tracking chambers measuring the deflection of muons in a magnetic field generated by the superconducting air-core toroidal magnets.
The field integral of the toroids ranges between \num{2.0} and \qty{6.0}{\tesla\metre}
across most of the detector.
Three layers of precision chambers, each consisting of layers of monitored drift tubes, cover the region \(|\eta| < 2.7\).
These were complemented in the innermost layer of the endcap region by cathode-strip chambers in \RunTwo,
which were replaced in \RunThr by layers of small-strip thin-gap chambers and Micromegas chambers, both providing precision tracking in the region \(1.3 < |\eta| < 2.7\).
The muon trigger system covers the range \(|\eta| < 2.4\) with resistive-plate chambers in the barrel region, thin-gap chambers in the endcap regions,
and, in \RunThr, the small-strip thin-gap chambers and Micromegas chambers in the innermost layer of the endcap, which provide higher redundancy and a large reduction in fake muon triggers.

Events are selected by the first-level trigger system implemented in custom hardware,
followed by the high-level trigger selections, based on algorithms implemented in software~\cite{TRIG-2016-01,TRIG-2022-01}.
The first-level trigger accepts events from the \qty{40}{\MHz} bunch crossings at a rate close to \qty{100}{\kHz},
which the high-level trigger further reduces to record complete events to disk with a rate of about \qty{1.25}{\kHz} and \qty{3}{\kHz} in \RunTwo and \RunThr, respectively.

The \RunThr detector configuration benefits from several upgrades compared with that of \RunTwo to maintain high detector performance at the higher average number of simultaneous collisions in \RunThr.
The trigger system benefits from new digital electronics readout of the LAr calorimeters with significantly increased granularity.
Further details of the \RunThr detector configuration are given in Ref.~\cite{GENR-2019-02}.

A software suite~\cite{SOFT-2022-02} is used in data simulation, in the reconstruction
and analysis of real and simulated data, in detector operations, and in the trigger and data acquisition
systems of the experiment.


%

%
\section{Data and simulated event samples }
\label{sec:sample}

Two proton--proton ($pp$) collision data samples are used for the analysis collected with the ATLAS detector during \RunTwo of the LHC from 2015--2018 ($\sqrt{s} = 13$~\TeV)
and partial \RunThr from 2022--2023 ($\sqrt{s} = 13.6$~\TeV) with a bunch-crossing separation of 25~ns. These correspond to integrated luminosities of $140.1 \pm 1.2~\mathrm{fb}^{-1}$ and $53.0 \pm 1.1~\mathrm{fb}^{-1}$, respectively~\cite{DAPR-2021-01,ATL-DAPR-PUB-2024-001}. The luminosity is measured mainly by the LUCID-2 detector~\cite{LUCID2}, which records Cherenkov light produced in the quartz windows of photomultipliers located close to the beampipe. A set of data quality requirements is applied to the data to ensure that the LHC beam conditions were stable and the ATLAS detector was fully functional during the data-taking~\cite{DAPR-2018-01}. The quoted integrated luminosity is computed after the rejection of data failing these requirements.

The channels with an ISR jet requirement rely on magnitude of missing transverse momentum (\met) triggers~\cite{TRIG-2016-01,TRIG-2022-01,TRIG-2019-01} to select events. The selection thresholds of these triggers increased during the data-taking from $70$~\GeV to $115$~\GeV.
The \staunoisr channel uses single-lepton triggers~\cite{TRIG-2018-01,TRIG-2018-05} to select the events. Their selection thresholds have changed between $20$~\GeV and $26$~\GeV during the data-taking periods considered.

The main SM backgrounds relevant for the analyses presented can be separated into two main categories. The first contains events with at least one jet misidentified as a hadronically decaying $\tau$-lepton (\enquote{fake-tau}). These provide the largest contribution to all channels and are modelled using a data-driven approach, described in Section~\ref{sec:faketau}. The other category contains SM processes with true $\tau$-leptons in the final states. They are modelled by Monte Carlo (MC) simulations, with the production cross-sections normalised to data in dedicated regions. In the \stau\stau channels, diboson production is the second most important background. In the intermediate \stau channel, the \ttbar and the single-top-quark production (collectively referred to as \enquote{Top}), diboson, and $Z(\rightarrow\tau\tau)$ production processes contribute significantly to the total SM background.

The details of the MC simulations and the cross-sections computation precision are summarised in Table~\ref{tab:samples_desc_run2} together with the parton distribution function (PDF) sets and
the set of parameters (tune) for the underlying event and parton shower models.
The same reconstruction and identification algorithms are applied to data and MC events.
The ATLAS simulation framework~\cite{SOFT-2010-01,SOFT-2022-02} was used for the production of all MC samples.
The detector response simulation is based on \GEANT~\cite{Agostinelli:2002hh}.
For SUSY signal Monte Carlo (MC) models, the calorimeter shower simulation (the most computationally-intensive process) is replaced with a faster \textsc{ATLFAST3}~\cite{SIMU-2018-04} approach.
The effect of multiple interactions in the same and neighbouring bunch crossings (\pileup) was modelled by overlaying each simulated hard-scattering event with inelastic $pp$ collision events.
The \pileup model combines collision events from \AtlasMC[2.0.1.4]{EPOS}~\cite{Werner:2005jf}, using the LHC tune~\cite{Pierog:2013ria}, and  \PYTHIA[8]~\cite{Sjostrand:2007gs,Bierlich:2022pfr}, using the A3 tune~\cite{ATL-PHYS-PUB-2016-017} and the \NNPDF[2.3lo] set of PDFs~\cite{Ball:2012cx}.

SUSY signal MC samples were simulated for different mass assumptions, following the simplified-model approach~\cite{Alwall:2008ve,Alwall:2008ag,Alves:2011wf}, with all SUSY particles not directly involved in the processes considered decoupled (i.e. having a much higher mass, typically $10$~\TeV). For direct \stau searches, the mass ranges considered are $m(\stau) \in [75,225]$~\GeV, and \masssplit $\in [10, 60]$~\GeV, motivated by models that predict a dark-matter abundance consistent with cosmological observations and experimental sensitivity, with a step-size of $10$~\GeV. For the intermediate \stau production, the mass ranges considered are $m(\chinoonepm / \ninotwo) \in [100,400]$~\GeV, and $m(\ninoone) \in [0,300]$~\GeV, with a step-size of $20-25$~\GeV. The matrix element (ME) calculation is performed at tree level and includes the emission of up to two additional partons. The ME--parton-shower merging uses the CKKW-L prescription~\cite{Lonnblad:2001iq,Lonnblad:2011xx}, with the merging scale set to one quarter of the mass of pair-produced particles. Signal cross-sections are calculated to next-to-leading-order (NLO) in the strong coupling constant, adding the resummation of soft gluon emission at next-to-leading-logarithm accuracy (NLO+NLL) in \RunTwo and to approximate next-to-next-to-leading-order in the strong coupling constant, adding the resummation of soft gluon emission at next-to-next-to-leading-logarithm accuracy (aNNLO+NNLL)~\cite{Beenakker:1999xh,Debove:2010kf,Fuks:2012qx,Fuks:2013vua,Fiaschi:2018hgm,Bozzi:2007qr,Fuks:2013lya,Fiaschi:2018xdm,newResummino} in \RunThr. The aNNLO+NNLL computation leads to cross-sections $1-5\%$ larger than the NLO+NLL one. The nominal cross-section and the uncertainty are taken from the cross-section predictions using different PDF sets and factorisation and renormalisation scales, as described in Refs.~\cite{Borschensky:2014cia,Ball:2022oua}.

\begin{table}[!h]
\centering
\caption{
Summary of the generators used to simulate the signal and background processes.
The symbol \enquote{$V$} stands for a massive vector boson. %
}
\label{tab:samples_desc_run2}
\begin{adjustbox}{width=1\textwidth}
\begin{tabular}{llllllll}
\toprule
Physics process     & $\sqrt{s}$~[TeV] & Generator     & Parton Shower         &  $\sigma$ accuracy & Tune                      & PDF (generator) & PDF (shower)\\
\midrule
$Z(\ell\ell )$+jets    & $13$ &  \SHERPA[2.2.11]~\cite{Bothmann:2019yzt}                        & \SHERPA[2.2.11]~\cite{Schumann:2007mg}         & NNLO~\cite{Anastasiou:2003ds}                 & SHERPA       & \NNPDF[3.0nnlo]~\cite{Ball:2014uwa,Carrazza:2015aoa} & \NNPDF[3.0nnlo] \\
& $13.6$ &  \SHERPA[2.2.14]  & \SHERPA[2.2.14]   & NNLO   & SHERPA       & \NNPDF[3.0nnlo] & \NNPDF[3.0nnlo] \\
$Z(\tau\tau )$+jets & $13/13.6$   & \SHERPA[2.2.14]  & \SHERPA[2.2.14] & NNLO  & SHERPA       & \NNPDF[3.0nnlo] & \NNPDF[3.0nnlo] \\
$W$+jets & $13$   & \SHERPA[2.2.11] & \SHERPA[2.2.11] & NNLO   & SHERPA     & \NNPDF[3.0nnlo] & \NNPDF[3.0nnlo] \\
& $13.6$   & \SHERPA[2.2.14] & \SHERPA[2.2.14] & NNLO   & SHERPA     & \NNPDF[3.0nnlo] & \NNPDF[3.0nnlo] \\
$VV$, $VVV$  & $13$  &  \SHERPA[2.2.1],\,2.2.2,\,2.2.14  &   \SHERPA[2.2.1],\,2.2.2,\,2.2.14 & NLO~\cite{Hoeche:2011fd,Hoeche:2012yf,Catani:2001cc,Hoeche:2009rj}                  &  SHERPA  &\NNPDF[3.0nnlo] & \NNPDF[3.0nnlo] \\
& $13.6$  &  \SHERPA[2.2.14] &   \SHERPA[2.2.14] & NLO  &  SHERPA  &\NNPDF[3.0nnlo] & \NNPDF[3.0nnlo] \\
$t\bar{t}$ & $13/13.6$ & \POWHEGBOX[v2]~\cite{Frixione:2007nw,Nason:2004rx,Frixione:2007vw,Alioli:2010xd}                      & \PYTHIA[8]~\cite{Sjostrand:2014zea}           & NNLO+NNLL~\cite{Beneke:2011mq,Cacciari:2011hy,Baernreuther:2012ws,Czakon:2012zr,Czakon:2012pz,Czakon:2013goa,Czakon:2011xx}           & A14~\cite{ATL-PHYS-PUB-2014-021} & \NNPDF[3.0nlo]& \NNPDF[2.3lo]~\cite{Ball:2012cx}\\
Single-top: \\
$~~~~Wt$  & $13/13.6$  & \POWHEGBOX[v2]         & \PYTHIA[8]    &NLO+NNLL~\cite{Kidonakis:2010ux,Kidonakis:2013zqa}     & A14   & \NNPDF[3.0nlo] & \NNPDF[2.3lo]    \\
~~~~$s$- and $t$-channels & $13/13.6$     & \POWHEGBOX[v2] & \PYTHIA[8] &NLO~\cite{Aliev:2010zk,Kant:2014oha}  & A14   & \NNPDF[3.0nlo]  & \NNPDF[2.3lo]      \\
$t\bar{t}{+}V$ & $13/13.6$    & \MGNLO[2.3.3]~\cite{Alwall:2014hca}      & \PYTHIA[8] & NLO~\cite{deFlorian:2016spz_fix}      & A14 & \NNPDF[3.0nlo] & \NNPDF[2.3lo] \\
$t\bar{t}{+}H$ & $13/13.6$ & \POWHEGBOX[v2]~\cite{Hartanto:2015uka}  & \PYTHIA[8]& NLO ~\cite{Alwall:2014hca} & A14 & \NNPDF[3.0nlo] & \NNPDF[2.3lo]\\

$V H (\tau\tau)$ & $13/13.6$ &\POWHEGBOX[v2]~\cite{Alioli:2010xd,Hamilton:2013fea,Hamilton:2015nsa} & \PYTHIA[8] &NNLO + NLO ~\cite{Ciccolini:2003jy,Brein:2003wg,Brein:2011vx,Altenkamp:2012sx,Denner:2014cla,Brein:2012ne,Harlander:2014wda}   &AZNLO~\cite{STDM-2012-23}      & \NNPDF[3.0nlo] & \NNPDF[2.3lo] \\
$ggH(\tau\tau)$ & $13$  & \POWHEGBOX v2 & \PYTHIA[8]    & NNLL+NNLO  & AZNLO & \NNPDF[3nlo] & \NNPDF[2.3lo]\\
& $13.6$  &  \POWHEGBOX[v2] & \PYTHIA[8] & NNLL+NNLO ~\cite{Hamilton:2012np,Campbell:2012am,Hamilton:2012rf,Catani:2007vq,Bozzi:2005wk,deFlorian:2011xf} & A14 & \NNPDF[3nlo]  & \NNPDF[2.3lo]\\
VBF $H(\tau\tau) $ & $13$ & \POWHEGBOX v2 & \PYTHIA[8] & NNLO+NLO & AZNLO & \NNPDF[3nlo]  & \NNPDF[2.3lo] \\
& $13.6$ & \POWHEGBOX[v2] & \PYTHIA[8]  & NNLO+NLO & A14 & \NNPDF[3nlo]   & \NNPDF[2.3lo] \\
SUSY signal & $13$ & \MADGRAPH[3.5] & \PYTHIA[8] &  NLO+NLL~\cite{Beenakker:2024nlf,Beenakker:2016lwe,Beenakker:2014sma,Beenakker:2013mva,Beenakker:2011sf,Beenakker:2009ha,Kulesza:2009kq,Kulesza:2008jb,Beenakker:1996ch} &  A14 & \NNPDF[3.0nlo]  &  \NNPDF[2.3lo]  \\
SUSY signal & $13.6$ & \MADGRAPH[3.5] & \PYTHIA[8] &  aNNLO+NNLL~\cite{Beenakker:2024nlf,Beenakker:2016lwe,Beenakker:2014sma,Beenakker:2013mva,Beenakker:2011sf,Beenakker:2009ha,Kulesza:2009kq,Kulesza:2008jb,Beenakker:1996ch,newResummino} &  A14 & \NNPDF[3.0nlo]  &  \NNPDF[2.3lo]  \\
\bottomrule
\end{tabular}
\end{adjustbox}
\end{table}


%

%
\section{Event reconstruction}
\label{sec:EvenRec}

In this Section, the reconstruction algorithms and selection criteria for the physics objects used in the analysis are described. These are harmonised across \RunTwo and \RunThr, with any exceptions highlighted. If an event contains several objects of the same type, they are ordered in decreasing transverse momentum, $\pt$. For all of the objects discussed below, dedicated scale factors are used to improve agreement between data and MC simulation.

Inner-detector tracks with \(\pt>\SI{500}{\MeV}\)~\cite{PERF-2015-01,ATL-PHYS-PUB-2019-015,IDTR-2021-01} are used to construct primary vertex candidates, with a minimum of two associated tracks. The candidate with the largest sum of squared track \pt associated with it is chosen as the hard-scatter (or primary) vertex. The primary vertex serves as the origin for all the objects described below (except for $\tau$-leptons, where a dedicated algorithm is used~\cite{ATL-PHYS-PUB-2022-044} to determine the $\tau$-lepton production vertex). Events with no primary vertex candidates are vetoed.

Jets are reconstructed from particle-flow objects~\cite{PERF-2015-09}. The \antikt\ clustering algorithm~\cite{Cacciari:2008gp,Fastjet} with a radius parameter of $R = 0.4$ is used to construct all jets considered by the analysis.
To improve the performance and the agreement between data and MC simulation, jets undergo a sequence of calibrations. These include corrections derived both from simulation and in situ measurements~\cite{JETM-2018-05}.
Jet candidates are required to have $\pt > 20$~\GeV\ and $|\eta| < 4.5$ to be considered in the analysis.
Jets with $\pt < 60$~\GeV\ are required to satisfy additional cleaning selections to suppress contributions from \pileup. Central ($|\eta| < 2.5$) jets are required to satisfy the \enquote{FixedEffPt} working point (\uppercase{WP}) of the neural-network-based jet vertex tagger (NNJVT)~\cite{PERF-2014-03}, which provides a flat hard-scatter jet selection efficiency across the full $\eta$ range. The forward jets ($2.5 < |\eta| < 4.5$) are required to satisfy the \enquote{Loose} \uppercase{WP} of the forward JVT algorithm~\cite{ATL-PHYS-PUB-2019-026}.
Jets originating from the hadronisation of $b$-quarks are tagged with the GN2 tagger~\cite{FTAG-2023-05,FTAG-2018-01,FTAG-2023-10,FTAG-2024-07}.
It provides six fixed-efficiency \uppercase{WP}s corresponding to selections that correctly identify a given fraction of the $b$-jets, as determined in a sample of simulated \ttbar\ events. The selection is applied only to jets in the $|\eta| < 2.5$ range, corresponding to the ID coverage. %

Electron candidates are first required to satisfy the \enquote{LooseAndBLayerLLH} identification criteria~\cite{PERF-2017-01,EGAM-2018-01,EGAM-2021-02,EGAM-2021-01}, and have $\pt > 4.5$~\GeV\ and $|\eta| < 2.47$. In addition, a $|z_0\sin\theta| < 0.5$~mm selection is applied to the longitudinal impact parameter $z_0$ to suppress non-prompt electrons.
Electron candidates satisfying the initial selection have to satisfy the overlap removal procedure (ORP, described below). Finally, the candidates that satisfy the \enquote{Tight\_VarRad} isolation criterion, the tighter \enquote{TightLLH} electron identification~\cite{PERF-2017-01,EGAM-2018-01,EGAM-2021-02,EGAM-2021-01}, and have the transverse impact parameter significance $\frac{d_{0}}{\sigma{(d_{0})}}$ (where $d_{0}$ is the distance of closest approach of the track to the primary vertex point) smaller than $5$ are used by the analysis.

Muon candidates are first required to satisfy the \enquote{Medium} identification criteria~\cite{MUON-2018-03,MUON-2022-01}, and have $\pt > 3$~\GeV, $|\eta| < 2.7$ in \RunTwo ($\pt > 5$~\GeV, $|\eta| < 2.5$ in \RunThr) and $|z_0\sin\theta| < 0.5$~mm. If it passes the ORP, a muon candidate is used in the analysis if it also satisfies the \enquote{PflowTight\_VarRad} isolation criteria~\cite{MUON-2018-03,MUON-2022-01} and has a $d_{0}$ significance smaller than $3$.

The reconstruction of hadronically decaying $\tau$-leptons starts with topological cell clusters (with the local hadronic cell-weighting calibration~\cite{PERF-2014-07,Barillari:1112035}) combined into $R = 0.4$ anti-$k_t$ jets. Tracks within $\Delta R = 0.2$ of the seed-jet axis are used to reconstruct the production vertex for the visible hadronic component of the $\tau$-lepton decay (\tauhadvis). Tracks and calorimeter-cell clusters are then used to build the $\tau$-lepton candidate, which then undergoes a sequence of calibrations (\enquote{$\tauhadvis$ Energy Scale calibration}~\cite{ATL-PHYS-PUB-2015-045,ATL-PHYS-PUB-2022-044}). A dedicated classifier is used to identify tracks originating from charged $\tau$-lepton decay products~\cite{ATL-PHYS-PUB-2022-044}. A $\tau$-lepton candidate is required to have exactly one or three such tracks, with the sum of their charge equal to $\pm1$, and to have $\pt > 20$~\GeV and $|\eta| < 2.5$. In the ISR channels the $\pt$ selection is relaxed to $15$~\GeV, using a dedicated low-$\pt$ calibration. Candidates reconstructed in the transition region between the barrel and endcap calorimeters, $|\eta|\in [1.37,1.52]$ are rejected. To further suppress background jets, a classifier based on a recurrent neural network (RNN)~\cite{ATL-PHYS-PUB-2022-044} is used. The analysis uses the \enquote{Medium} and \enquote{Tight} $\tauhadvis$ identification (\enquote{tau-ID}) \uppercase{WP}s, corresponding, respectively, to $75\%$ and $60\%$ efficiency for $1$-track $\tau$-leptons, and $60\%$ and $45\%$ efficiency for $3$-track $\tau$-leptons. To suppress backgrounds from electrons passing the tau-ID selection, the $\tau$-lepton candidate has to satisfy the \enquote{Loose} \uppercase{WP} of the dedicated electron-veto algorithm, corresponding to $95\%$ efficiency, based on an RNN classifier~\cite{ATL-PHYS-PUB-2022-044}.
Additionally, $\tau$-lepton candidates that fail to satisfy the tau-ID selection, but satisfy the much looser \enquote{Very Loose} \uppercase{WP}, corresponding to $95\%$ efficiency, are used in the analysis to estimate the fake-tau background. These objects are referred to as \enquote{anti-ID taus}, and are discussed further in Section~\ref{sec:faketau}.

A single energy deposit in the calorimeter or a track in the inner detector can be used in the reconstruction of several analysis objects. To avoid double-counting and resolve the ambiguity, the ORP is performed for every event. It ensures that every detector signature is matched to at most one reconstructed object. The ORP used follows the one described in Ref.~\cite{HIGG-2019-09}.

An important kinematic variable used in the analysis is the missing transverse momentum \met. It is defined as the magnitude of $\pTmiss$, the negative vector sum of the transverse momenta of electrons, muons, jets, and a \enquote{soft term} composed of tracks from the primary vertex not associated with any reconstructed object~\cite{JETM-2020-03}. The hadronically decaying $\tau$-leptons are not included in the \met computation and are treated like jets instead, ensuring that the results with the different tau-ID \uppercase{WP}s used in the analysis are compatible. The \enquote{Tight} \met\ \uppercase{WP} is used, defining the selection on the jets used for the calculation. Three additional variables related to \met are used in the analysis:
\begin{itemize}
\item The transverse mass \mt, defined as
$\mt^2(\vec{p}_{\text{T}}^{\,a}, \vec{p}_{\text{T}}^{\,b})
= m_a^2 + m_b^2 + 2\left(p_\text{T}^{a}p_\text{T}^{b} - \vec{p}_\text{T}^{\,a} \cdot \vec{p}_\text{T}^{\,b}\right)$,
where $a$ and $b$ represent any two particles. When the massless $\pTmiss$ vector is used as one of the objects in the computation of the transverse mass, it is omitted from the notation for brevity (e.g. $\mt(\vec{p}_{\text{T}}^{\tau},\pTmiss)$ is written as \mttau). For the case of one particle, the transverse mass is defined as $\mt^2 = E^2 - p_z^2$.

\item The stransverse mass \mttwo~\cite{Lester:1999tx,Lester:2014yga} is used to discriminate between processes where two leptons originate from a resonant decay and processes where they come from non-resonant decays.
The \mttwo\ variable is computed as %
\begin{equation}\nonumber
\mttwo = \min_{\vec{p}_\text{T}^{a} + \vec{p}_\text{T}^{b} = \pTmiss} \left(\max\left[\mt(\vec{p}_{\text{T}}^{\,\text{lep}_1},\vec{p}_\text{T}^{\,a}) , \mt(\vec{p}_{\text{T}}^{\,\text{lep}_2},\vec{p}_\text{T}^{\,b})   \right]\right),
\end{equation}
where $\vec{p}_\text{T}^{\,\text{lep}_1}$ and $\vec{p}_\text{T}^{\,\text{lep}_2}$ correspond to the transverse momenta of the two leading leptons, at least one of which is a $\tau$-lepton,
and $(a,b)$ refers to two invisible particles assumed to be produced with transverse momenta $\vec{p}_\text{T}^{\,a}$ and $\vec{p}_\text{T}^{\,b}$. Several different assumptions on the masses of $a,b$ are used in the analysis ($m_a = m_b$ is always assumed). This is denoted by, e.g. \mttwo ($m_\chi=70$) for the case of $m_a = 70$~\GeV.

\item Object-based \met significance~\cite{ATLAS-CONF-2018-038} is a variable designed to separate events where the measured \met is consistent with being due to detector resolution alone from events where \met is likely to be the result of neutrinos or SUSY particles not interacting with the detector material.
\end{itemize}
Another important variable used in the analysis is \mreco, the invariant mass of the di-$\tau$-lepton system reconstructed in the collinear approximation~\cite{Ellis:1987xu}. The assumption is that the neutrinos from the $\tau$-lepton decays are propagating collinearly with the visible decay products, and that the \met of the event is only due to these neutrinos.
Finally, multiplicities of various physics objects are widely used, in particular $N_\ell$, $N_\tau$, $N_\text{jet}$, and $N_\text{b-jet}$ for the number of light leptons, hadronically decaying $\tau$-leptons, jets, and $b$-jets in an event, respectively.


%

%
%
\section{General analysis strategy}
\label{sec:strategy}
%
%
%

The general analysis strategy is similar for all four channels. Signal regions (SRs) are first defined, which optimise the sensitivity to the signal models, using Asimov signal significance as the figure of merit~\cite{Cowan:2010js}.
Various machine learning (ML) techniques, including deep neural networks (DNN) and boosted decision trees (BDT), are used to enhance the sensitivity. DNNs are used for direct $\tau$-slepton production as the signal events are hard to distinguish from background, requiring fine-tuned ML approaches. For the intermediate $\tau$-slepton channels, the kinematic differences between signal and backgrounds are more pronounced, and the preference is given to BDT architecture, which is generally more robust and easier to interpret. Then control regions (CRs) are used to normalise the event yields of the major SM backgrounds with true $\tau$-leptons to data. Validation regions (VRs) are defined to be kinematically close to SRs, but with the signal contribution suppressed, to evaluate the modelling of the SM backgrounds. Backgrounds with jets misidentified as $\tau$-leptons are modelled in a data-driven way, described in Section~\ref{sec:faketau}. All ISR channels are split into two channels: the \lephad channel (one leptonic and one hadronic $\tau$-lepton decay) and the \hadhad channel (two hadronic $\tau$-lepton decays), while only \lephad final states are considered for the \staunoisr channel. Direct \stau production analyses are described in Section~\ref{sec:Dstau} (ML-based). The direct \stau production in association with an ISR jet is also explored with a cut-and-count analysis, described in Section~\ref{sec:Dstau_cnc}, providing a more re-interpretable result. The intermediate \stau channel analyses are summarised in Section~\ref{sec:IMstau}. A summary of all the channels considered is shown in Table~\ref{tab:exsum}.

\begin{table}[!h]
\centering
\caption{Summary of the different analysis channels.}
\begin{tabular}{l|cc|ccc}
\toprule
Channel & \lephad & \hadhad & ISR Topology & Trigger & Approach \\
\midrule
\staunoisr & $\checkmark$  & $\times$ & $\times$ & single-lepton & multi-class DNN \\
\stauisr & $\checkmark$  & $\checkmark$  & $\checkmark$ & \met & multi-class DNN \\
\stauisr cut-and-count & $\checkmark$  & $\checkmark$ & $\checkmark$ & \met & cut-and-count \\
\midrule
\ConepConem & $\checkmark$  & $\checkmark$  & $\checkmark$ & \met & Binary BDT \\
\ConepmNtwo  & $\checkmark$  & $\checkmark$  & $\checkmark$ & \met & Binary BDT \\
\bottomrule
\end{tabular}
\label{tab:exsum}
\end{table}

\FloatBarrier

\subsection{Statistical analysis}
\label{sec:fitStrategy}
The statistical analysis of the data is performed using the \textsc{pyhf}~\cite{pyhf_joss} library, the Python implementation of \textsc{HistFactory}~\cite{Cranmer:2012sba}. The \textsc{cabinetry} package~\cite{Cranmer:2021oxr} is used for visualisation and diagnostic applications.
A parametric model describing the background and signal prediction, the observed data, and their uncertainties is built for all CRs and SRs (described in Sections~\ref{sec:Dstau}-\ref{sec:IMstau}).
A binned likelihood function $L$ is constructed, a product of Poisson distributions corresponding to the yields in regions of interest and constraint terms describing the systematic uncertainties. A likelihood fit is performed, which minimises the value of $-\log(L)$, simultaneously for all regions of interest included. Signal strength and background normalisation factors (NFs) are allowed to float freely.

Three types of fits are performed, distinguished by the regions included in the fit and the assumptions about the signal, following the \histfitter workflow~\cite{Baak:2014wma}. For all of the fits, \RunTwo and \RunThr data samples are first treated separately in individual fits as a consistency check. Combined fits are then performed on both of the data samples to obtain the most stringent limits, using a common signal strength (where relevant), but separate NFs for \RunTwo and \RunThr data samples.
While the NFs are found to agree between \RunTwo and \RunThr within their uncertainties, this is a more conservative choice allowing for small differences between the setups.

\begin{itemize}
\item
The \textit{background-only} fit is run over the CRs only. The signal is assumed to be absent in the regions, and the signal strength is thus decoupled from the fit.
The free parameters in the fit are the NFs of the SM processes. They are used to extrapolate the results of the fit to the VRs and SRs. The main purpose of this fit is to provide a conservative (and model-independent) evaluation of the background estimate. All direct comparisons of data and MC modelling shown in this \paper are done using background-only fits.

\item In the \textit{model-dependent} (or \textit{exclusion}) fit both the CRs and SRs are included.
The SUSY signal for the particular model considered is taken into account in all regions and is scaled by a floating signal normalisation factor.
The background NFs are also free in the fit, determining the normalisation of the SM backgrounds. These fits are used to construct the exclusion contours that are presented in Section~\ref{sec:result}.

\item
The \textit{model-independent} fit combines a single-bin SR with the CRs. The signal is assumed to only contribute in the SR, allowing this set-up to be treated as a very simple counting experiment that does not depend on the assumptions of the model. The free parameters of the fit are the NFs of the SM backgrounds and the normalisation of the signal. The signal strength derived from the fit is interpreted as the allowed/excluded cross-section of any new physics. These fits are run for each of the single-bin SRs considered individually, with the selection and results further discussed in Section~\ref{sec:result}.
\end{itemize}


%
%
%

%
\section{Direct $\tau$-slepton production, ML-based analysis}
\label{sec:Dstau}

Two orthogonal channels target direct \stau production. The low \met ($<200$~\GeV) phase space is targeted by the \staunoisr channel, while the \stauisr channel is used for $\met \geq 200$~\GeV.
The \staunoisr channel, shown in Figure~\ref{fig:feynmanslepton_DSNoISR}, relies on single-lepton triggers, so only the \lephad final state is considered, while for the \stauisr channel, shown in Figure~\ref{fig:feynmanslepton_DSISR}, both the \lephad and \hadhad final states are considered.
An ML-based approach is employed to enhance the sensitivity to the SUSY signal and to improve the separation between signal and SM backgrounds, leading to CRs with less than $1\%$ signal contamination.
The ML techniques are described in Section~\ref{sec:variables0}, followed by the background estimate in Section~\ref{sec:bgest0}.
The strategy for the cut-and-count approach, which serves as a cross-check, is reported separately in Section~\ref{sec:Dstau_cnc}.

\subsection{Event selection \label{sec:variables0}}
The event selection proceeds in two stages. First a preselection is applied, to ensure orthogonality between channels and to reduce the most obvious background contributions while preserving the signal yields. The summary of the preselection requirements is presented in Table~\ref{tab:stau_presel}. The \staunoisr channel is split based on the presence (\wjet) or lack (\nojet) of at least one jet with $\pt > 40$~\GeV in the event, which was optimised by maximising the Asimov signal significance. The \stauisr selection is split into \lephad and \hadhad channels. After the preselection criteria are applied, the signal sensitivity is further optimised by using ML techniques. The orthogonality between \stauisr and \staunoisr channels is achieved by the requirement on the \met to be larger or smaller than $200$~\GeV, respectively. Requirements on \met significance, stransverse mass, and $\Delta R(\tau,\ell)$ are applied to remove parts of the phase space where no signal is expected. The two-dimensional selections on \mttau and \mtlep, and on angular variables, strongly suppress \Ztautau and \Wjets backgrounds. The selection on \mreco strongly reduces the \Ztautau contributions in the \stauisr channel.

\begin{table}[ht]
\caption{Preselection in the direct \stau channels. The \staunoisr is separated into \wjet and \nojet channels based on the presence of at least one jet with $\pt > 40$~\GeV. The two variables $\phi_1$ and $\phi_2$ are defined as: $\phi_1 = \max(\phi_\ell,\phi_\tau) - \phi_\text{MET}$, $\phi_2 = \phi_\text{MET} - \min(\phi_\ell,\phi_\tau)$. }
\label{tab:stau_presel}
\centering
\footnotesize
\begin{tabular}[ht]{lccc}
\toprule
& \staunoisr  & \multicolumn{2}{c}{\stauisr}  \\
& \nojet / \wjet & \lephad & \hadhad \\
\midrule
Trigger & Single-lepton & \met & \met \\
\midrule
$N_\ell$ & $=1$ & $=1$ & $=0$\\
$N_\tau$ & $\geq 1$ & $\geq 1$ & $\geq 2$ \\
Leading jet \pt[GeV] & $</\geq 40$ & \multicolumn{2}{c}{$\geq 50$} \\
tau-ID & Tight &  \multicolumn{2}{c}{Medium}\\
$N_\text{b-jet}$~(85\% \uppercase{WP}) & $=0$ & \multicolumn{2}{c}{$=0$}\\
Charge $(\tau,\tau) / (\ell,\tau)$ & opposite-sign & \multicolumn{2}{c}{opposite-sign}\\
\midrule
\met[GeV] & $\in [15,200]$ & \multicolumn{2}{c}{$\geq 200$} \\
\met significance & $>2$ &  \multicolumn{2}{c}{--}  \\
\mttwo ($m_\chi=0$) [GeV] & $<90$ &  \multicolumn{2}{c}{--} \\
\mttwo ($m_\chi=70$) [GeV] & -- & $<90$ & --\\
$\Delta R(\tau,\ell)$ & $<3.6$ & $\in[0.6,3.5]$ & -- \\
\mreco[GeV] & -- & -- & $\notin [40,130]$ \\
$\mttau + \mtlep$ [GeV] & $\geq 70$ & \multicolumn{2}{c}{--} \\
$(|\phi_1|-\pi)^2 + (|\phi_2|-\pi)^2$ & $\geq (\pi/2)^2$ & \multicolumn{2}{c}{--} \\
\midrule
(\mttau AND \mtlep) [GeV] & $>20$   & \multicolumn{2}{c}{--} \\
OR (\mttau OR \mtlep) [GeV] & $>90$ & \multicolumn{2}{c}{--} \\
\bottomrule
\end{tabular}
\end{table}

For each of the four channels a separate DNN is trained, with multi-class logistic loss~\cite{068cb646-8936-3df0-aa16-e97573eda53c} used as the loss function. The six classes considered are: the SUSY signal; Top (\ttbar and single-top-quark production combined); \Ztautau; diboson production; various Higgs boson production modes and other sources of true $\tau$-leptons (such as \tth or \ttz); an inclusive sample with fake-tau contributions from all considered backgrounds. The individual weights of the events within one class vary based on the generator weights, cross-sections of the considered processes, and any additional corrections improving data/MC agreement. The total sums of weights of the six classes are then reweighted to the same value, to avoid bias during training (balanced dataset). In the \stauisr channel an additional importance factor of $0.5$ is applied to the \ttbar and \Ztautau datasets. The softmax function~\cite{softmax} is used to combine the results of the multi-class categorisation into six probabilities for each event, such that they are always positive and sum up to one.

Each DNN is optimised using the five-fold cross-validation~\cite{crossval} approach. The Adam optimiser~\cite{DBLP:journals/corr/KingmaB14} is used for all DNNs. The ReLU activation function~\cite{relu_cite} is used in the \staunoisr channel, while the leaky ReLU~\cite{relu_cite} is used in the \stauisr. The hyperparameters of the models (number of layers, number of nodes per layer, learning rate, dropout rate, L2 regularisation, exponential decay rates) and the lists of input parameters are optimised for each channel separately and are presented in Appendix~\ref{app:ml_app}. High signal scores ($>0.9$ for \staunoisr channels, $>0.7$ for ISR channels) are used to define multi-bin SRs, while the lower signal score selection is used to define CRs and VRs using the other classes' scores. The binning of the \RunTwo and \RunThr SRs is determined separately to take into account the difference between the expected yields. In the \stauisr channel, 4 (3) bins are defined in LH SRs, and 11 (2) bins in HH SRs in \RunTwo (\RunThr), optimised to maximise the significance while keeping the MC statistical uncertainties under control. In the \staunoisr channel, the signal distribution is fairly flat against the DNN score, while backgrounds still decrease exponentially with higher score. In each of the sub-channels, 4 (3) bins with optimised width are defined at the high signal score for \RunTwo (\RunThr), approximately in signal score range $[0.98,1.0]$. Then five equidistant bins are defined in the signal score range $[0.9,0.95]$. Finally, the remaining signal score range is divided into bins with width 0.01 (0.02), to fully utilise the difference between signal and background distributions.

\subsection{Background estimation \label{sec:bgest0} }

The key variable used to define CRs and VRs is \bestbkg. The DNN scores of the SM background classes in an event are compared, and \bestbkg is the class corresponding to the largest score. For example, an event with \bestbkg $=$ \enquote{Top} implies that the largest non-signal DNN score for this event corresponds to the \enquote{Top} class. %

\begin{figure}[htbp]
\centering
\subfloat[][]{\includegraphics[width=0.48\linewidth]{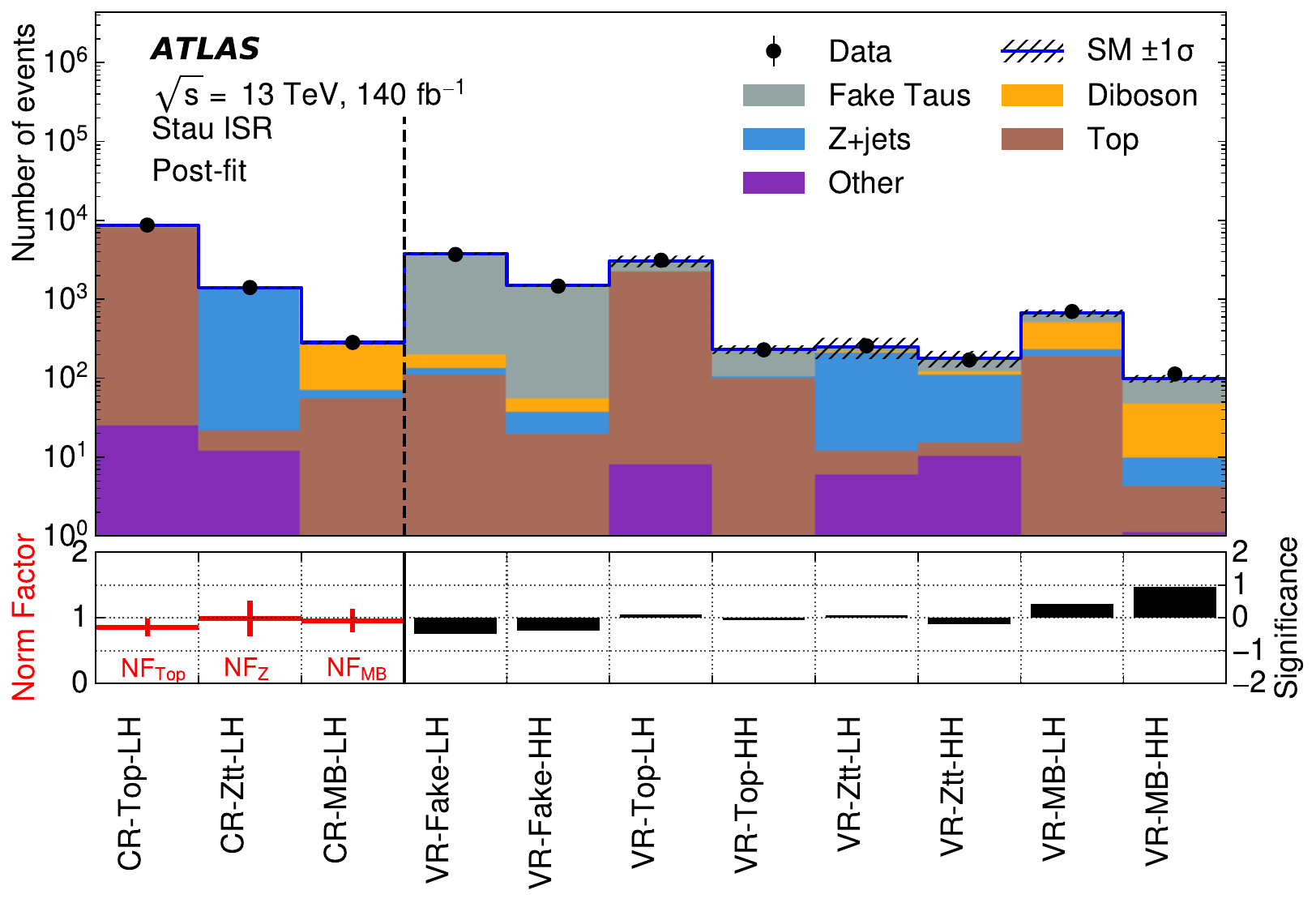}}
\subfloat[][]{\includegraphics[width=0.48\linewidth]{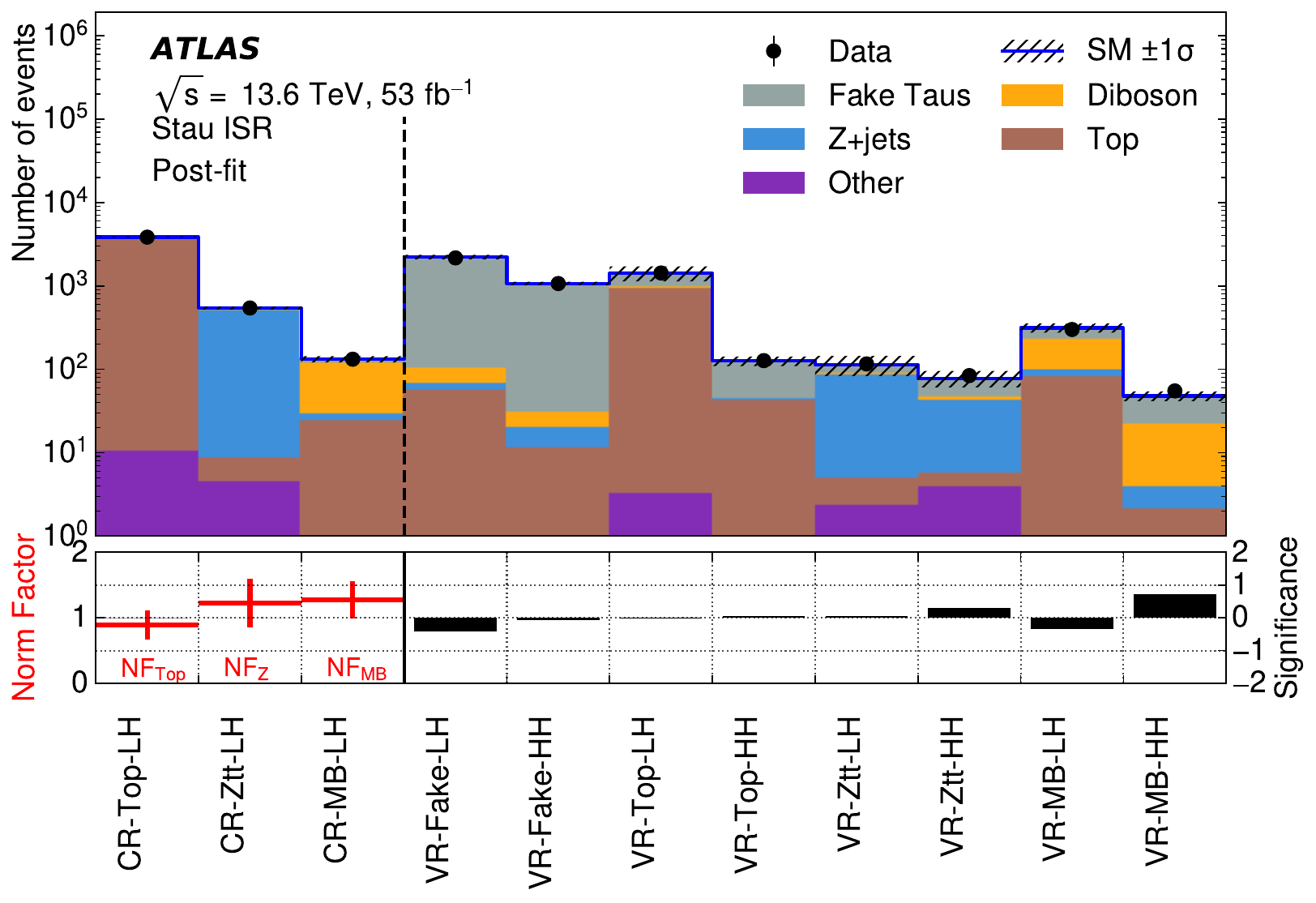}}\\
\subfloat[][]{\includegraphics[width=0.48\linewidth]{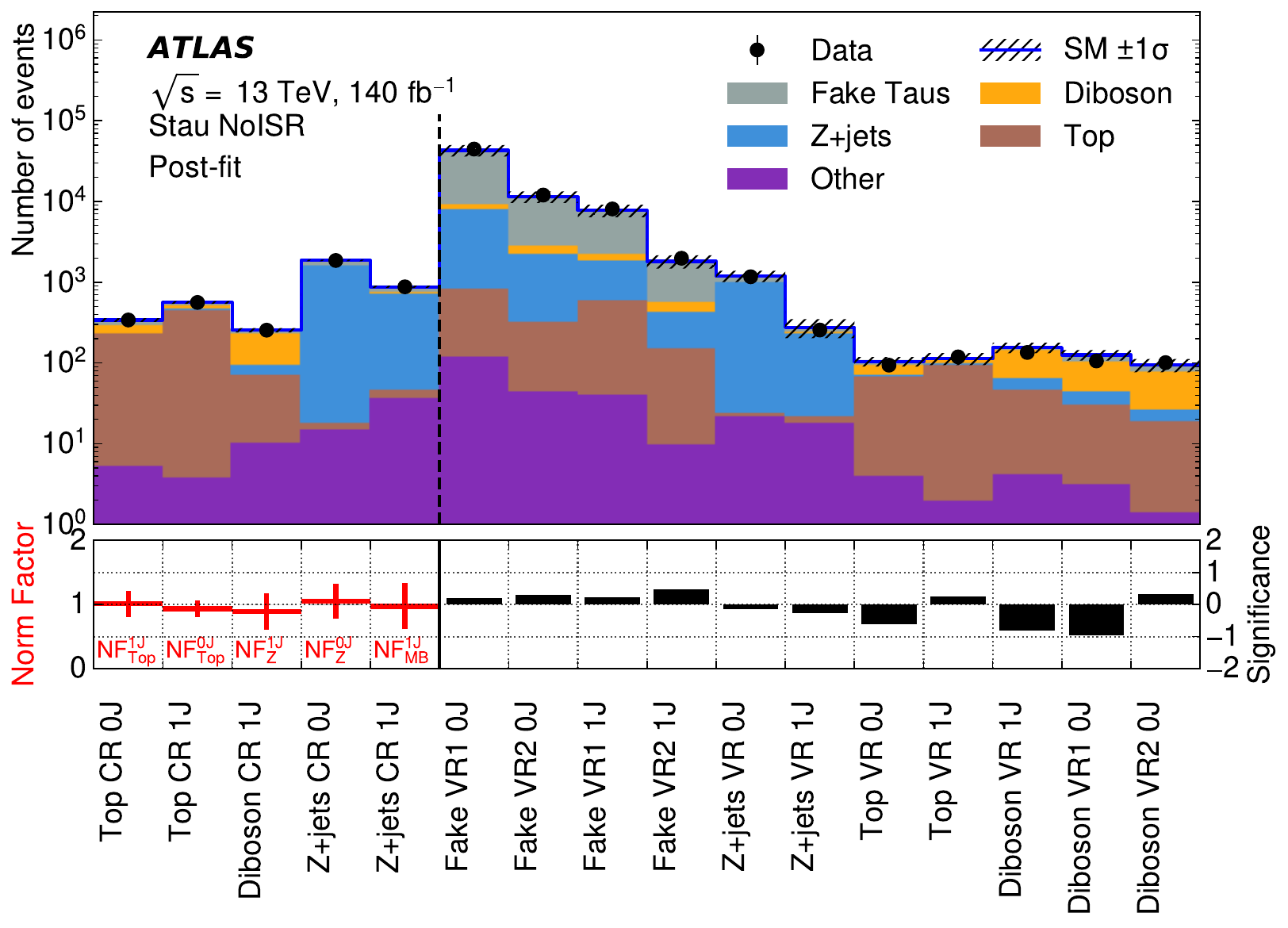}}
\subfloat[][]{\includegraphics[width=0.48\linewidth]{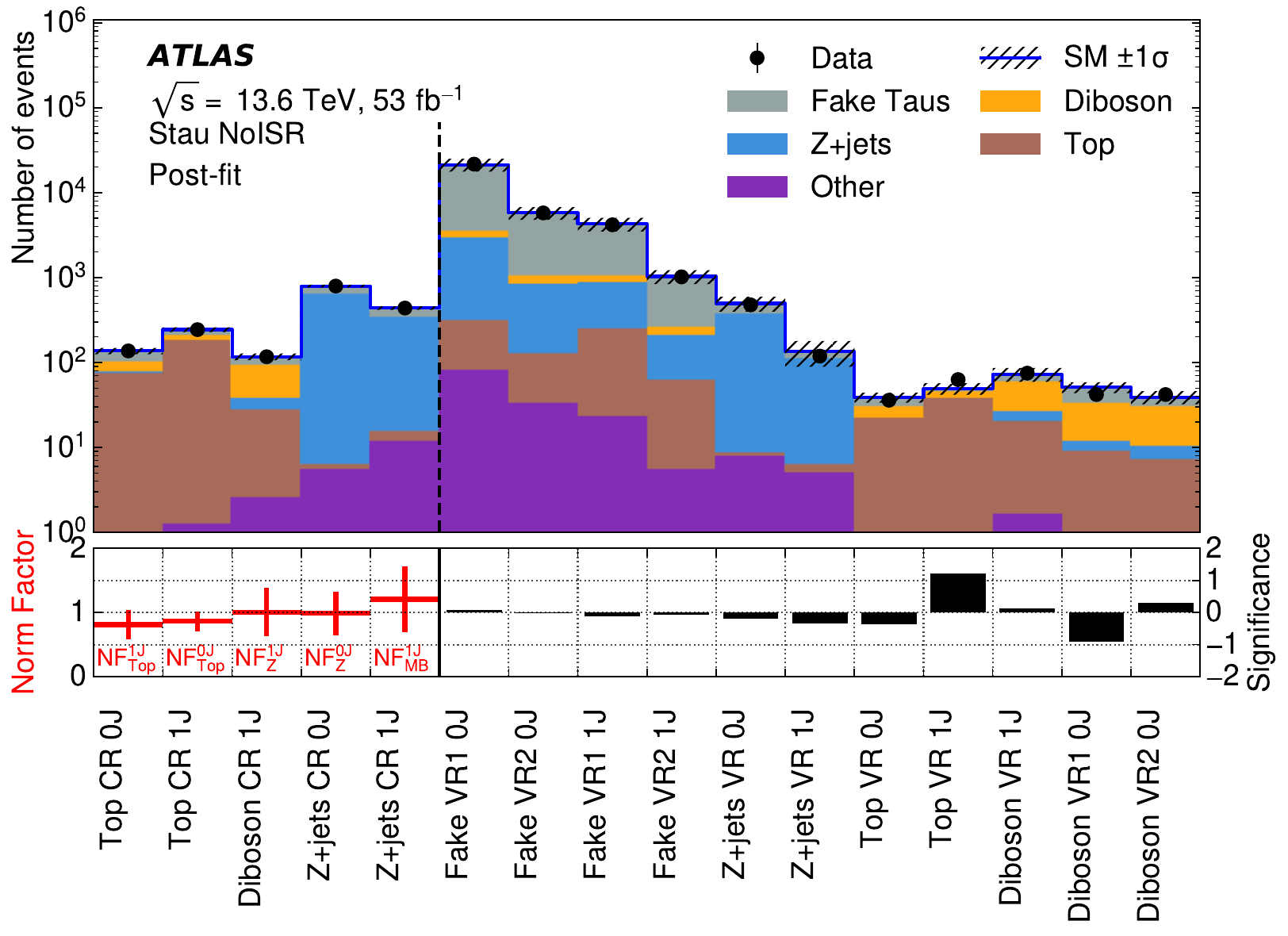}}
\caption{A comparison between the observed and expected yields in the CRs and VRs of the (a) \stauisr \RunTwo channel, (b) \stauisr \RunThr channel, (c) \staunoisr \RunTwo channel, and (d) \staunoisr \RunThr channel. \Others
\BottomPanelSigPlus
\Postfit
\BandIsStatSyst}
\label{fig:noisrstau_crvr}
\end{figure}

In the \stauisr channels, three CRs for each of \RunTwo and \RunThr are used to derive NFs for Top, \Ztautau, and multi-boson (predominantly diboson) MC backgrounds (CR-Top, CR-Ztt, CR-MB, respectively, with \enquote{MB} referring to multi-boson). Only the \lephad channel is used to derive the NFs, as the number of events available in the \hadhad channel is lower, leading to larger statistical uncertainties. When using the \hadhad CRs to derive a set of NFs, these are found to agree with the NFs obtained using the \lephad regions only.
Validation regions are defined for both of the \lephad and \hadhad channels to ensure the quality of the modelling of all main backgrounds and the extrapolation from the CRs to the SRs. The definitions of all regions are gathered in Table~\ref{tab:stauisr_crs}. For the regions enriched in Top events, the preselection criteria are relaxed and the $b$-veto requirement is inverted. Figure~\ref{fig:noisrstau_crvr} shows predicted and observed yields in all the CRs and VRs.

\begin{table}
\centering
\caption{Analysis regions for the \stauisr channels. The selections are applied on top of the preselection criteria of the \stauisr \lephad and \hadhad channels listed in Table~\ref{tab:stau_presel}. For the definition of the $\ttbar$ and single-top enhanced region, some of the preselection requirements are relaxed. The class score always refers to the class identified by \bestbkg, the class corresponding to the highest background DNN score. The $85\%$ \uppercase{WP} is used to count the number of $b$-tagged jets $N_\text{b-jet}$. }
\label{tab:stauisr_crs}
\footnotesize
\begin{adjustbox}{width=1\textwidth}
\begin{tabular}{lcccccccc}
\toprule
& Channel & \bestbkg  & Class Score & Signal score & $N_\text{b-jet}$ & \mttwo ($m_\chi=70$) [GeV] & $\Delta R(\tau,\ell)$ & \mreco[GeV] \\
\midrule
HH SR & \hadhad & -- & -- & $>0.7$ & $=0$ & -- &  -- & $\notin [40,130]$\\
LH SR & \lephad & -- & -- & $>0.7$ & $=0$ & $<90$ & $[0.6,3.5]$ & --\\
\midrule
CR-Ztt-LH & \lephad & \Ztautau & $> 0.6$ & $< 0.7$ & $=0$ & $<90$ & $[0.6,3.5]$ & --\\
CR-MB-LH & \lephad & Diboson & $> 0.7$ & $< 0.7$ & $=0$ & $<90$ & $[0.6,3.5]$ & --\\
CR-Top-LH & \lephad & Top & $> 0.9$  & -- & $\geq 1$ & -- & -- & $\notin [40,130]$\\
\midrule
VR-Ztt-LH & \lephad & \Ztautau &$\in [0.5,0.6]$ & $< 0.7$ & $=0$ &  $<90$ & $[0.6,3.5]$ & --\\
VR-Ztt-HH & \hadhad & \Ztautau & $<0.5$ & $< 0.7$ & $=0$ & -- &  -- & $\notin [40,130]$\\
VR-MB-LH & \lephad & Diboson & $\in [0.45,0.7]$ & $< 0.7$ & $=0$ & $<90$ & $[0.6,3.5]$ & --\\
VR-MB-HH & \hadhad & Diboson & $>0.6$ & $< 0.7$ & $=0$ & -- &  -- & $\notin [40,130]$\\
VR-Top-LH & \lephad & Top & $\in[0.8,0.9]$ & -- & $\geq 1$ & -- &  -- & -- \\
VR-Top-HH & \hadhad & Top & $\in [0.85, 0.9]$ & -- & $\geq 1$ & -- &  -- & -- \\
VR-Fake-LH & \lephad & Fake & $>0.5$ & $< 0.7$ & $=0$ & $<90$ & $[0.6,3.5]$ & --\\
VR-Fake-HH & \hadhad & Fake & -- & $< 0.7$ & $=0$ & -- &  -- & $\notin [40,130]$\\
\bottomrule
\end{tabular}
\end{adjustbox}
\end{table}

\FloatBarrier

\begin{table}
\centering
\caption{Analysis regions for the \staunoisr channels. The selections are applied on top of the preselection criteria of the \staunoisr channels listed in Table~\ref{tab:stau_presel}.}
\label{tab:staunoisr_crs}
\footnotesize
\begin{adjustbox}{width=1\textwidth}
\begin{tabular}{lcccccc}
\toprule
& \bestbkg & Signal score & \Ztautau score & Top score & Diboson score & Fake score \\
\midrule
\wjet SR & -- & $>0.9$ & -- & -- & --  & -- \\
\nojet SR & -- & $>0.9$ & -- & -- & --  & -- \\
\midrule
\Zjets CR \wjet & \Ztautau  &  $ [0.5,0.7]$ & $\geq 0.28$ & -- & --  & -- \\
Top CR \wjet & Top & $ [0.5,0.7]$ & -- & $\geq 0.27$ &  -- & -- \\
Diboson CR \wjet  & Diboson & $ [0.0,0.7]$ & -- & -- & --  & -- \\
\Zjets CR \nojet &\Ztautau  &  $ [0.5,0.7]$ & $\geq 0.35$ & -- & --  & -- \\
Top CR \nojet & Top & $ [0.0,0.85]$ & -- & $\geq 0.05$ &  -- & -- \\
\midrule
\Zjets VR \wjet & \Ztautau & $[0.7,0.9]$ & $\geq 0.165$ & -- & -- & -- \\
Top VR \wjet& Top & $[0.7,0.9]$ & --& $\geq 0.2$ & --  & --\\
Diboson VR \wjet & Diboson & $[0.7,0.9]$ & --  & -- & $\geq 0.06$ & -- \\
Fake VR1 \wjet & Fake & $[0.5,0.7]$ & -- &  -- & -- & $\geq 0.3$ \\
Fake VR2 \wjet  & Fake & $[0.7,0.9]$ &  -- & -- & -- & $\geq 0.2$ \\
\Zjets VR \nojet & \Ztautau & $[0.7,0.9]$ & $\geq 0.165$ & -- & -- & -- \\
Top VR \nojet& Top & $[0.85,0.9]$ & --& $\geq 0.05$ & --  & --\\
Diboson VR1 \nojet & Diboson & $[0.0,0.85]$ & -- & -- & --  & -- \\
Diboson VR2 \nojet & Diboson & $[0.85,0.9]$ & --  & -- & $\geq 0.05$ & -- \\
Fake VR1 \nojet & Fake & $[0.5,0.7]$ & -- &  -- & -- & $\geq 0.3$ \\
Fake VR2 \nojet & Fake & $[0.7,0.9]$ &  -- & -- & -- & $\geq 0.2$ \\
\bottomrule
\end{tabular}
\end{adjustbox}
\end{table}

Five CRs are defined for each of \RunTwo and \RunThr in the \staunoisr channels. The \Ztautau and Top backgrounds are normalised separately in the \wjet and \nojet channels, because of the different kinematic regions selected.
The normalisation of the diboson background is taken from the \wjet only and shared between both of the channels, as the diboson purity in the \nojet channel is very low. Several VRs are used to check the modelling of the main backgrounds and the validity of the extrapolation from the CRs (defined at lower signal score) to the SRs (defined at signal score $>0.9$). All the region definitions are gathered in Table~\ref{tab:staunoisr_crs}. These also include regions enriched in fake $\tau$-leptons, used to validate the data-driven modelling approach. The predicted and observed yields in all CRs and VRs are shown in Figure~\ref{fig:noisrstau_crvr}.

\FloatBarrier


\FloatBarrier
\section{Direct $\tau$-slepton production, cut-and-count analysis}
\label{sec:Dstau_cnc}

An additional cut-and-count analysis is performed in the \stauisr channel, to serve as a cross-check of the ML approach. While the expected sensitivity is weaker than the ML-based approach described in Section~\ref{sec:Dstau}, the advantage of the cut-and-count approach is its ease of interpretability. The presented analysis serves as a proof of concept that with improved calibration and identification algorithms, and using full \RunTwo and \RunThr datasets, it is possible to achieve sensitivity to the compressed region with pure cut-and-count methods. However, it is also clear that ML-based methods outperform cut-and-count in this kinematic regime.

Events used in the cut-and-count analysis are first required to satisfy a preselection similar to that used in the ML-based \stauisr. It is applied to ensure that the \met triggers operate at full efficiency. The preselection criteria are summarised in Table~\ref{tab:stau_cnc}.
A set of additional kinematic requirements are applied on top of the preselection to define a set of SRs.
The SR-specific selections are described in the lower part of Table~\ref{tab:stau_cnc}.
After the preselection, the dominant background is events with fake-taus. To suppress these events, selections on transverse masses of $\tau$-leptons and $\tau$-lepton $\pt$ are effective.
In the \hadhad channel, four SR bins are defined based on \mttauone and \mttautwo values (HH CnC SR), targeting the intermediate $30<\masssplit<50$~\GeV region.
In the \lephad channel, two sets of orthogonal SRs are defined, four SR bins targeting the intermediate scenarios (LH CnC SR), based on \mttauone and $\pt^{\tau_1}$ selections, and two SR bins targeting $\masssplit = 10$~\GeV (LH Comp CnC SR), characterised by large \met significance and large imbalance between \met and the light lepton \pt.

\begin{table}[ht]
\caption{Definition of SRs in the \stauisr cut-and-count analysis. The upper half of the table describes the preselection requirements in the cut-and-count approach, \stauisr channel. The lower half describes additional selection requirements used to define SRs. When the kinematic selections of LH Comp CnC SR and LH CnC SR overlap, the preference is given to the Comp SR, and such events are vetoed from LH CnC SR.}
\label{tab:stau_cnc}
\footnotesize
\centering
\begin{adjustbox}{width=1\textwidth}
\renewcommand{\arraystretch}{1.1}
\begin{tabular}[ht]{l|c|c|c|c|c|c|c|c|c|c}
\toprule
& \multicolumn{4}{c|}{HH CnC} & \multicolumn{2}{c}{LH Comp CnC} & \multicolumn{4}{c}{LH CnC}\\
\midrule
& SR1 & SR2 & SR3 & SR4 & SR1 & SR2  & SR1 & SR2 & SR3 & SR4 \\
Target signal                                                       & \multicolumn{4}{c|}{$30<\masssplit<50$~\GeV} & \multicolumn{2}{c|}{$\masssplit = 10$~\GeV} & \multicolumn{4}{c}{$30<\masssplit<50$~\GeV}\\
\midrule
Preselection \\
\midrule
Trigger                                                & \multicolumn{4}{c|}{\met} & \multicolumn{6}{c}{\met}\\
\met[GeV]                                             & \multicolumn{4}{c|}{$\ge 200$} & \multicolumn{6}{c}{$\ge 200$}\\
tau-ID                                                 & \multicolumn{4}{c|}{Medium} & \multicolumn{6}{c}{Medium}\\
$N_\text{b-jet}$~(85\% \uppercase{WP})                             & \multicolumn{4}{c|}{$=0$} & \multicolumn{6}{c}{$=0$}\\
Charge $(\tau,\tau) / (\ell,\tau)$ & \multicolumn{4}{c|}{opposite-sign} & \multicolumn{6}{c}{opposite-sign}\\
$N_\ell$                                               & \multicolumn{4}{c|}{$=0$} & \multicolumn{6}{c}{$=1$}\\
$N_\tau$                                               & \multicolumn{4}{c|}{$=2$} & \multicolumn{6}{c}{$=1$}\\
\midrule
SRs \stauisr cut-and-count analysis \\
\midrule
$N_\text{jets}$                                         & \multicolumn{4}{c|}{$\in[1,4]$} & \multicolumn{2}{c|}{$\in[1,4]$} & $=1$ & $\in [2,4]$ & $=1$ & $\in [2,4]$ \\
\mreco[GeV]                                        & \multicolumn{4}{c|}{$\geq 250$} & \multicolumn{2}{c|}{$ > 150$} & $>200$ & $>180$ &  \multicolumn{2}{c}{$> 150$}\\
Leading jet \pt[GeV]    		               & \multicolumn{4}{c|}{$> 200$} & \multicolumn{2}{c|}{$> 200$} & $> 150$ & $> 200$ & \multicolumn{2}{c}{$> 150$} \\
$\tau$ \pt[GeV]                               & \multicolumn{4}{c|}{$> 20$} & \multicolumn{2}{c|}{$> 15$} &  \multicolumn{2}{c}{$> 50$} &  \multicolumn{2}{c}{$\in[25,50]$}\\
\mttwo ($m_\chi=70$) [GeV]                               & \multicolumn{4}{c|}{$\in [71,85]$} & \multicolumn{2}{c|}{$\in[71,75]$} &  \multicolumn{4}{c}{$<85$} \\
\mttauone[GeV]                                        & \multicolumn{2}{c}{$\in [80,140]$}  & \multicolumn{2}{c|}{$>140$} & \multicolumn{2}{c|}{--} & \multicolumn{4}{c}{$>40$}\\
\mttautwo[GeV]                                        & $\in [60,100]$ & $>100$ & $\in [60,90]$ & \multicolumn{1}{c|}{$>90$} &\multicolumn{2}{c|}{--} & \multicolumn{4}{c}{--}\\
$\Delta\phi(\tau, \ell)$                               & \multicolumn{4}{c|}{--} & \multicolumn{2}{c|}{$< 2.5$} & \multicolumn{4}{c}{--}\\
$\Delta R(\tau, \ell)$                                 & \multicolumn{4}{c|}{--} & \multicolumn{2}{c|}{--} & $<2.6$ & \multicolumn{3}{c}{$<3.0$}\\
$\Delta\phi(\tau, \text{jet}_1)$                       & \multicolumn{4}{c|}{--} & \multicolumn{2}{c|}{--} &  \multicolumn{4}{c}{$>1.5$}\\
$\Delta \eta(\tau_1, \tau_2)$                          & \multicolumn{4}{c|}{$<1.7$} & \multicolumn{2}{c|}{--} & \multicolumn{4}{c}{--}\\
$\min \Delta \phi(\tau_{1,2}, \text{jets}) $           & \multicolumn{4}{c|}{$>0.5$} & \multicolumn{2}{c|}{--} & \multicolumn{4}{c}{--}\\
$\min \Delta \phi(\met, \text{jets}) $                 & \multicolumn{4}{c|}{$>0.4$} & \multicolumn{2}{c|}{--} & \multicolumn{4}{c}{--}\\
$\min \Delta \phi(\met, \ell) $                        & \multicolumn{4}{c|}{--} & \multicolumn{2}{c|}{$>0.1$} & \multicolumn{4}{c}{--}\\
$\min \Delta R(\tau, \text{jets}) $                    & \multicolumn{4}{c|}{--} & \multicolumn{2}{c|}{--} & \multicolumn{4}{c}{$>1.2$} \\
\met $/ \sqrt{\pt^{\ell}}$ [\GeV$^{0.5}$]    & \multicolumn{4}{c|}{--} & \multicolumn{2}{c|}{$> 50$} & \multicolumn{4}{c}{--} \\
\met significance                                      & \multicolumn{4}{c|}{--} & $\in[14,16]$  & \multicolumn{1}{c|}{$> 16$} & -- & $>13$ & -- & $>10$ \\
$(\pt^{\ell} + \pt^{\tau})/$\met & \multicolumn{4}{c|}{--} & $< 0.3$ & \multicolumn{1}{c|}{$< 0.15$}  & \multicolumn{4}{c}{--}\\
$\pt^{\ell} / \pt^{\tau}$        & \multicolumn{4}{c|}{--} & \multicolumn{2}{c|}{$< 1$} & \multicolumn{4}{c}{--}\\
\bottomrule
\end{tabular}
\end{adjustbox}
\end{table}

\begin{figure}[htbp]
\centering
\subfloat[][]{\includegraphics[width=0.48\linewidth]{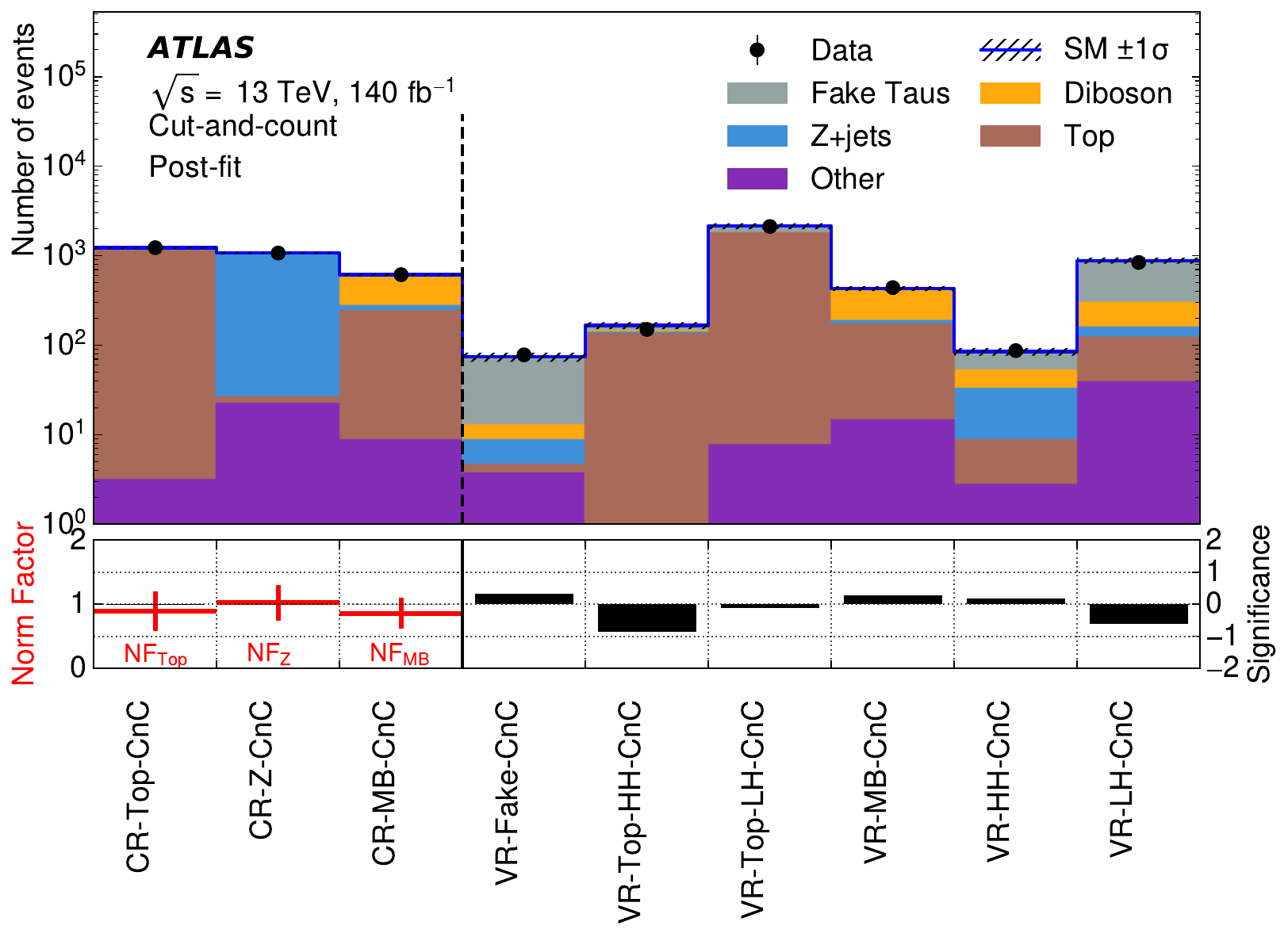}}
\subfloat[][]{\includegraphics[width=0.48\linewidth]{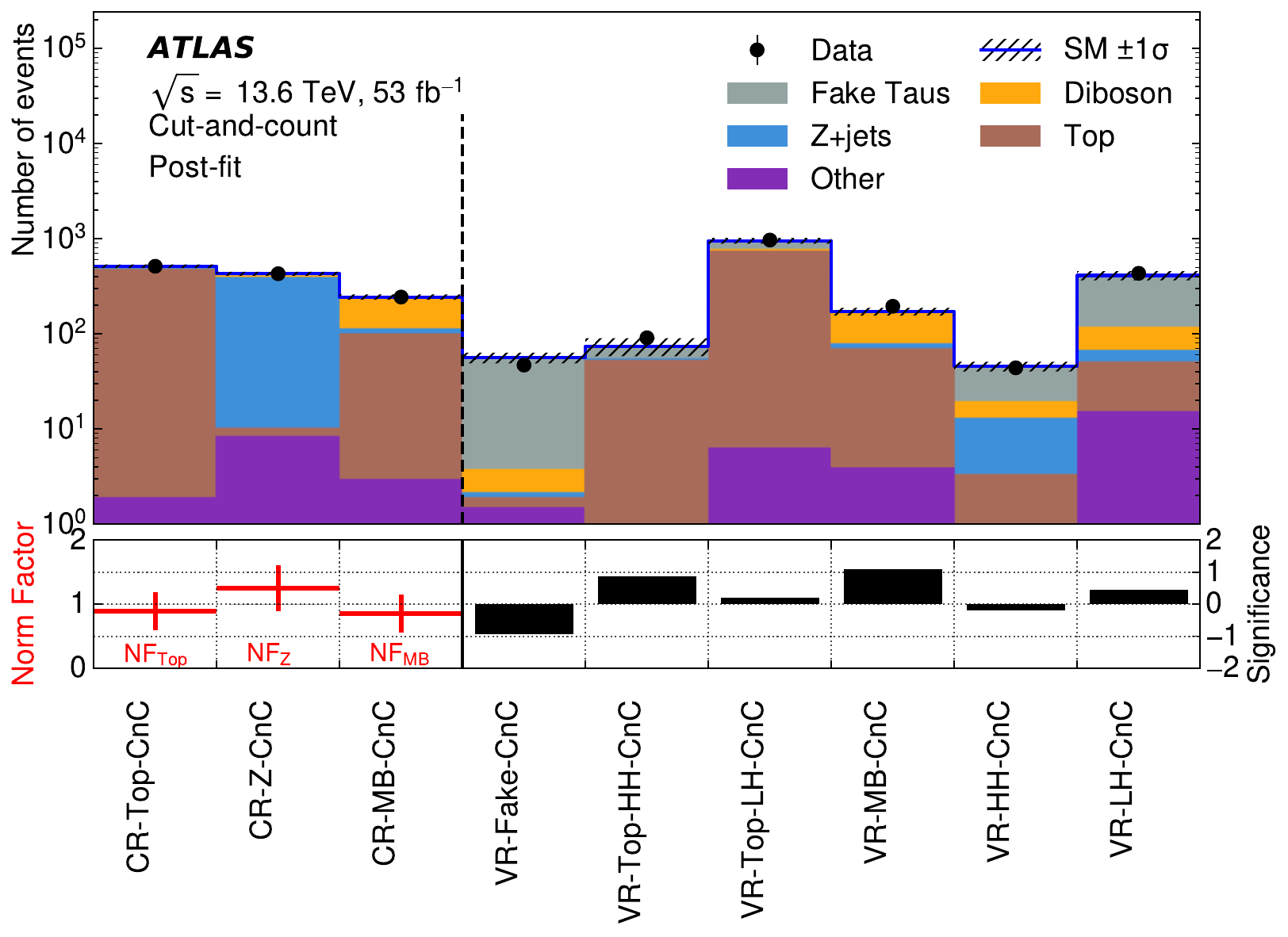}}
\caption{A comparison between the observed and expected yields in the CRs and VRs of the \stauisr cut-and-count channel with the (a) \RunTwo and (b) \RunThr datasets. \Others
\BottomPanelSigPlus
\Postfit
\BandIsStatSyst}
\label{fig:cnc_crvr}
\end{figure}

\FloatBarrier
The background estimate strategy in the cut-and-count channel is generally similar to the \stauisr channel. Top and \Ztautau backgrounds are normalised using CRs in low \mttauone phase space of the \hadhad and \lephad channels, and extrapolated across the channels. In the \leplep channel, used to enhance diboson selection purity, diboson CRs are defined similarly using low \mtlep selection.
The modelling is then validated with VRs aligned to the SRs in high \mttauone/\mtlep regions. All of the selections used are described in Table~\ref{tab:stauisr_cac_crvr}.
For the Top CR, the \lephad channel is used. The $b$-jet veto requirement is inverted, and events with \mreco within $40$~\GeV of the $Z$ boson mass are rejected. Finally, an \mttauone selection is used to separate the Top-enriched region into a CR and a VR\@. Another Top VR is defined in the \hadhad channel, by selecting events with high \mttauone and \mreco. The \Ztautau CR is defined in the \hadhad channel selecting events with di-$\tau$-lepton mass consistent with the $Z$ boson mass and low \mttauone.
Because of the low purity of diboson events and their kinematic similarity to the signal in the \lephad and the \hadhad channels, a two-lepton (\leplep) selection is used.
Events are required to have a pair of same-flavour light leptons, and are separated into a CR and a VR by \mtlep selection. For the diboson CR, events with high transverse mass are selected to suppress $Z$ events. A fake-tau enriched region (VR-Fake) is defined in the \hadhad channel by requiring that the two $\tau$-leptons have the same electric charge. Finally, two general-purpose validation regions are defined in the \hadhad and the \lephad channels (VR-HH and VR-LH), by selecting a phase space similar to the SRs, but with the \mttwo ($m_\chi=70$) selection inverted. A good agreement between data and prediction is observed in all regions, shown in Figure~\ref{fig:cnc_crvr}. The rate at which fake lepton candidates appear in the selection regions (mostly affecting the \leplep channel) is corrected in Monte Carlo samples to improve the agreement with the data.

\begin{table}[!ht]
\centering
\caption{The definition of the control and validation regions in the \stauisr cut-and-count analysis. %
}
\label{tab:stauisr_cac_crvr}
\footnotesize
\begin{adjustbox}{width=1\textwidth}
\renewcommand{\arraystretch}{1.1}
\begin{tabular}{lccccccccc}
\toprule
Region & VR-Fake & CR-Z & CR-MB & VR-MB & CR-Top & VR-Top-LH & VR-Top-HH & VR-HH & VR-LH \\
Channel & \hadhad & \hadhad & \leplep & \leplep & \lephad & \lephad & \hadhad & \hadhad & \lephad \\
\midrule
Charge $(\tau,\tau) / (\ell,\tau) / (\ell,\ell)$ & same-sign & \multicolumn{8}{|c}{opposite-sign}\\
$N_\text{b-jet}$~(85\% \uppercase{WP})    &  \multicolumn{4}{c|}{= 0} & \multicolumn{3}{c|}{$\geq 1$} & \multicolumn{2}{c}{= 0} \\
Leading $\tau$ \pt[GeV] & \multicolumn{2}{c|}{> 20} & \multicolumn{2}{c|}{--} & \multicolumn{4}{c|}{> 20} & > 25 \\
\mttauone[GeV] & > 80 & \multicolumn{1}{c|}{$< 90$} & \multicolumn{2}{c|}{--} & < 40 & > 40 & \multicolumn{2}{|c|}{> 80} & > 40 \\
\mtlep[GeV] & \multicolumn{2}{c|}{--} & $<90$ &\multicolumn{1}{c|}{$> 90$} & \multicolumn{5}{c}{--} \\
\mreco[GeV] & $> 250$ &  $\in[50, 130]$ & $> 150$ & -- & \multicolumn{2}{c|}{$\notin[50,130]$} & \multicolumn{3}{c}{> 150} \\
\mttwo ($m_\chi=70$) [GeV] & \multicolumn{7}{c|}{--} & \multicolumn{2}{c}{$> 85$} \\
\bottomrule
\end{tabular}
\end{adjustbox}
\end{table}

%
%
%
%
%
%
%
%
%
%
%
%
%
%
%
%
%
%
%
%
%
%
%
%
%
%


\FloatBarrier
\section{Intermediate $\tau$-slepton channel}
\label{sec:IMstau}

%
The intermediate stau channel targets two different production mechanisms of electroweak SUSY,
\ConepmNtwoFancy and \ConepConemFancy, in association with an ISR jet, shown in Figures~\ref{fig:feynmanslepton_C1C1}-\ref{fig:feynmanslepton_C1N2}.
Mass-degenerate \ConepmNtwo production implies that the \ConepConem pair production is possible as well, so both of the production modes are considered in the \ConepmNtwo channel. The decay into the lightest neutralino is assumed to proceed only through an intermediate \stau or \tausneut, as shown in Figure~\ref{fig:feynmanslepton_C1C1} and Figure~\ref{fig:feynmanslepton_C1N2}, giving signatures with multiple $\tau$-leptons and \met.
BDT discriminants are trained to target compressed regions of the $m(\chinoonepm/\ninotwo), m(\ninoone)$
phase space. The \stau and \tausneut masses are set to the same value of the average mass of $m(\chinoonepm)$ and $m(\ninoone)$.
The event selection and BDT training procedure are described in Section~\ref{sec:variables1}, followed by the background estimate in Section~\ref{sec:bgesti1}.

\subsection{Event selection \label{sec:variables1}}
Events are selected using an $\met$ trigger, with the requirement that the reconstructed \met must be larger than $200$~\GeV to ensure that the trigger operates at maximum efficiency.
At least one jet with $\pt \geq 100$~\GeV is required to be present in the event, to enhance the sensitivity to the ISR signal topologies. Events with $b$-tagged jets are vetoed to suppress top-quark-induced backgrounds. The reconstructed mass \mreco is required to lie outside of the $[40,130]$~\GeV range to suppress \Zjets backgrounds.
For the \lephad channels, events are required to have at least one light lepton (i.e. an electron or a muon) and $\tau$-lepton candidate of opposite sign (OS). For the \hadhad channels, events must have at least two OS $\tau$-lepton candidates and no light leptons. All $\tau$-leptons are required to satisfy the Medium tau-ID criteria. These selection criteria are summarised in Table~\ref{tab:IMstauSel}. BDT discriminants are trained to separate SM backgrounds from SUSY signal using the Gradient Boosted Decision Trees (GBDT) package~\cite{GBDT}, separately for the \ConepConem and \ConepmNtwo processes, using the events that satisfy these criteria. Input variables used for the training and hyperparameters of the models are summarised in Appendix~\ref{app:ml_app}.

\begin{table}[!h]
\centering
\caption{Preselection requirements in the intermediate stau channel SRs.
\label{tab:IMstauSel}}
\footnotesize
\begin{tabular}{l|cccc}
\toprule
Channel       & \multicolumn{2}{c}{\ConepConem}  & \multicolumn{2}{c}{\ConepmNtwo}  \\
& \lephad & \hadhad & \lephad & \hadhad \\
\midrule
Trigger        & \multicolumn{4}{c}{$\met$ Trigger} \\
$\met$ [GeV] & \multicolumn{4}{c}{$ > 200$ }         \\
$N_\ell$      &  $\ge$ 1 &  = 0   & $\ge$ 1   & = 0 \\
$N_\tau$  &  $\ge$ 1  & $\ge$ 2 &  $\ge$ 1   & $\ge$ 2 \\
tau-ID    & \multicolumn{4}{c}{Medium} \\
Charge $(\tau,\tau) / (\ell,\tau)$ & \multicolumn{4}{c}{opposite-sign}\\
$N_\text{jets}$      & \multicolumn{4}{c}{$\ge$ 1}       \\
Leading jet \pt[GeV] & \multicolumn{4}{c}{$\ge$ 100}                              \\
$N_\text{b-jet}$~(77\% \uppercase{WP})    & \multicolumn{4}{c}{= 0}                                        \\
\mreco[GeV] & \multicolumn{4}{c}{$\notin [40, 130] $ }         \\
Signal score & \multicolumn{2}{c}{$ > 0.9$} & \multicolumn{2}{c}{$>0.7 $} \\
\bottomrule
\end{tabular}
\end{table}

A signal region is defined for each of the channels (\ConepConem and \ConepmNtwo) and sub-channels (\lephad and \hadhad) by applying a requirement on the BDT discriminant score (defined on the $[-1,1]$ interval).
In the \ConepConem channels the SRs are required to have signal score $\geq 0.9$ and three equidistant bins are defined in each SR\@. In the \ConepmNtwo scenario the SRs are required to have signal score $\geq 0.7$ and then split into 9 (7) bins in the \lephad channel and 7 (5) bins in the \hadhad channel for \RunTwo (\RunThr) data samples. The bin width is optimised for best sensitivity for each run individually.
The event selections of the \lephad and \hadhad signal regions are orthogonal and are therefore be statistically combined for interpretation.
In contrast, the \ConepConem and \ConepmNtwo signal regions show significant overlap. Therefore only the \ConepmNtwo SRs are used for the combined \ConepmNtwo and \ConepConem production modes interpretation.

\FloatBarrier
\subsection{Background estimation \label{sec:bgesti1} }

Events containing only true $\tau$-leptons are estimated using MC simulations, normalised to data in dedicated CRs, with signal contamination required to be below $1\%$ for all signal points considered. The extrapolation of the normalisation factors from the CRs to the SRs are tested in VRs. Contributions from events with one or two misidentified $\tau$-leptons are evaluated using a data-driven fake factor method detailed in Section~\ref{sec:faketau} and validated in fake-tau-enriched validation regions.  The full list of selection requirements is summarised in Table~\ref{tab:IMstau_CRVR}. The small selection differences between \RunTwo and \RunThr are driven by the lower integrated luminosity and larger fake-tau contribution of the \RunThr data sample due to the higher \pileup conditions during the data-taking.

\begin{table}[ht!bp]
\centering
\caption{Control and validation regions for the Intermediate stau channel for the different background processes and channels. Common preselection is applied to all channels.}
\label{tab:IMstau_CRVR}
\footnotesize
\resizebox{\textwidth}{!}{
\begin{tabular}{l|ccc|ccc|ccc|cc}
\toprule
\multicolumn{12}{c}{Common preselection}\\
\midrule
Process & \multicolumn{3}{c}{Top} & \multicolumn{3}{c}{Z+jets} & \multicolumn{3}{c}{Diboson} & \multicolumn{2}{c}{Fake} \\
Region  & CR-Top-LH & VR-Top-LH & VR-Top-HH & CR-Ztt-LH & VR-Ztt-LH & VR-Ztt-HH & CR-MB-LH & VR-MB-LH  & VR-MB-HH & VR-Fake-LH & VR-Fake-HH  \\
\midrule
Trigger        & \multicolumn{11}{c}{$\met$ Trigger} \\
$\met\,[GeV]$ & \multicolumn{11}{c}{$ > 200$ }         \\
$N_\text{jets}$      & \multicolumn{11}{c}{$\ge$ 1}       \\
Leading jet \pt[GeV] & \multicolumn{11}{c}{$\ge$ 100}                              \\
$N_\ell$      &  \multicolumn{2}{c}{$\ge$ 1} &  = 0   & \multicolumn{2}{c}{$\ge$ 1}   & = 0 & \multicolumn{2}{c}{$\ge$ 1}   & = 0 &   $\ge$ 1 & = 0\\
$N_\tau$      &  \multicolumn{2}{c}{$\ge$ 1} & $\ge$ 2 &  \multicolumn{2}{c}{$\ge$ 1}   & $\ge$ 2 &  \multicolumn{2}{c}{$\ge$ 1}   & $\ge$ 2 & $\ge$ 1 & $\ge$ 2 \\
Charge $(\tau,\tau) / (\ell,\tau)$ & \multicolumn{3}{c|}{opposite-sign} &  \multicolumn{3}{c|}{opposite-sign} &  \multicolumn{3}{c|}{opposite-sign} &  \multicolumn{2}{c}{opposite-sign} \\
$N_\text{b-jet}$~(77\% \uppercase{WP}) & \multicolumn{3}{c|}{$\geq$ 1} & \multicolumn{3}{c|}{= 0} & \multicolumn{3}{c|}{= 0} & \multicolumn{2}{c}{= 0} \\
$m(\tau_1, \tau_2)$ [GeV] & \multicolumn{3}{c|}{$\notin [40, 130] $ } & \multicolumn{3}{c|}{-- } & \multicolumn{3}{c|}{$\notin [40, 130] $} &  \multicolumn{2}{c}{$\notin [40, 130] $} \\
\midrule
\multicolumn{12}{c}{\ConepmNtwo \RunTwo}\\
\midrule
$\mtlep$ [GeV] & \multicolumn{2}{c}{$\geq 110$} & -- & \multicolumn{3}{c|}{--}  & \multicolumn{3}{c|}{--}   & \multicolumn{2}{c}{--}  \\
$\mttauone$ [GeV] & \multicolumn{2}{c}{--} & $\geq 120$ & \multicolumn{3}{c|}{--}   & \multicolumn{3}{c|}{--}   & \multicolumn{2}{c}{--}  \\
$\Delta R(\tau_1,\tau_2)$ & \multicolumn{3}{c|}{-- } & \multicolumn{2}{c}{$\leq$ 0.6 } & $\leq$ 1.0 & \multicolumn{3}{c|}{-- }  & \multicolumn{2}{c}{--} \\
$\Delta \phi(\tau_1,\met)$ & \multicolumn{3}{c|}{--} & \multicolumn{3}{c|}{--} & \multicolumn{3}{c|}{--}  & \multicolumn{2}{c}{$\geq 2.0$}\\
Signal score & [-1.0,-0.8]& [-0.8,1.0] & [-0.1,1.0] & [-1.0,-0.9]& [-0.9,0.7] & [-0.4,0.4] &\multicolumn{3}{c|}{$\leq 0.7$}  &\multicolumn{2}{c}{$\leq 0.7$} \\
MB score & \multicolumn{3}{c|}{-- } & \multicolumn{2}{c}{$\leq 0$} & $\leq 0.6$ & [0.8,1] &[0.75,0.8] &$\geq$ 0.85 &  \multicolumn{2}{c}{--}\\
\midrule
\multicolumn{12}{c}{\ConepmNtwo \RunThr}\\
\midrule
$\mtlep$ [GeV] & \multicolumn{2}{c}{$\geq 110$} & -- & \multicolumn{3}{c|}{--} & \multicolumn{3}{c|}{--} & \multicolumn{2}{c}{--} \\
$\mttauone$ [GeV] & \multicolumn{2}{c}{--} & $\geq 150$ & \multicolumn{3}{c|}{--}   & \multicolumn{3}{c|}{--}   & \multicolumn{2}{c}{--}  \\
$\Delta R(\tau_1,\tau_2)$ & \multicolumn{3}{c|}{-- } & \multicolumn{2}{c}{$\leq$ 0.8 } & $\leq$ 1.0 & \multicolumn{3}{c|}{-- } & \multicolumn{2}{c}{-- } \\
$\Delta \phi(\tau_1,\met)$ & \multicolumn{3}{c|}{--} & \multicolumn{3}{c|}{--} & \multicolumn{3}{c|}{--}  & \multicolumn{2}{c}{$\geq 2.0$}\\
Signal score & [-1.0,-0.8]& [-0.8,1.0] & [-0.1,1.0] & [-1.0,-0.9]& [-0.9,0.7] & [-0.8,0.4] &\multicolumn{3}{c|}{$\leq 0.7$} &\multicolumn{2}{c}{$\leq 0.7$} \\
MB score & \multicolumn{3}{c|}{-- } & \multicolumn{2}{c}{$\leq 0$} & $\leq 0.6$ & [0.8,1] &[0.75,0.8] &$\geq$ 0.8 &  \multicolumn{2}{c}{--}\\
\midrule
\multicolumn{12}{c}{\ConepConem \RunTwo and \RunThr} \\
\midrule
$\mtlep$ [GeV] & \multicolumn{2}{c}{$\geq 120$} & -- & \multicolumn{3}{c|}{--} & \multicolumn{3}{c|}{--} & \multicolumn{2}{c}{--}  \\
$\mttauone$ [GeV] & \multicolumn{2}{c}{--} & $\geq 200$ & \multicolumn{3}{c|}{--} & \multicolumn{3}{c|}{--} & \multicolumn{2}{c}{--}   \\
$\Delta R(\tau_1,\tau_2)$ & \multicolumn{3}{c|}{--} & \multicolumn{2}{c}{--} & $<1$ & \multicolumn{3}{c|}{--} & \multicolumn{2}{c}{--}  \\
$\Delta R(\tau_1,\ell)$ &  \multicolumn{3}{c|}{--} &  \multicolumn{2}{c}{$< 1$} & -- & \multicolumn{3}{c|}{--} & \multicolumn{2}{c}{--} \\
Signal score & $\in [-1.0,-0.8]$ & $\in [-0.8,1.0]$ & $\in [0,1.0]$ & $\in [-1.0,-0.8]$ & $\in [-0.8,0.9]$ & $\in [-0.9,0.9]$ &\multicolumn{3}{c|}{$\in [-1,0.9]$} &\multicolumn{2}{c}{$\in [-1,0.9]$}  \\
MB score & \multicolumn{3}{c|}{--} & \multicolumn{3}{c|}{--} &  $\in [0.65,1]$ & $\in[0.5,0.65]$ & $[0.7,1.0]$ & $\in [-0.6,-0.1]$ & $\in[-0.2,0.6]$ \\
\bottomrule
\end{tabular}
}
\end{table}

The dominant SM backgrounds with true $\tau$-leptons in both the \hadhad and \lephad channels originate from diboson production. A similar strategy is used for both the \ConepConem and \ConepmNtwo scenarios to improve the purity of the CRs. A separate binary BDT discriminant is trained to identify diboson events, with a training strategy nearly identical to that employed for the signal BDT score in the \lephad and \hadhad channels, but now treating diboson as the signal against the remainder of the SM backgrounds. The classifier output (MB score) is used to define diboson-enriched CRs in the \lephad channel (CR-MB-LH) and VRs in both the \lephad and \hadhad channels (VR-MB-LH/HH). The CR-MB-LH and VR-MB-LH are separated using the MB score, and a signal score below 0.7 (0.9) is required to ensure orthogonality with the \ConepmNtwo (\ConepConem) SRs.

Top-enriched CRs (CR-Top-LH) require at least one $b$-tagged jet ($77\%$ \uppercase{WP}) to enhance top purity and ensure orthogonality with the SRs, and the control and validation regions of other background processes. Requirements on the transverse mass of the light lepton (\lephad channel) or subleading $\tau$-lepton (\hadhad channel) are applied to further increase the purity of Top backgrounds. The modelling of these processes is validated in two top-enriched validation regions. The CRs and VRs are separated using a requirement on the signal score. %

The \Ztautau contribution is negligible in both of the signal regions. However, it contributes to some VRs and therefore dedicated \Ztautau CRs are defined. Two \Ztautau VRs are also defined, VR-Ztt-LH and VR-Ztt-HH, similar to the other backgrounds. A requirement on $\Delta R(\tau_1,\tau_2/\ell)$ is applied to enhance the purity of the \Ztautau selection (as the $Z$ boson is expected to be Lorentz-boosted in the ISR topologies considered), while a signal score below 0.7 (0.9) ensures orthogonality with the \ConepmNtwo (\ConepConem) SRs. The MB score selection is used to suppress diboson contributions and ensure orthogonality with the diboson CRs and VRs. %

\begin{figure}[htbp]
\centering
\subfloat[][]{\includegraphics[width=0.48\linewidth]{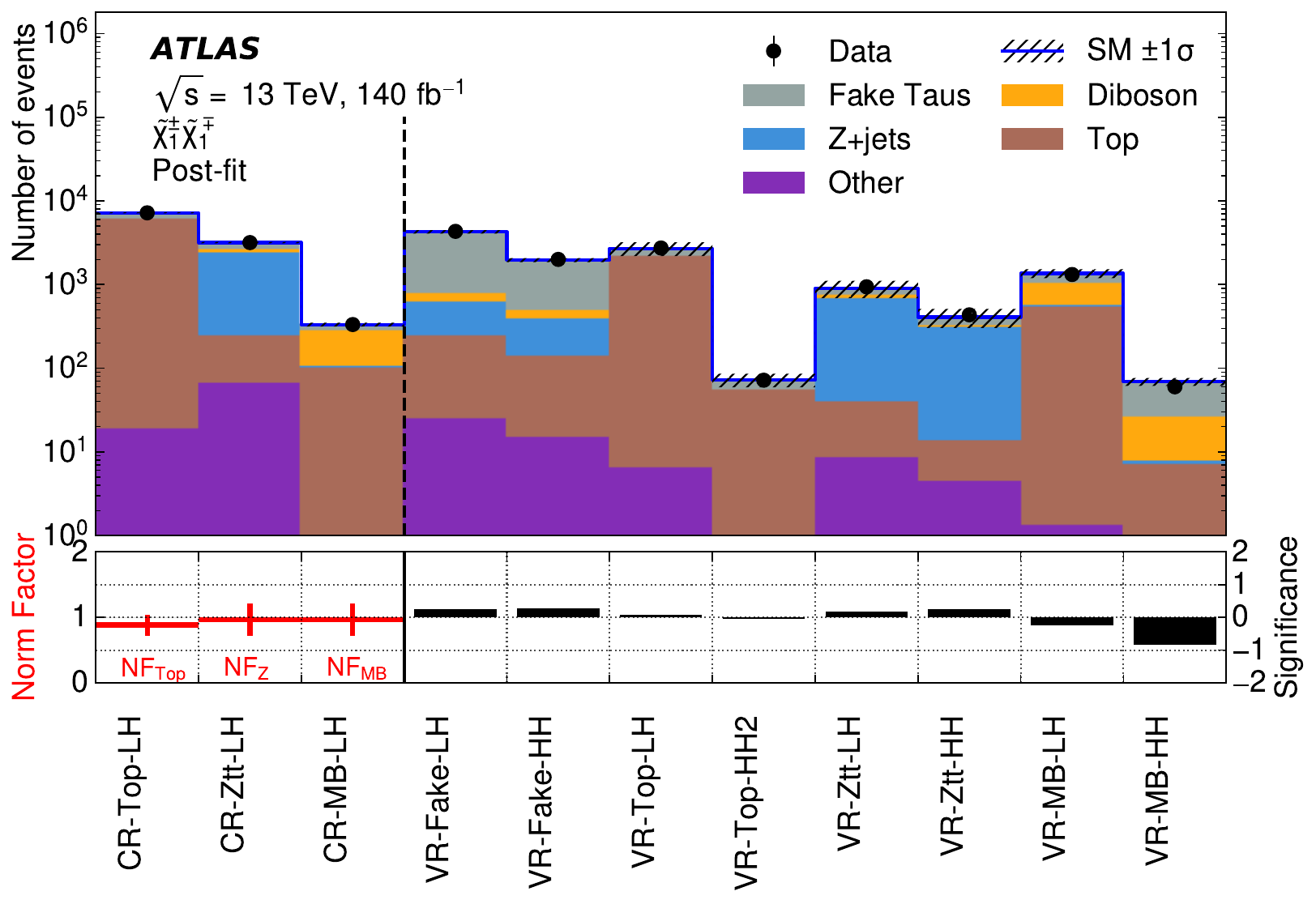}}
\subfloat[][]{\includegraphics[width=0.48\linewidth]{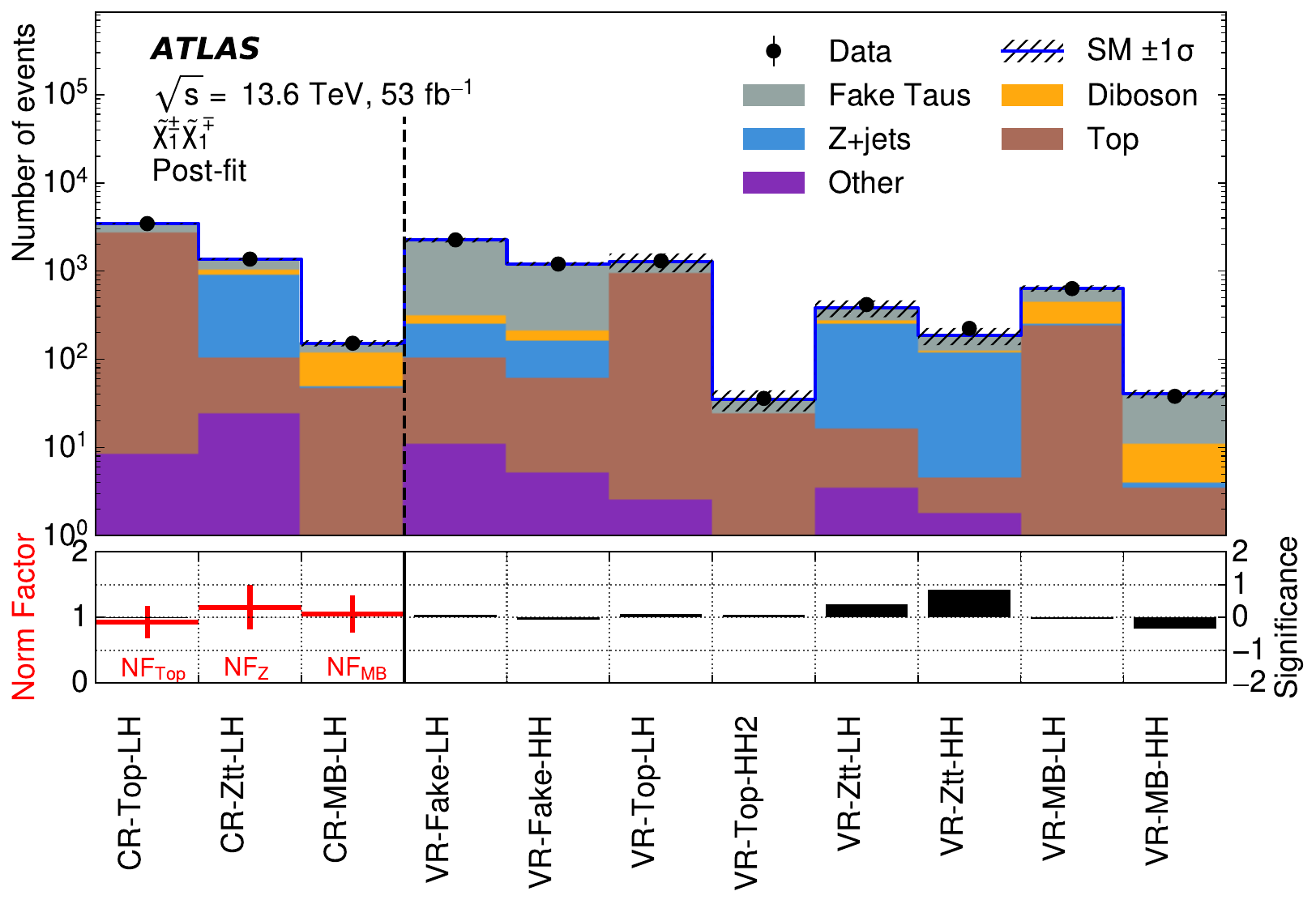}}\\
\subfloat[][]{\includegraphics[width=0.48\linewidth]{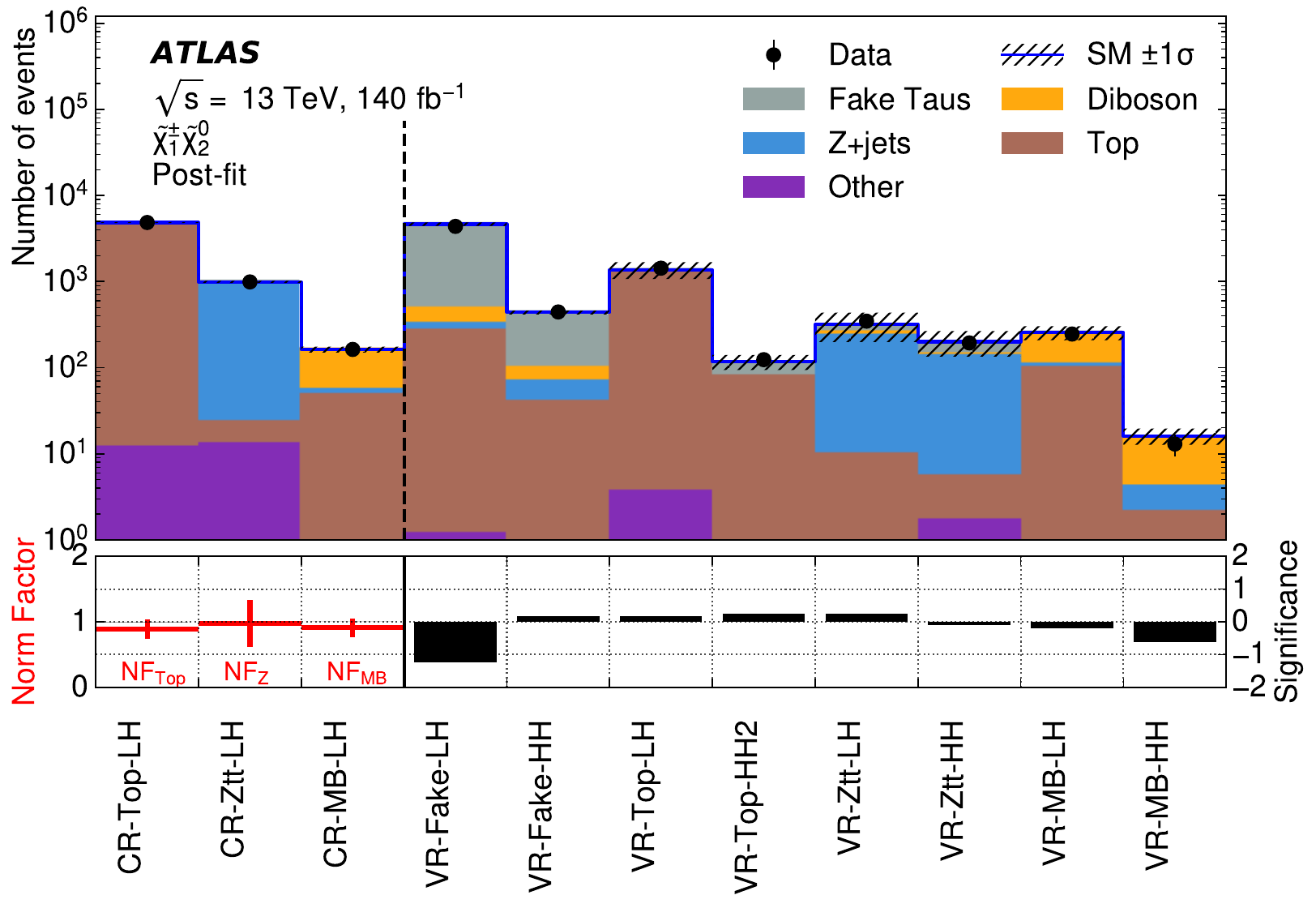}}
\subfloat[][]{\includegraphics[width=0.48\linewidth]{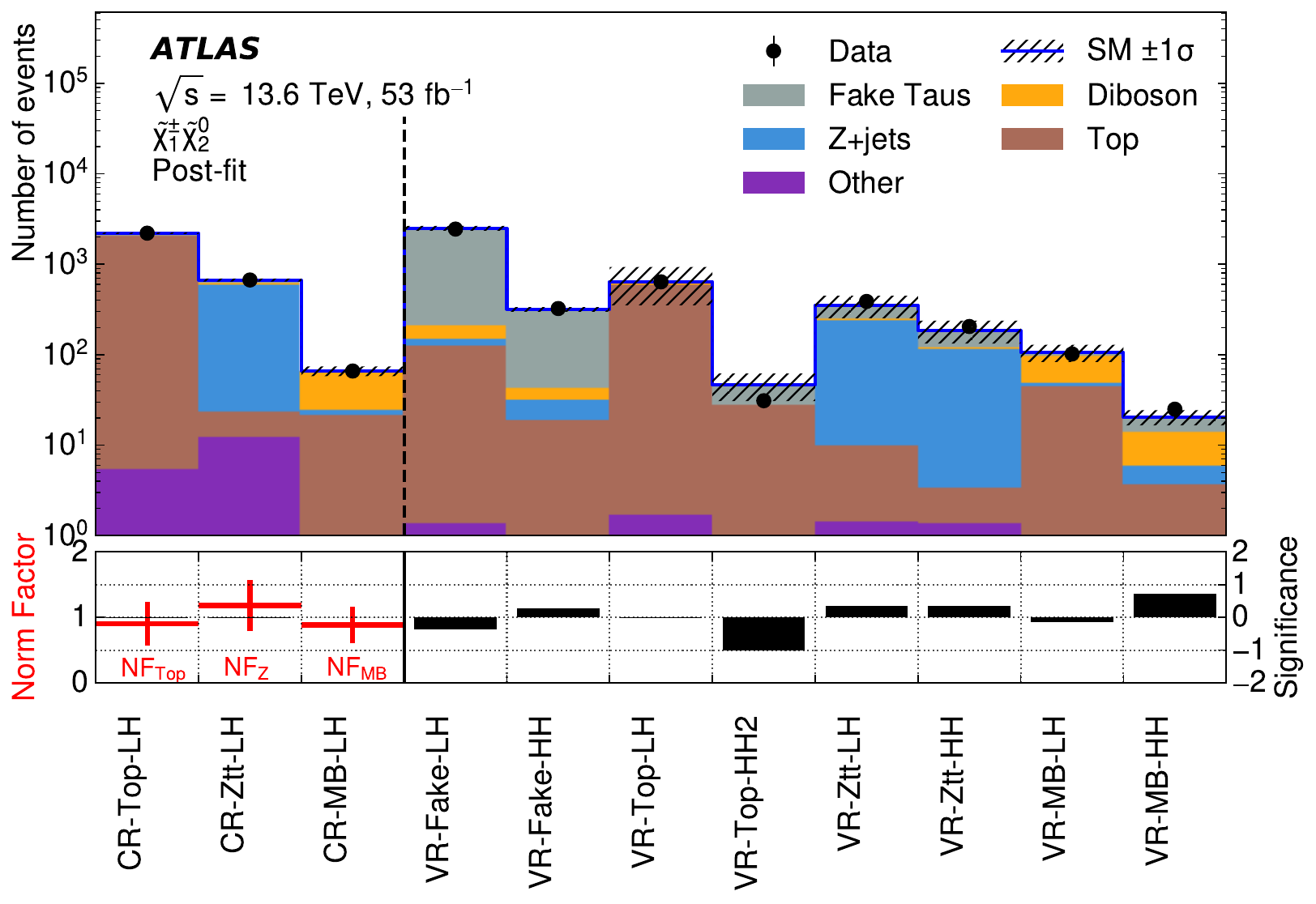}}
\caption{A comparison between the observed and expected yields in the CRs and VRs of the (a) \ConepConem \RunTwo channel, (b) \ConepConem \RunThr channel, (c) \ConepmNtwo \RunTwo channel, and (d) \ConepmNtwo \RunThr channel. \Others
\BottomPanelSigPlus
\Postfit
\BandIsStatSyst}
\label{fig:c1n2_crvr}
\end{figure}

Events with at least one jet misidentified as a $\tau$-lepton are the largest background in all SRs. Two fake-tau validation regions, VR-Fake-LH and VR-Fake-HH, are defined to validate the estimations obtained from the fake factor method, as described in Section~\ref{sec:faketau}. A requirement of $\Delta \phi(\tau_1,\pTmiss) \geq 2.0$ is applied to enhance the purity of the fake-tau background, and signal score below 0.7 (0.9) is imposed to ensure orthogonality with the \ConepmNtwo (\ConepConem) SRs.

The comparisons between SM backgrounds and observed data yields are shown in Figure~\ref{fig:c1n2_crvr}, following a background-only fit to the data as detailed in Section~\ref{sec:fitStrategy}. Overall, the background estimation procedure provides a good description of the data in all validation regions for the intermediate stau channel.

\FloatBarrier


\FloatBarrier
\section{Fake $\tau$-lepton modelling}
\label{sec:faketau}

%
In this Section, the data-driven estimation of the fake-tau background is described. This background is composed of events with at least one quark- or gluon-induced jet that satisfy the tau ID selection requirements. They are modelled through the fake factor method~\cite{HIGG-2017-07}, irrespective of which process originated the fake tau. In the analysis phase space, the vast majority of the fake taus (around $90\%$) come from $W$+jets production, with the $W$ boson decaying leptonically, and one of the jets being misidentified as a tau. While multijet processes have a very large cross-section and can in principle mimic the di-$\tau$ events if both jets are misidentified as $\tau$-leptons, such events are effectively suppressed by the preselection requirements (described further in Sections~\ref{sec:Dstau} and~\ref{sec:IMstau}).

The fake factors (FFs) are defined as:
$$\text{FF} = \frac{\text{Number of fake taus passing tau ID criteria}}{\text{Number of fake taus passing anti-ID criteria}}$$
as a function of several $\tau$-lepton kinematic variables (\pt, $|\eta|$) and number of charged tracks. The tau ID criteria change between channels considered in this analysis (Tight for \staunoisr, Medium for the ISR channels) and the FFs are computed accordingly. The $\tau$-leptons that satisfy the Very Loose WP, but not the tighter ID selection, are referred to as anti-ID $\tau$-leptons. For each analysis region, an anti-ID region is defined. All selections remain the same, except that at least one of the $\tau$-leptons has to be an anti-ID $\tau$-lepton. The contributions from true $\tau$-leptons are modelled through MC simulation and subtracted. The resulting data-minus-simulation mixture is scaled by the FFs as a function of anti-ID $\tau$-lepton kinematic properties, estimating the fake-tau background. In the \hadhad channels, the contributions from events with either one or two anti-ID taus are considered, with opposite signs to avoid double-counting, following the combinatorial formula described in Ref.~\cite{TAUP-2023-01}.

\begin{table}[ht]
\centering
\caption{The selection criteria of the fake factor measurement region (FF CR).}
\footnotesize
\label{tab:stauisr_ffcr}
\begin{tabular}{lcc}
\toprule
& Anti-ID FF CR & ID FF CR \\
\midrule
Medium $\tau$-leptons & 0 &  1 \\
Anti-ID $\tau$-leptons & 1 & 0 \\
$N_\ell$ & \multicolumn{2}{c}{1} \\
$N_\text{b-jet}$~(85\% \uppercase{WP}) &  \multicolumn{2}{c}{$=0$}\\
Charge $\ell$,$\tau$ & \multicolumn{2}{c}{same-sign} \\
\met trigger & \multicolumn{2}{c}{pass} \\
\met[GeV] & \multicolumn{2}{c}{$\geq 200$}\\
Leading jet \pt[GeV] & \multicolumn{2}{c}{$\geq 50$}\\
\bottomrule
\end{tabular}
\end{table}

The final states considered by the analyses in this \paper require that the two leptons have opposite charges. This requirement is inverted in the ISR channels to measure the fake factors in a region that is kinematically close to the analysis phase space, but with no expected signal contribution. The full selection is defined in Table~\ref{tab:stauisr_ffcr}. The seed partons of the fake taus were studied using simulated events and found to be consistent between the opposite-sign and same-sign selections within the uncertainties. In particular, the high $\met$ selection suppresses the contribution from gluon-originating jets.

In the \staunoisr phase space, characterised by the lower $\met$ selection, the contribution of fake taus from gluons differs significantly between the same-sign and opposite-sign regions, and a different approach is used to measure the FFs. The fake factors are derived from simulated events in analysis regions, binned in $\pt$, $|\eta|$, $\met$, and number of charged tracks, using the information of the fake tau origin. These FFs are then combined according to the expected distribution of the fake taus origins, computed for each region. An additional data-driven correction is derived in the relaxed preselection region, with only \met, number of $b$-jets, lepton, and classifier score selections applied, to account for any differences between tau ID performance between data and MC\@. The resulting FFs are applied to the data in the anti-ID regions, in the same way as in the ISR channels.


\FloatBarrier
\section{Systematic uncertainties}
\label{sec:systematics}

%
Systematic uncertainties have an impact on the background and signal estimates in all analysis regions. Three main sources of uncertainties are considered:
\begin{itemize}
\item Uncertainties arising from experimental effects, affecting the reconstruction, identification, and calibration applied to the physical objects considered by the analysis;
\item Theoretical uncertainties related to the MC generators and PDF sets used for background and signal modelling;
\item Uncertainties related to the limited number of data and MC events available, affecting primarily the data-driven fake-tau estimate.
\end{itemize}
Detector uncertainties are correlated between \RunTwo and \RunThr whenever a common approach was used to derive them.

The experimental uncertainties are propagated consistently to all objects to which they apply. Jet uncertainties include effects from the energy scale calibration~\cite{PERF-2016-04} and resolution~\cite{JETM-2018-05}, and the NNJVT jet-vertex-tagging uncertainties~\cite{ATLAS-CONF-2014-018}. The jet flavour-tagging efficiency uncertainties~\cite{FTAG-2023-10,FTAG-2023-05,FTAG-2024-07} are also considered. Experimental uncertainties for $\tau$-leptons consist of uncertainties from reconstruction and identification algorithms, energy scale calibrations, and the electron veto~\cite{ATL-PHYS-PUB-2015-045,ATLAS-CONF-2017-029}. The experimental uncertainties for electrons and muons include uncertainties from reconstruction, calibration, isolation, and identification algorithms~\cite{EGAM-2018-01,EGAM-2021-02,MUON-2018-03,MUON-2022-01}.
The resolution and energy scale uncertainties for all the objects used in the analysis are propagated into the \met calculations, and uncertainties in the modelling of the underlying event and tracks not matched to any of the reconstructed objects are used to compute dedicated soft-term uncertainties~\cite{JETM-2020-03}.
The uncertainty in the combined \RunTwo(\RunThr) integrated luminosity is $0.83\%~(2.0\%)$~\cite{LUCID2,DAPR-2021-01,ATL-DAPR-PUB-2024-001}, obtained from luminosity measurements with the LUCID-2 detector. A variation in the \pileup reweighting of the MC simulated event samples is included to cover the uncertainty in the ratio of the predicted and measured inelastic cross-section~\cite{Aaboud:2016mmw}.

Theoretical uncertainties affecting the main true $\tau$-leptons backgrounds are estimated by varying the generator parameters. Uncertainties due to the missing higher-order corrections are estimated by using the $7$-point scale-variation of renormalisation and factorisation scales (\muR and \muF) , where $\{\muR, \muF\}$ are varied from their nominal values by factors of $\{0.5,0.5\}$, $\{0.5,1\}$, $\{1,0.5\}$, $\{2,1\}$, $\{1,2\}$, and $\{2,2\}$. For the \NNPDF sets the PDF4LHC prescription is followed~\cite{Butterworth:2015oua}. Theoretical uncertainties related to final-state radiation variations, variations of the A14 tune due to different \alphas values for initial-state radiation, and variations of the merging parameter and resummation scale in \SHERPA-generated samples~\cite{Catani:2001cc,Hoeche:2009rj} are considered where appropriate. For \ttbar direct comparisons with alternative samples (produced with \POWHER~\cite{Bahr:2008pv,Bellm:2015jjp}, and with nominal \POWPY with \textsc{pThard} and $h_\mathrm{damp}$ values varied) are used as additional uncertainties. For the single-top-quark production, a comparison is made between samples produced using the diagram removal (nominal) and diagram subtraction schemes.

The fake-tau modelling uncertainties include the effect of the limited data yields in the measurement and application regions, and the uncertainties in the subtraction of the true $\tau$-lepton contribution. The difference between the quark/gluon contributions to the fake-tau between measurement and signal regions is studied and found to be in agreement within the considered uncertainties, and no additional ad hoc uncertainties are used. In the \staunoisr channel the statistical modelling uncertainty in the MC samples used to derive the FFs and process fractions of the fake-tau sources are taken into account as well.

\begin{figure}[htbp]
\centering
\subfloat[\stauisr uncertainties]{\includegraphics[width=0.95\linewidth]{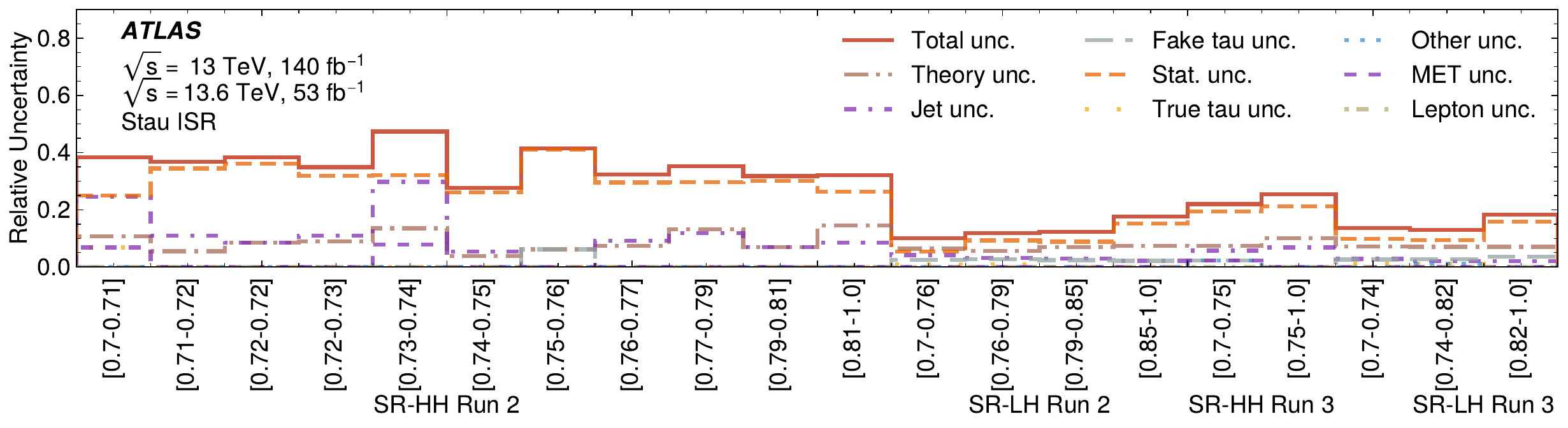}}\\
\subfloat[Cut-and-count uncertainties]{\includegraphics[width=0.95\linewidth]{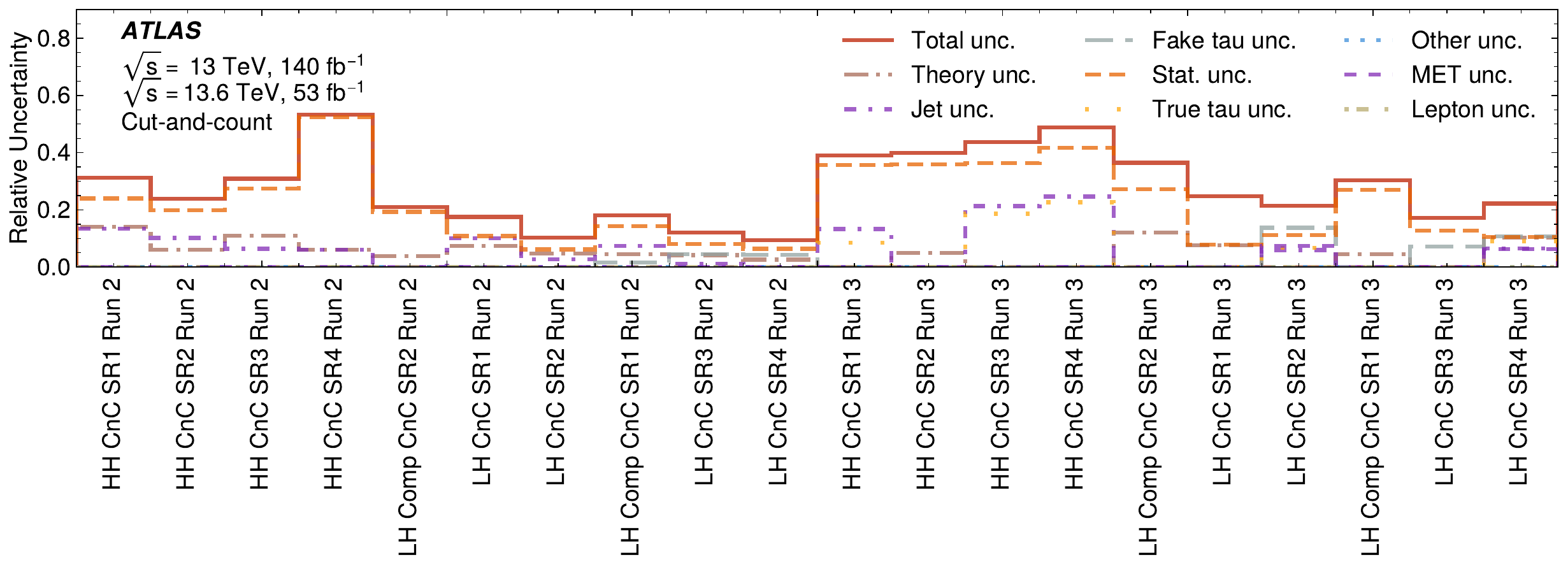}}\\
\subfloat[\ConepConem uncertainties]{\includegraphics[width=0.6\linewidth]{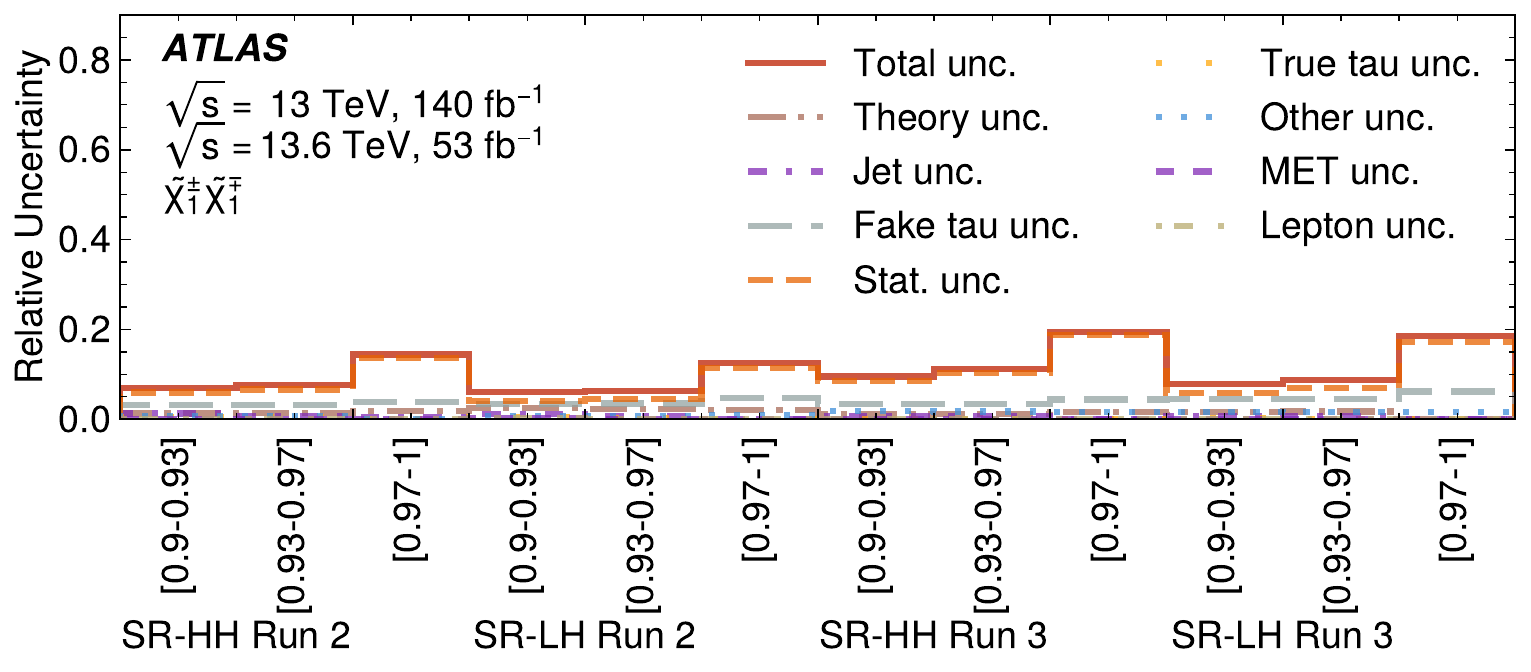}}\\
\subfloat[\ConepmNtwo uncertainties]{\includegraphics[width=0.95\linewidth]{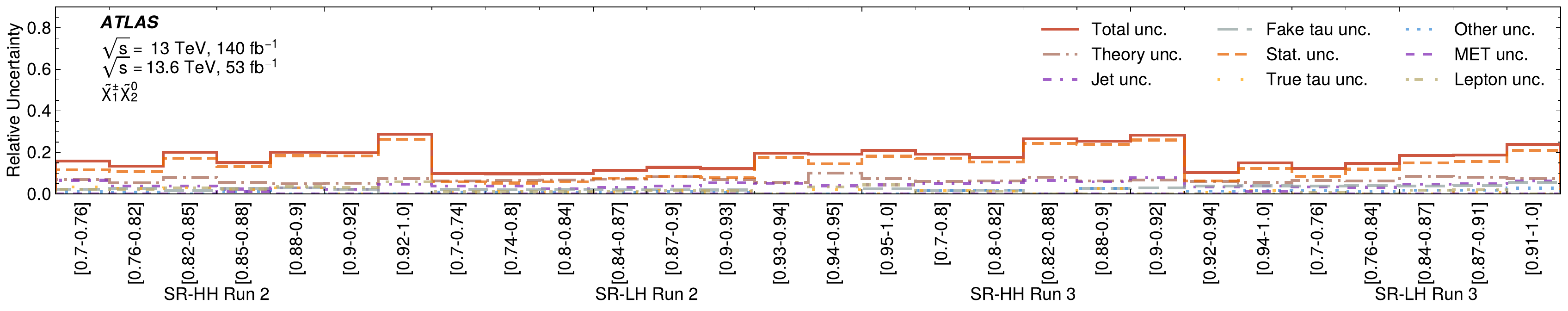}}
\caption{Comparison of the dominant groups of systematic uncertainties in the total background prediction after the background-only fit in the analysis SRs, relative to the total yields. The total uncertainty is calculated using all the available systematic and statistical uncertainties. \enquote{Other} refers to $b$-tagging and \pileup uncertainties. \enquote{True tau} refers to the experimental uncertainties for $\tau$-leptons. The individual uncertainties can be correlated and do not necessarily add in quadrature to the total uncertainty. Statistical uncertainty refers only to the uncertainties of the backgrounds and does not include the effects of the data Poisson uncertainty.}
\label{fig:systematics}
\end{figure}

\begin{figure}[htbp]
\centering
\subfloat[\nojet SR \RunTwo]{\includegraphics[width=0.95\linewidth]{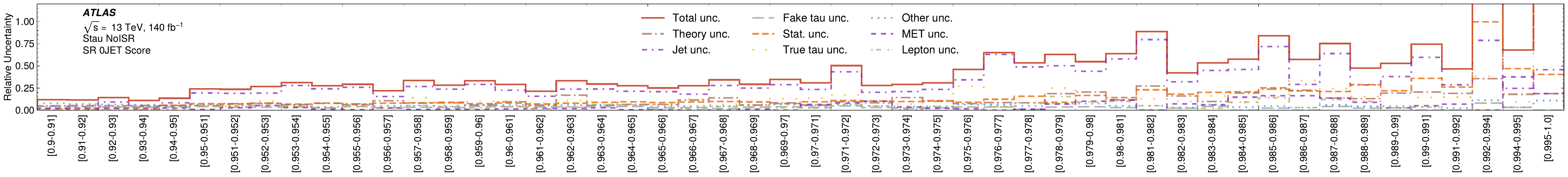}}\\
\subfloat[\wjet SR \RunTwo]{\includegraphics[width=0.95\linewidth]{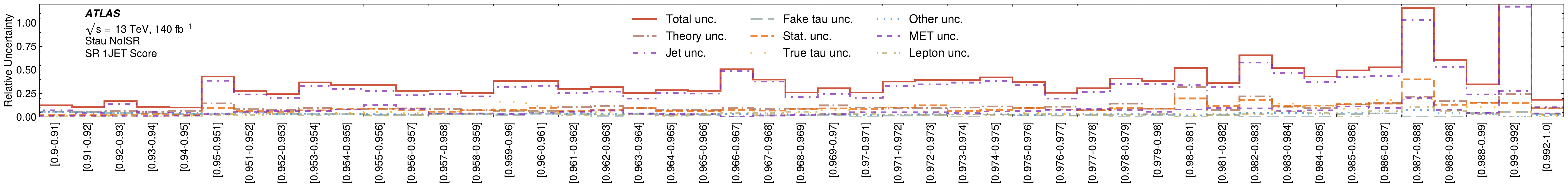}}\\
\subfloat[\nojet SR \RunThr]{\includegraphics[width=0.95\linewidth]{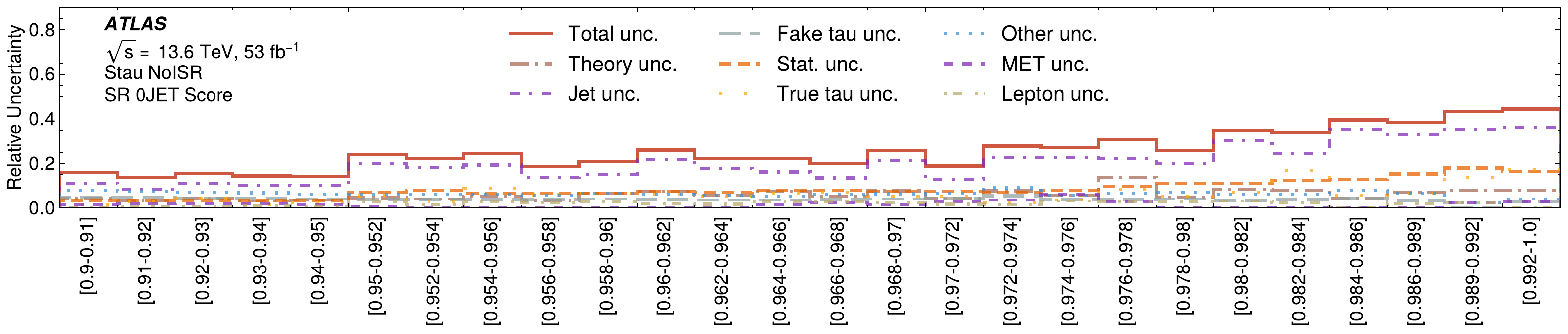}}\\
\subfloat[\wjet SR \RunThr]{\includegraphics[width=0.95\linewidth]{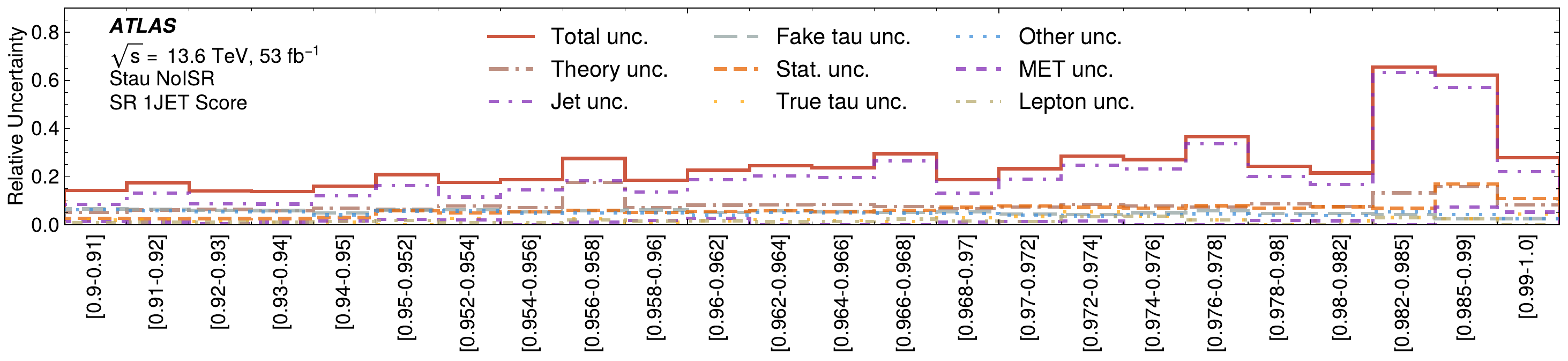}}
\caption{Comparison of the dominant groups of systematic uncertainties in the total background prediction after the background-only fit in the \staunoisr SRs, relative to the total yields. The total uncertainty is calculated using all the available systematic and statistical uncertainties. \enquote{Other} refers to $b$-tagging and \pileup uncertainties. \enquote{True tau} refers to the experimental uncertainties for $\tau$-leptons. The individual uncertainties can be correlated and do not necessarily add in quadrature to the total uncertainty. Statistical uncertainty refers only to the uncertainties of the backgrounds and does not include the effects of the data Poisson uncertainty.}
\label{fig:systematics2}
\end{figure}

The effects of different sources of systematic uncertainty in the SRs are shown in Figures~\ref{fig:systematics}-\ref{fig:systematics2}. To find the effect of a group of uncertainties, the square root of the difference between the squares of the total uncertainty and the uncertainty computed without this group is used.
The main sources of experimental systematic uncertainty in the SM backgrounds are related to the jet-energy calibration and resolution. The uncertainties in modelling fake-taus are relevant in regions with significant fake-tau contributions. Theoretical uncertainties are significant for regions enriched in $Z(\tau\tau)$ and diboson backgrounds. Finally, the statistical uncertainty becomes dominant in most SR bins.

%
%


\FloatBarrier
\section{Results}
\label{sec:result}

In this Section, the results of the fits for all channels are presented. For all the results shown in this Section, the \RunTwo and \RunThr data samples are fitted together. Individual sub-channels (\lephad/\hadhad for ISR channels, \wjet/\nojet for \staunoisr) are also fitted together. In all presented results, the asymptotic approximation of the profile likelihood ratio~\cite{Cowan:2010js} is assumed. The validity of this approximation is confirmed for all SRs by randomly generating many pseudo-datasets and repeating the fit.

\subsection{Direct $\tau$-slepton analysis}
\begin{figure}[htbp]
\centering
\subfloat[][]{\includegraphics[width=0.48\linewidth]{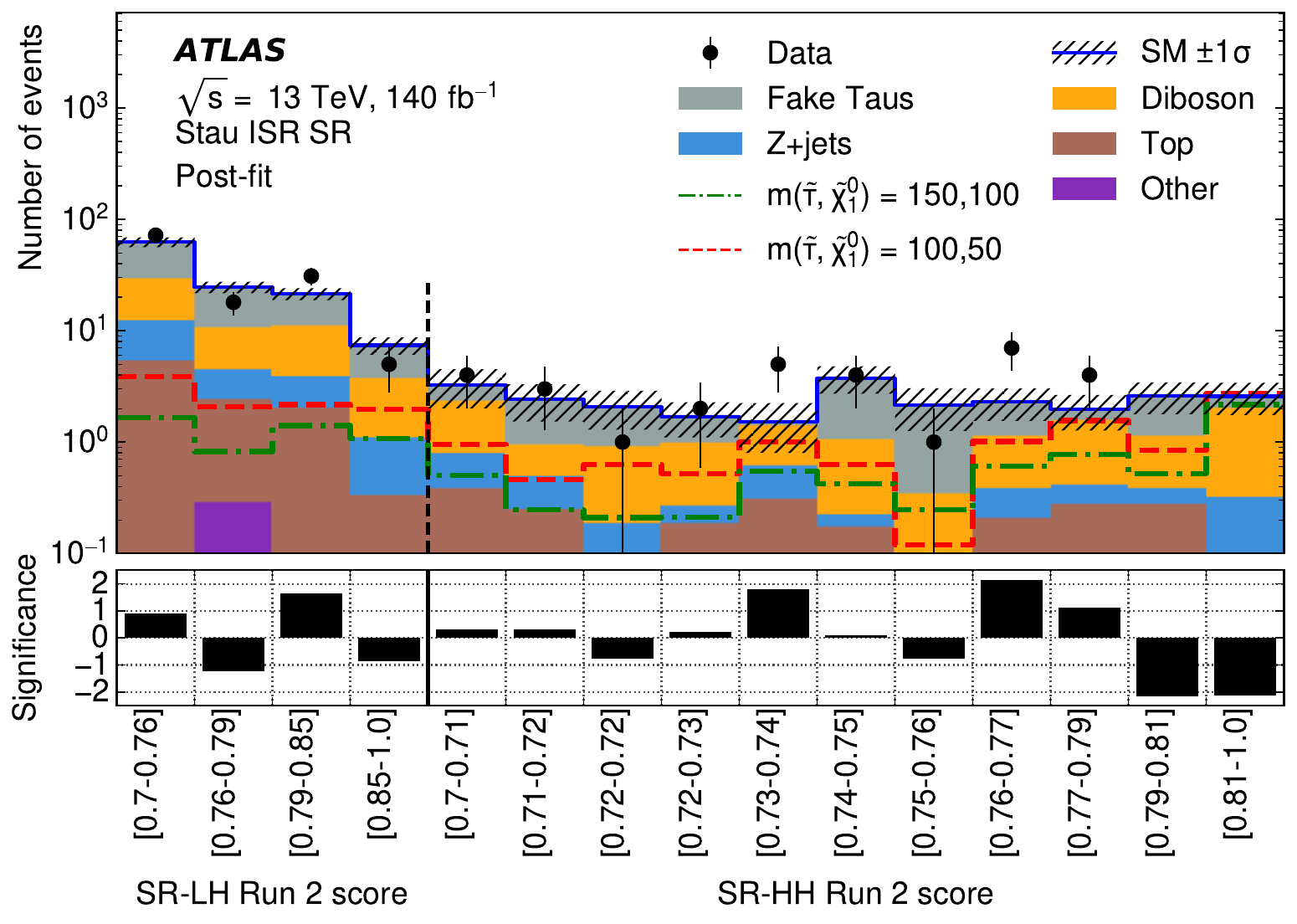}}
\subfloat[][]{\includegraphics[width=0.48\linewidth]{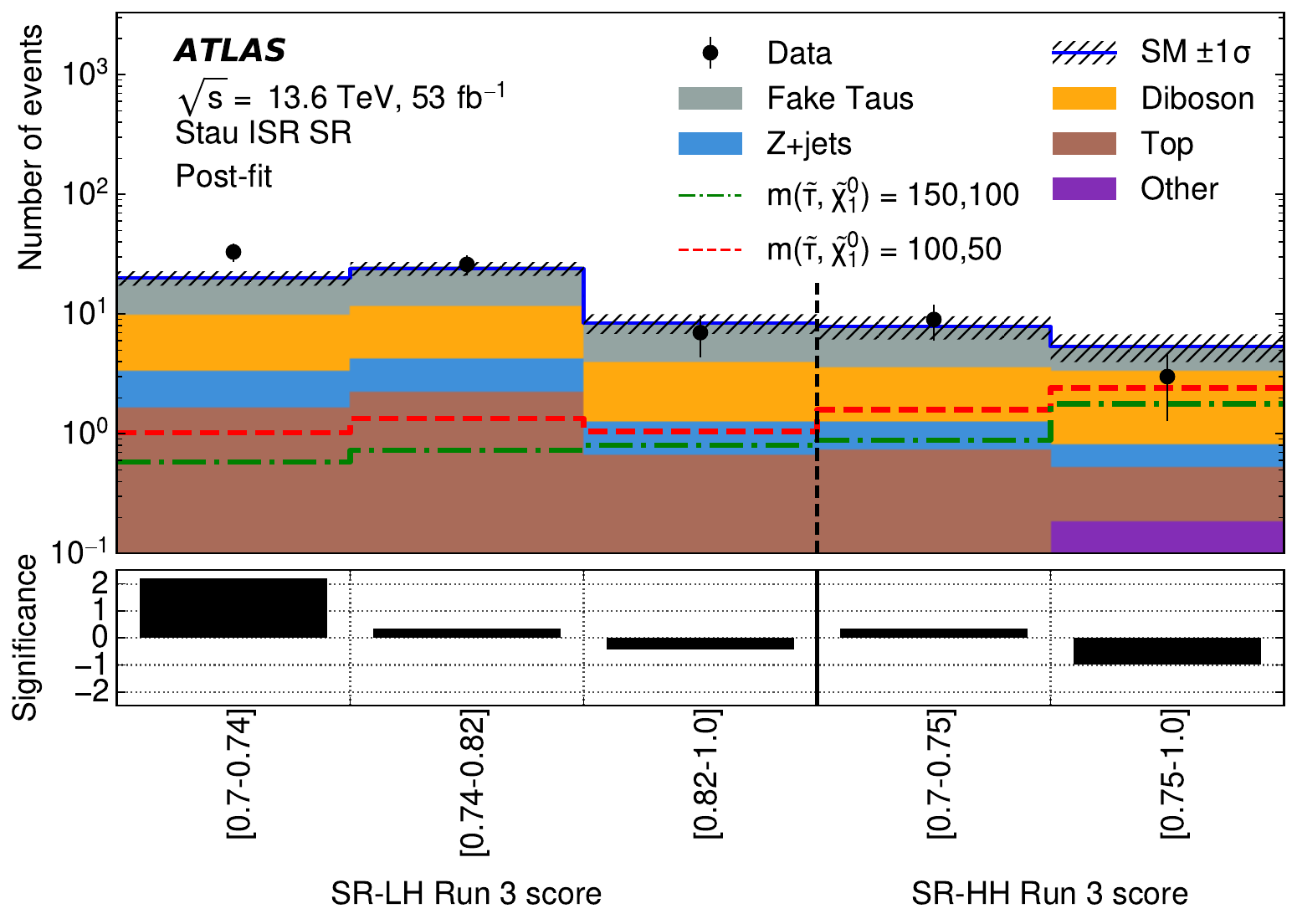}}\\
\subfloat[][]{\includegraphics[width=0.48\linewidth]{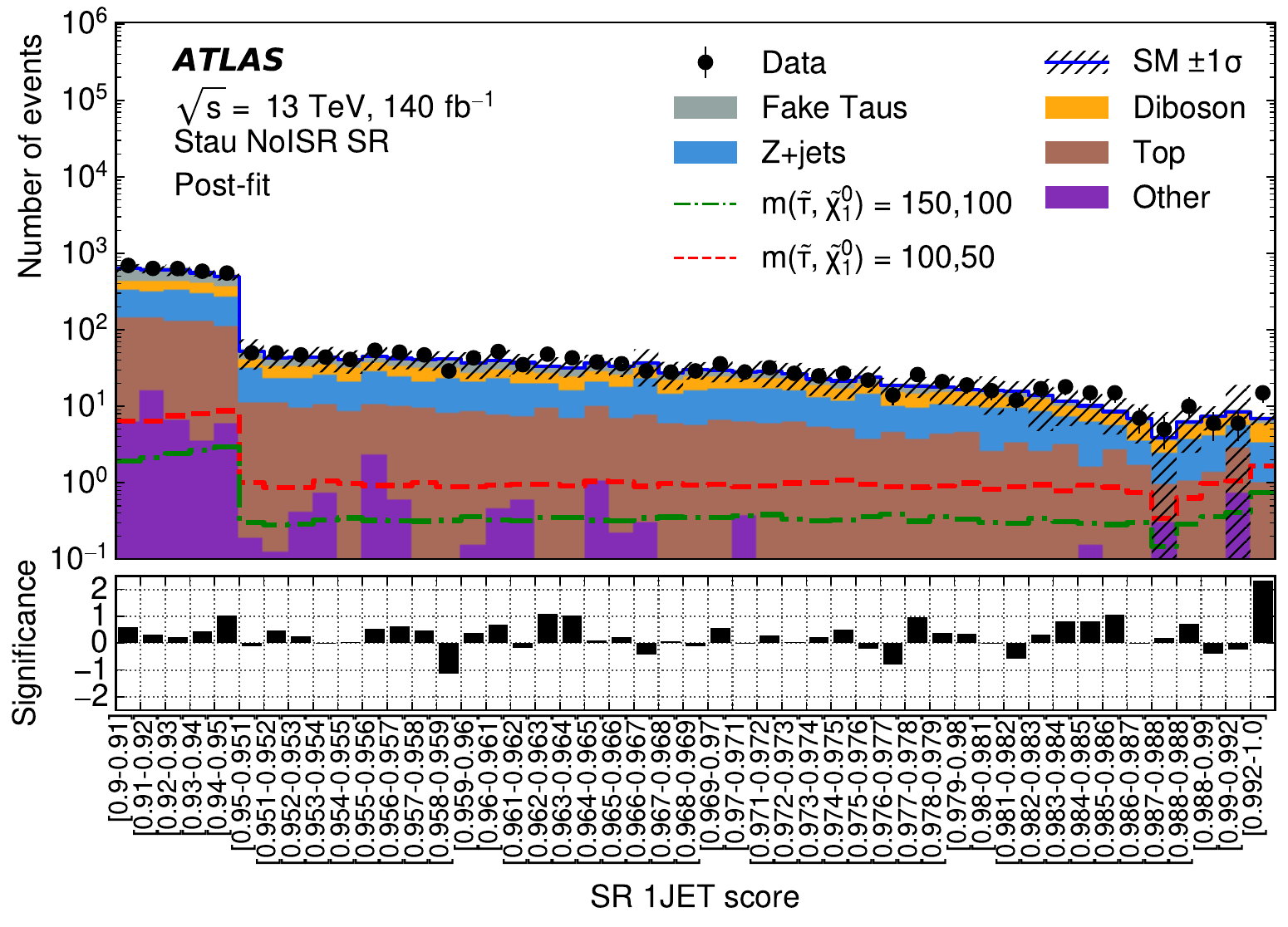}}
\subfloat[][]{\includegraphics[width=0.48\linewidth]{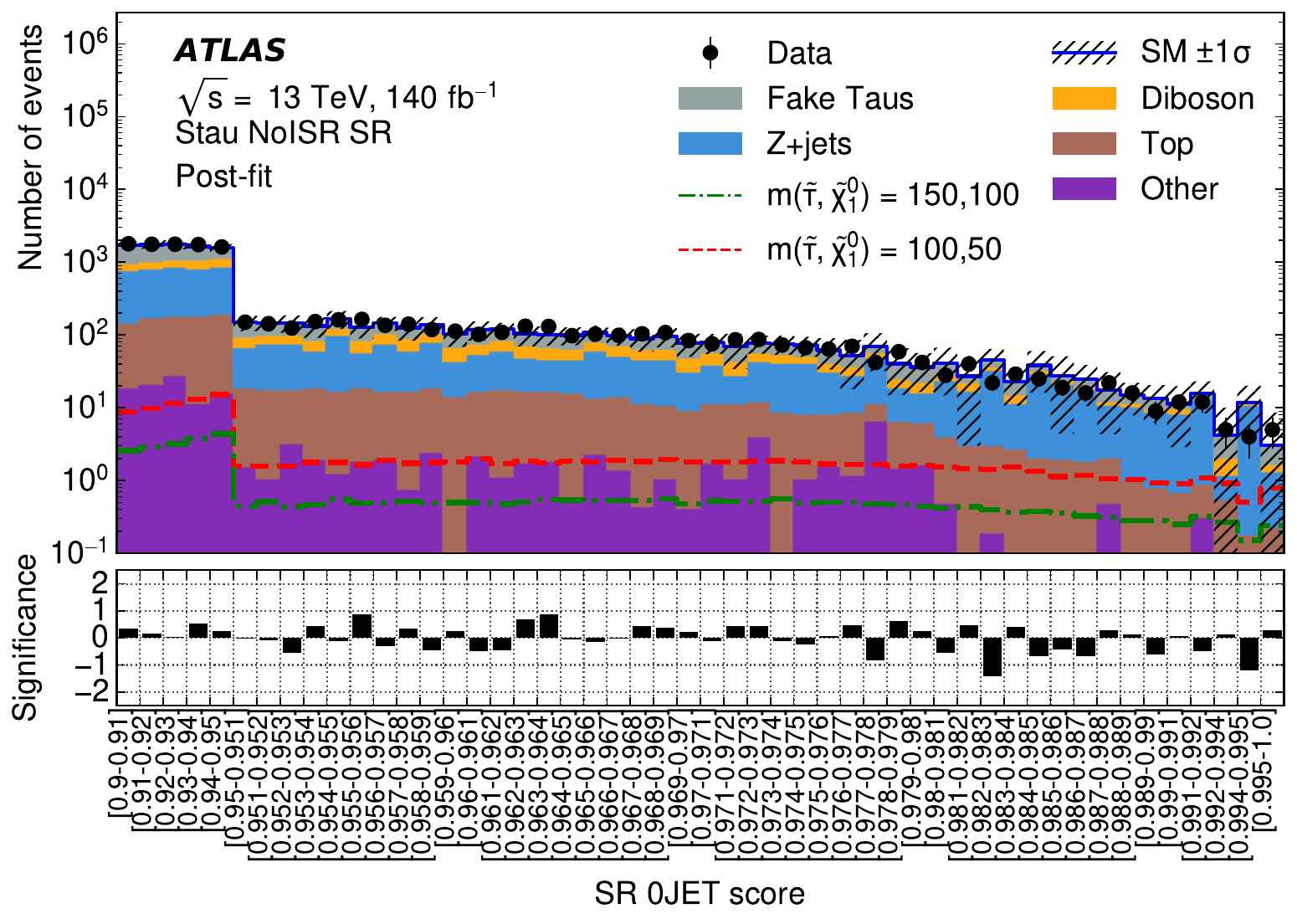}}\\
\subfloat[][]{\includegraphics[width=0.48\linewidth]{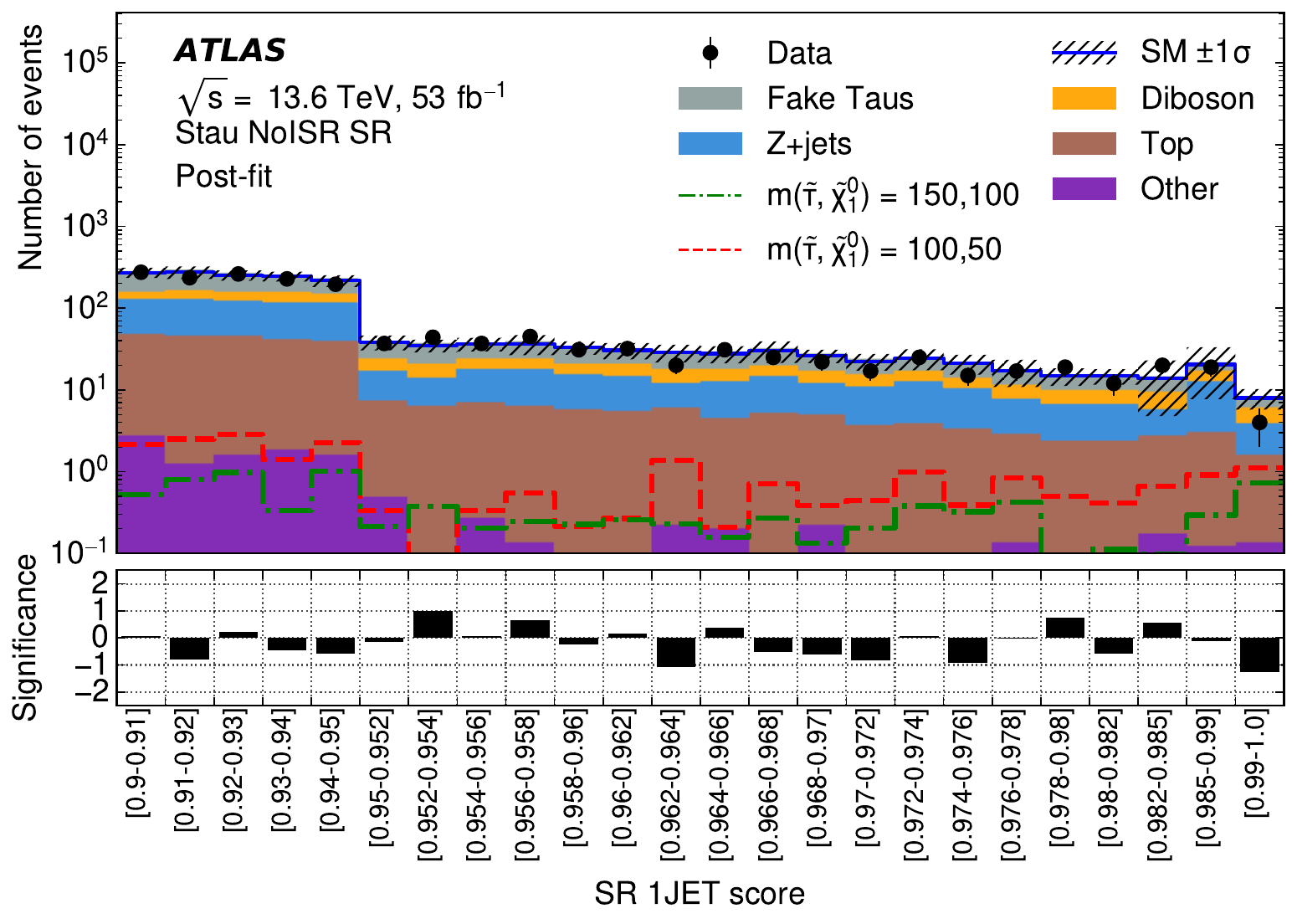}}
\subfloat[][]{\includegraphics[width=0.48\linewidth]{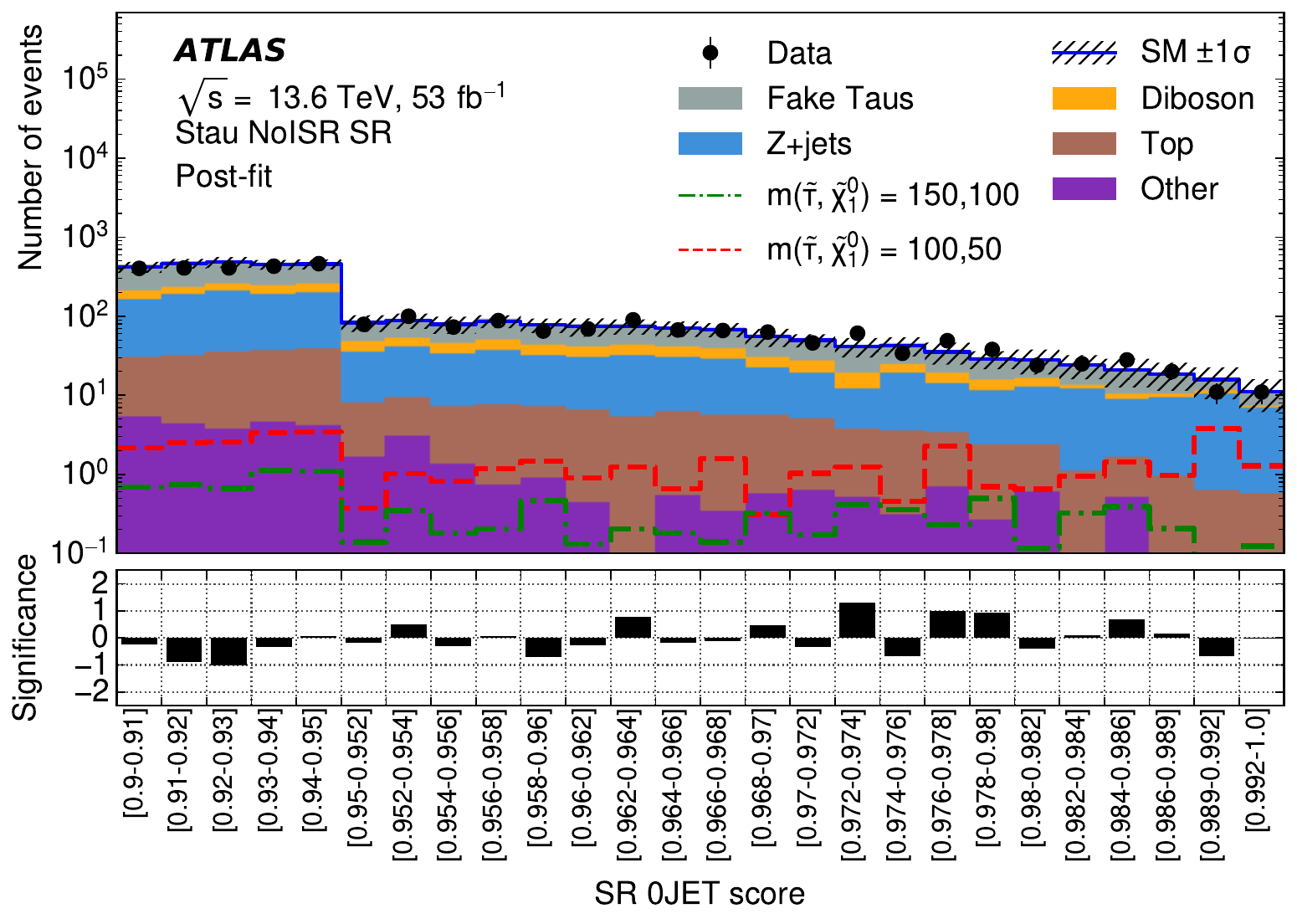}}
\caption{A comparison between the observed and expected yields in the SRs of the (a) \stauisr \RunTwo channel, (b) \stauisr \RunThr channel, (c), (d) \staunoisr \RunTwo channel, and (e),(f) \staunoisr \RunThr channel. Equal bin width is used for the presentation. \Others
\BottomPanelSig
\Postfit
\BandIsStatSyst
Predicted yields from two benchmark signal models, one with $m(\stau) = 100~\GeV$, $m(\ninoone) = 50~\GeV$ and the other with $m(\stau) = 150~\GeV$, $m(\ninoone) = 100~\GeV$, are overlaid for comparisons.
}
\label{fig:noisrstau_sr1}
\end{figure}

The comparison between the observed and expected yields in all of the \stauisr and \staunoisr SRs is shown in Figure~\ref{fig:noisrstau_sr1}.  %
The binning shown in the presented distributions is the same as the binning used in the statistical analysis.
The SM backgrounds are scaled by the NFs determined in the background-only fits and propagated to the SRs, as described in Section~\ref{sec:fitStrategy}. The largest excess over the SM prediction has a local significance of $2.3 \sigma$, in the last bin of the \RunTwo \wjet SR\@. The largest deficit of events is observed in the last two bins of the \RunTwo HH SR, corresponding to $2.2$ and $2.1 \sigma$ local significance. %

\begin{figure}[htbp]
\centering
\subfloat[]{\includegraphics[width=0.48\textwidth]{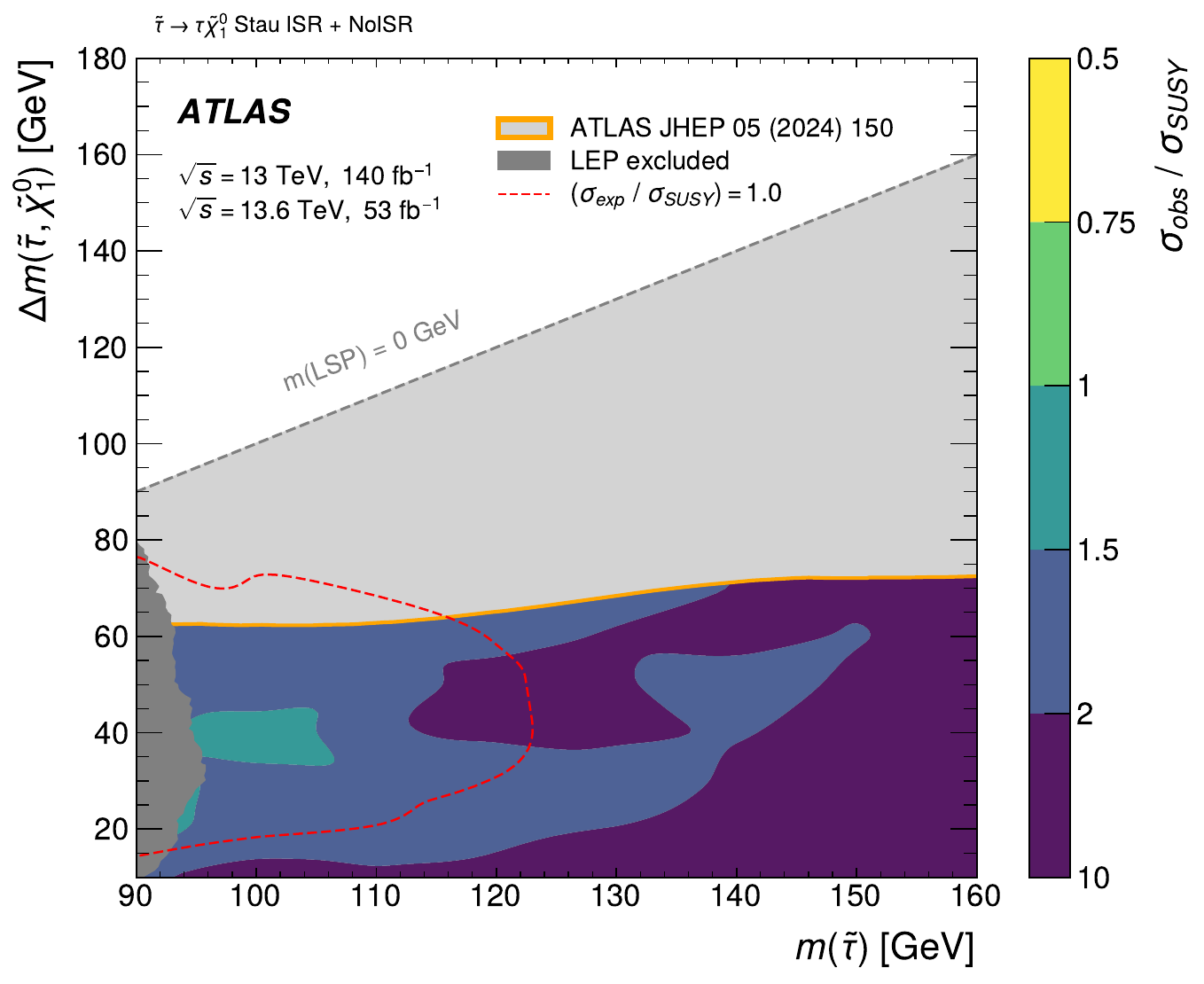}}
\subfloat[]{\includegraphics[width=0.48\textwidth]{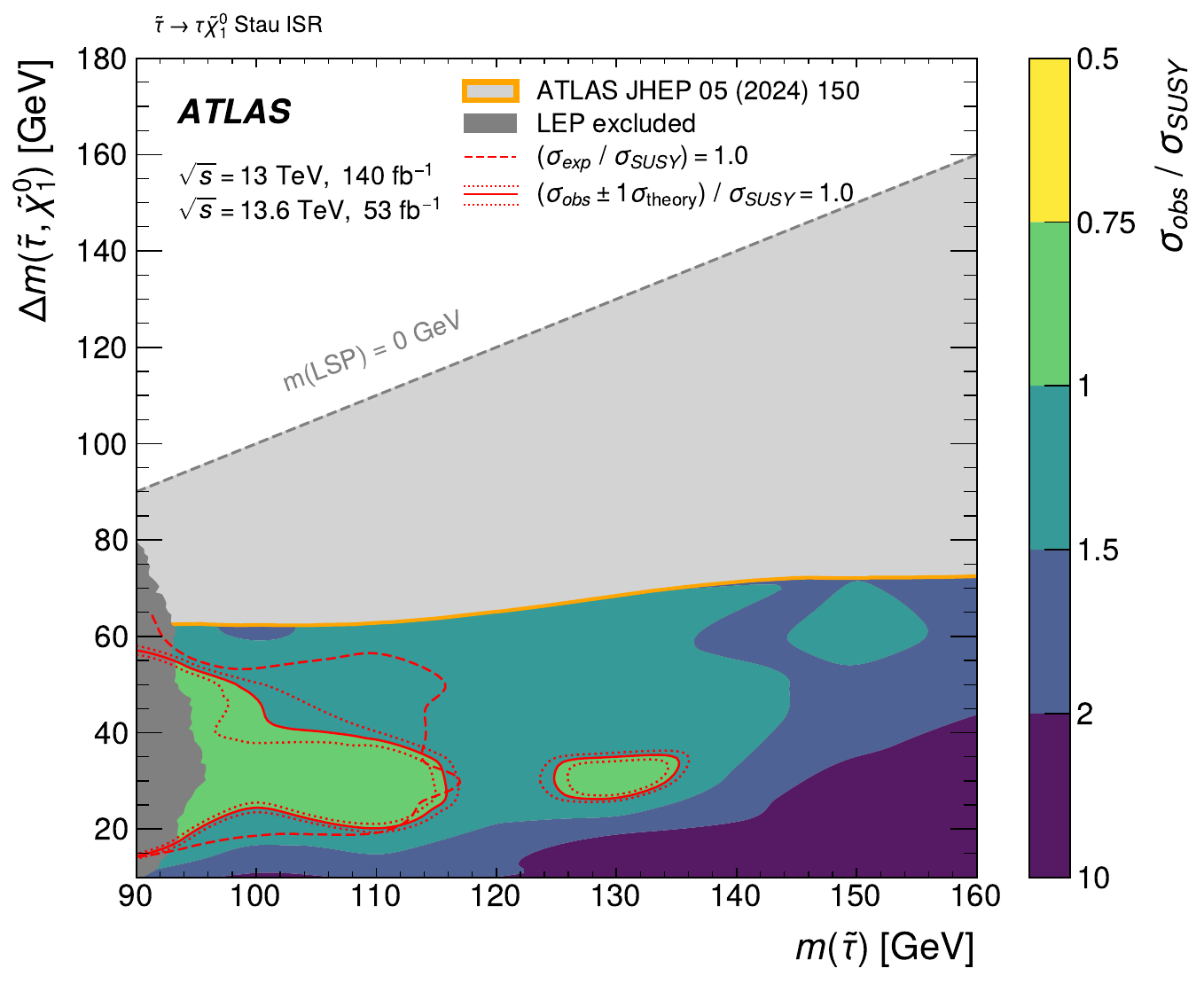}}
\caption{Upper limit on the observed cross-section normalised to the theory-predicted SUSY cross-section, obtained from (a) the combination of the \stauisr and \staunoisr channels and (b) \stauisr channel alone. The solid gray area at higher \masssplit is excluded by the previous ATLAS searches. Dashed red line shows the expected exclusion limits at $95\%$ CL.
}
\label{fig:stau_limits}
\end{figure}

For each of the simplified SUSY models considered, i.e. for each set of $m(\stau)$ and $m(\ninoone)$, an exclusion fit is performed. Both the \staunoisr and \stauisr channels are included in the simultaneous fits, with a common signal strength. The background-only and background-plus-signal hypotheses are tested and the SUSY models are rejected if the upper limit at \percent{95} CL
of the signal normalisation factor obtained in this fit is smaller than one, following the \CLs prescription~\cite{CLs:2002}.
The exclusion contours achieved in this way are shown in Figure~\ref{fig:stau_limits}. The combined fit is unable to exclude any of the models due to the excesses in the observed data. The expected limits of the combination reach $m(\stau)$ above $120$~\GeV with $\masssplit = 30$~\GeV, and completely cover the gap with the previous results, focused on the non-compressed final states, at higher $\masssplit$. The signal model most compatible with the combination of \stauisr and \staunoisr channels is $m(\stau) = 90~\GeV$, $m(\ninoone) = 40~\GeV$, with local $p$-value of $0.0033$ ($2.7 \sigma$ local significance).

In the \stauisr channel alone, which has the best individual expected limits, $\tau$-sleptons with masses up to $115$~\GeV are excluded for lightest neutralinos with $\masssplit = 30$~\GeV. The signal model most compatible with the observed data in \staunoisr channel, mostly driven by the excess in the tail of the \RunTwo \wjet SR, is $m(\stau) = 130~\GeV$, $m(\ninoone) = 90~\GeV$, with local $p$-value of $0.00020$ ($3.5 \sigma$ local significance). However, this model is strongly disfavoured by the combination of the \stauisr and \staunoisr channels, with local $p$-value of $0.12$.

\begin{figure}[htbp]
\centering
\subfloat[][]{\includegraphics[width=0.48\linewidth]{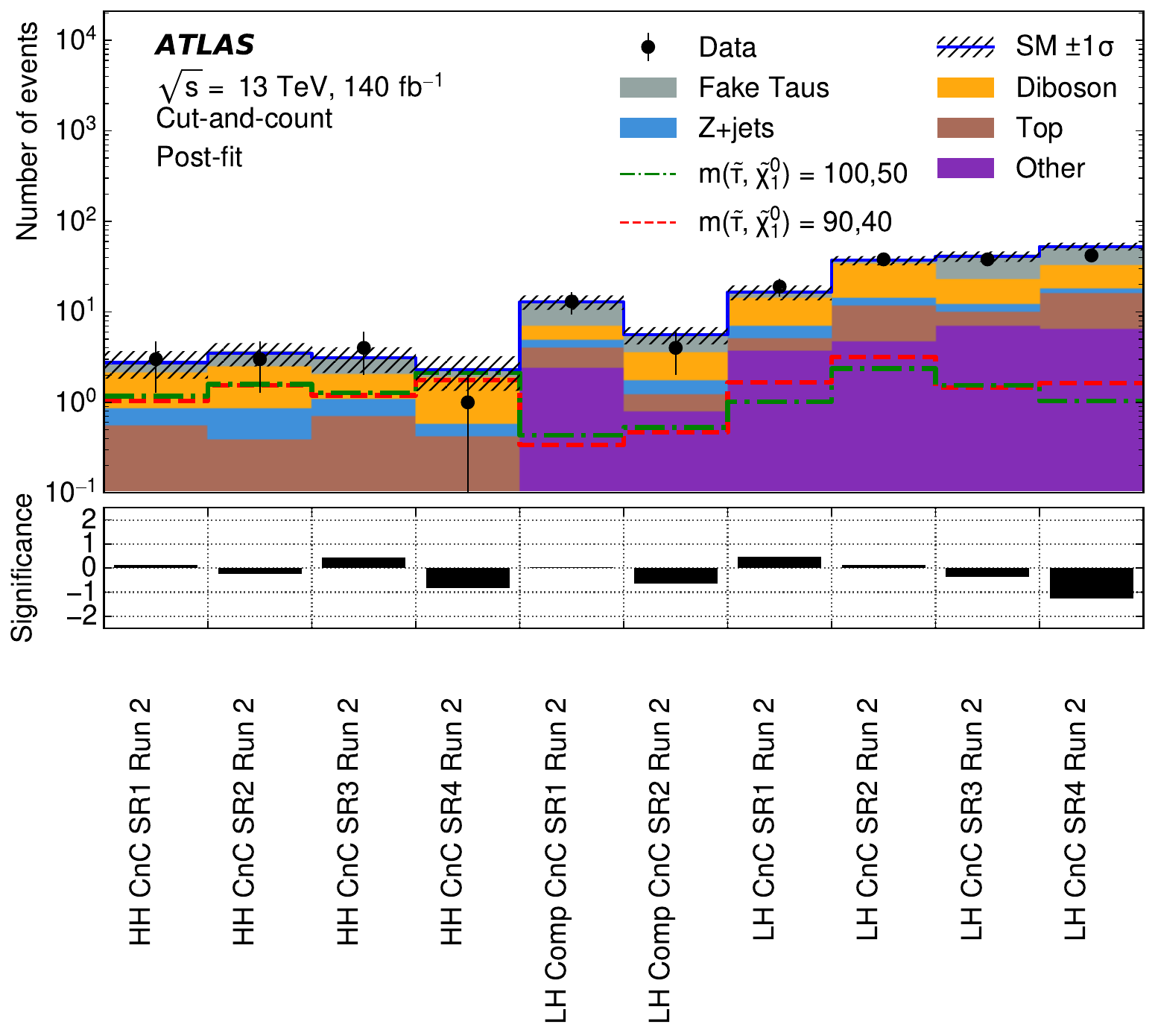}}
\subfloat[][]{\includegraphics[width=0.48\linewidth]{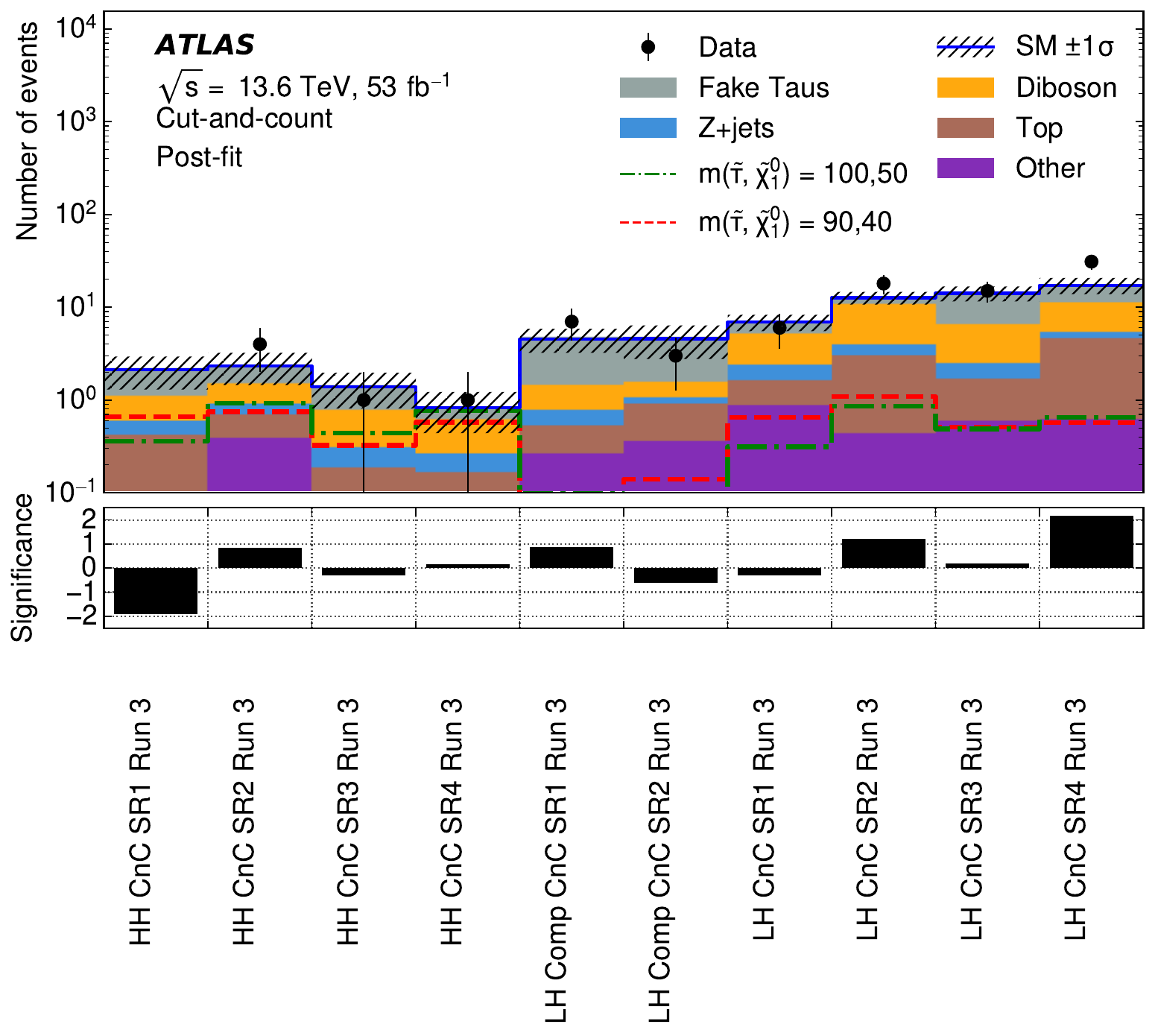}}
\caption{A comparison between the observed and expected yields in the SRs of the \stauisr cut-and-count channel. Equal bin width is used for the presentation. \Others
\BottomPanelSig
\Postfit
\BandIsStatSyst
\OverlaidSignal}
\label{fig:cncstau_sr}
\end{figure}

\begin{figure}[htbp]
\centering
\subfloat[]{\includegraphics[width=0.48\textwidth]{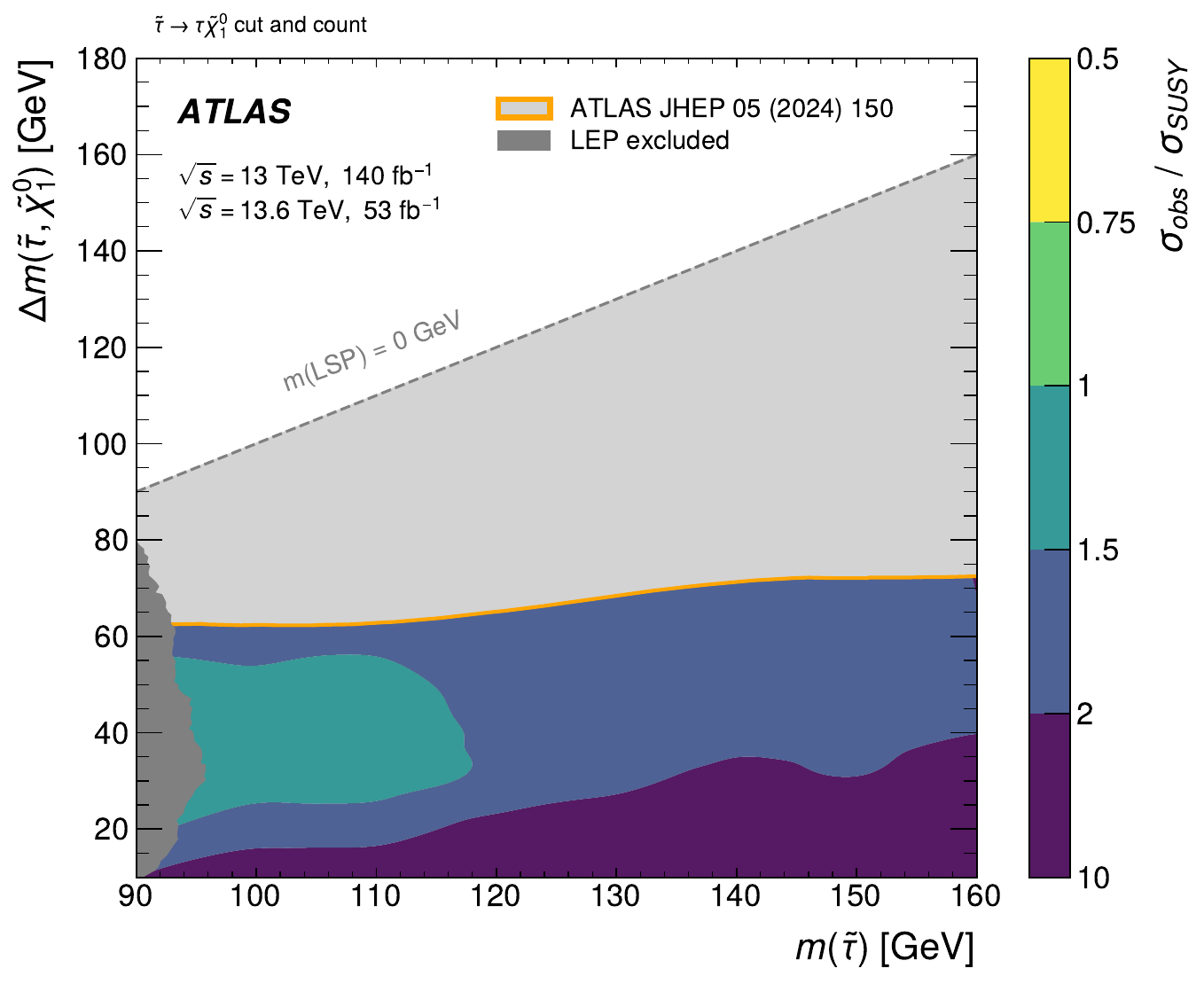}}
\caption{Upper limit on the observed cross-section normalised to the theory-predicted SUSY cross-section, obtained from the \stauisr cut-and-count channel. The solid gray area at higher \masssplit is excluded by the previous ATLAS searches.
}
\label{fig:cnc_limits}
\end{figure}

The same set of results is presented for the cut-and-count approach. First, the data yields in every SR are compared with the prediction in background-only fits. These are shown in Figure~\ref{fig:cncstau_sr}. The largest excess is $2.2 \sigma$ local significance, in the \RunThr LH CnC SR4, while the largest deficit is $1.9 \sigma$ local significance, in the \RunThr HH CnC SR1. %
While no simplified SUSY models can be excluded with the current analysis, upper limits are set on the predicted SUSY cross-section, as shown in Figure~\ref{fig:cnc_limits}. %

\FloatBarrier
\subsection{Intermediate $\tau$-slepton analysis}

\begin{figure}[htbp]
\centering
\subfloat[][]{\includegraphics[width=0.48\linewidth]{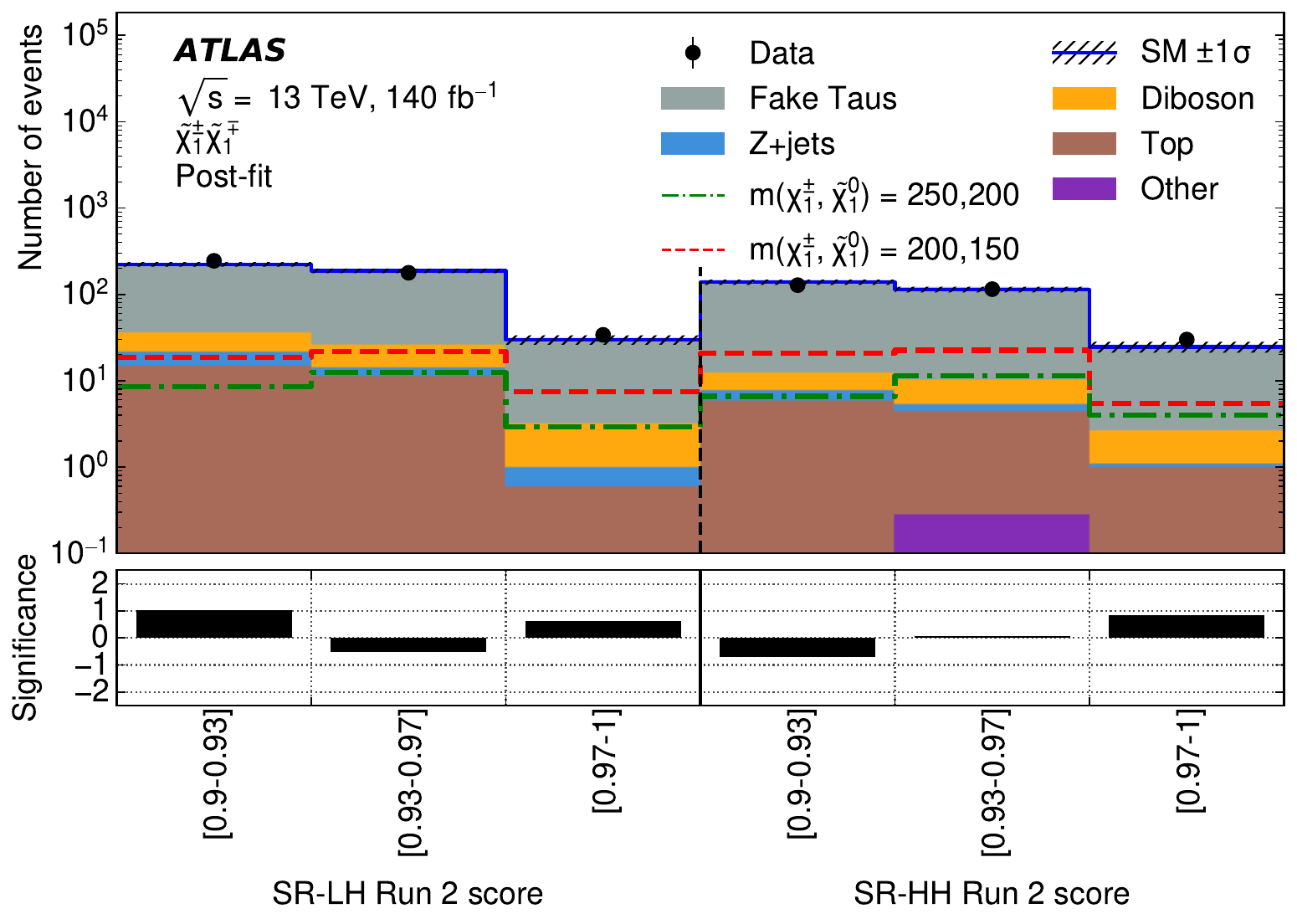}}
\subfloat[][]{\includegraphics[width=0.48\linewidth]{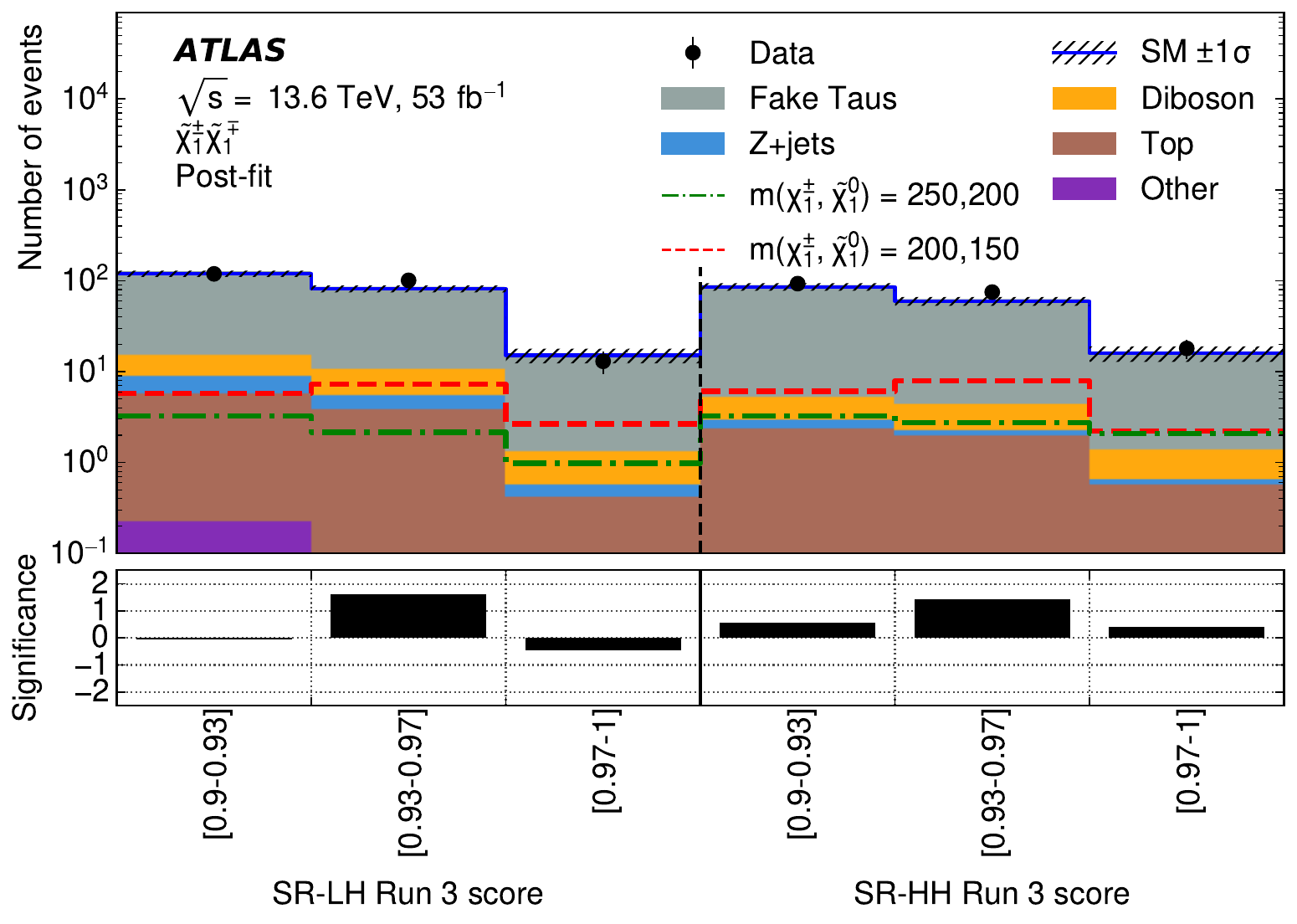}} \\
\subfloat[][]{\includegraphics[width=0.48\linewidth]{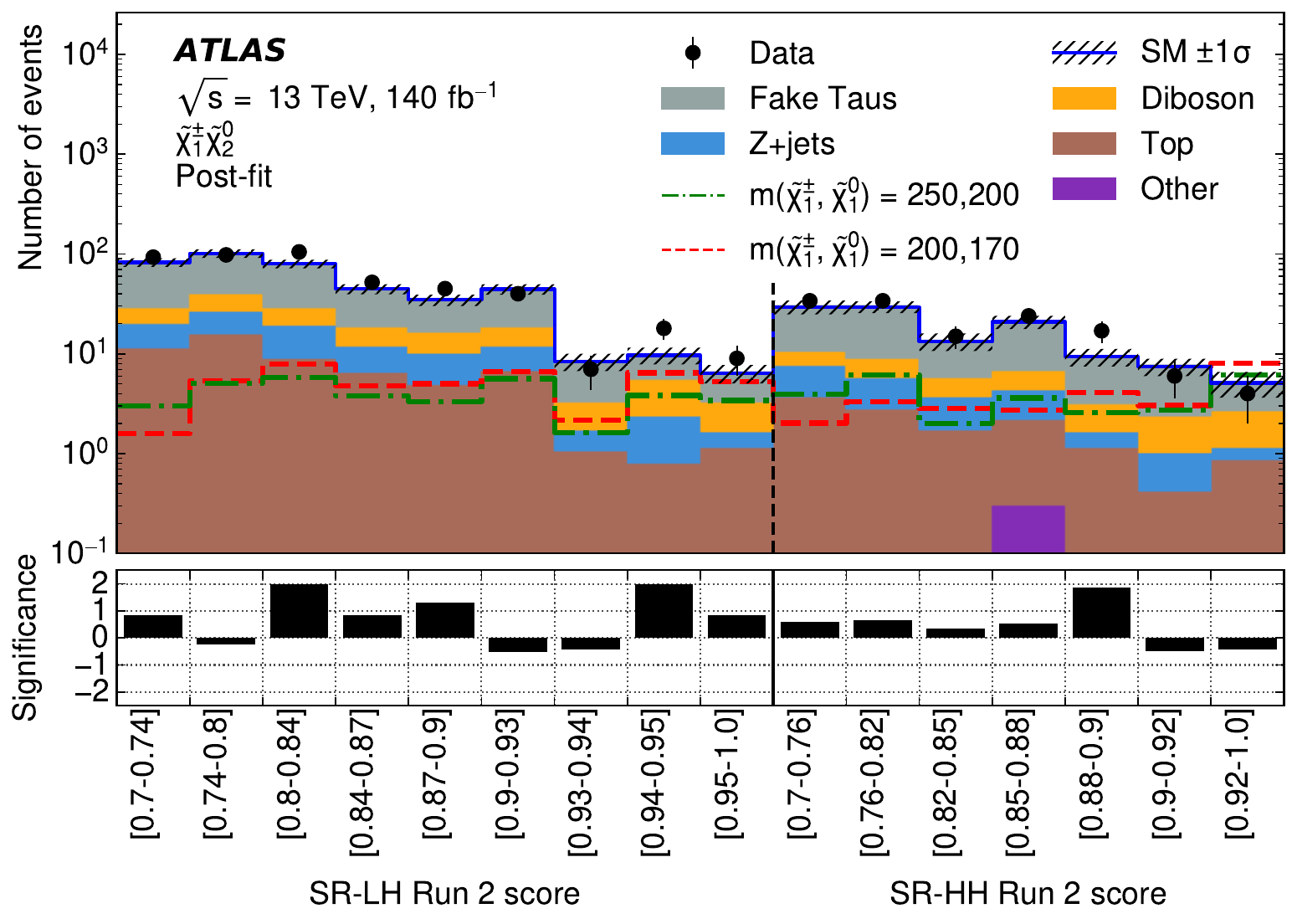}}
\subfloat[][]{\includegraphics[width=0.48\linewidth]{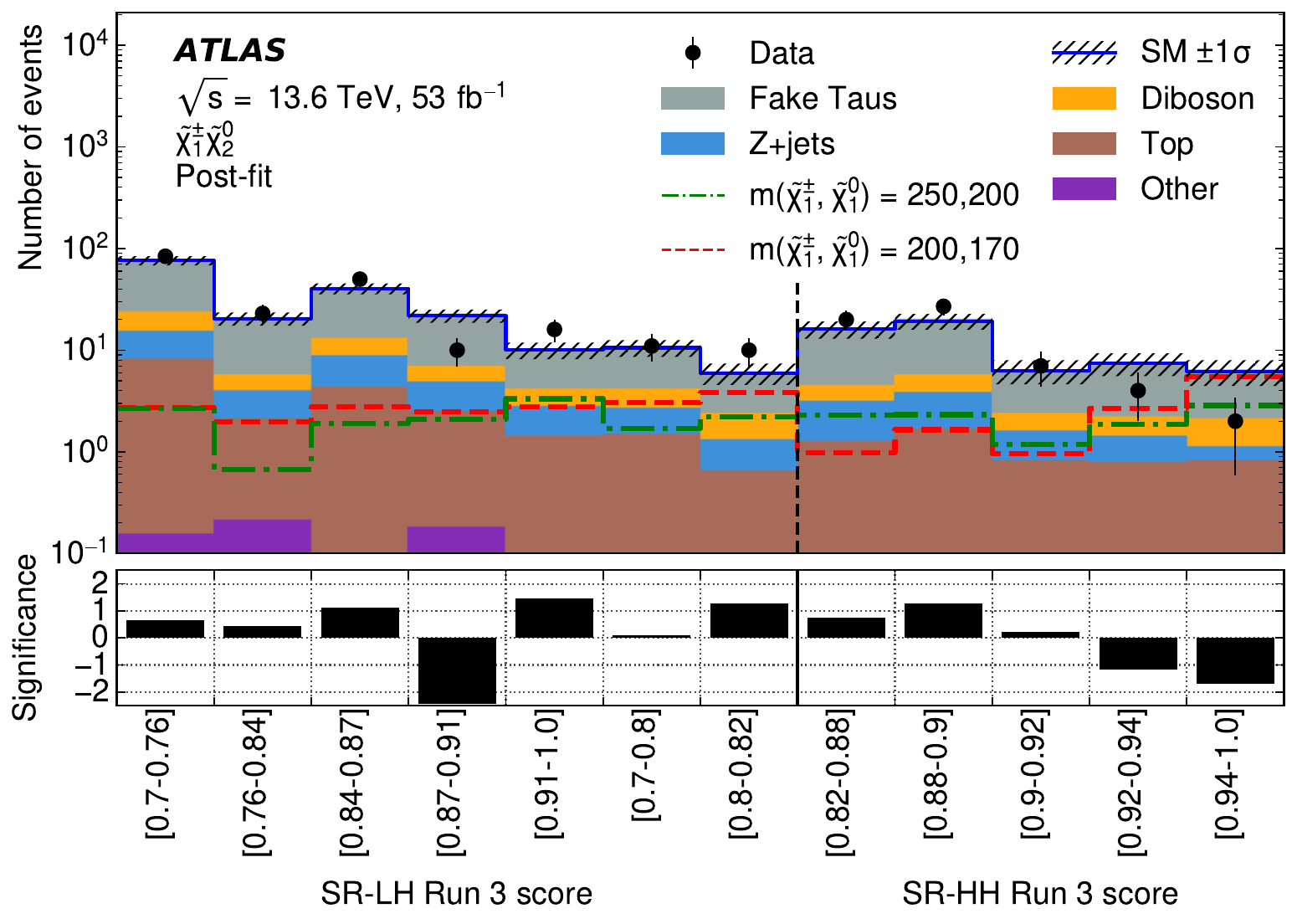}}
\caption{A comparison between the observed and expected yields in the SRs of the (a) \ConepConem \RunTwo channel, (b) \ConepConem \RunThr channel, (c) \ConepmNtwo \RunTwo channel, and (d) \ConepmNtwo \RunThr channel. Equal bin width is used for the presentation. \Others
\BottomPanelSig
\Postfit
\BandIsStatSyst
Predicted yields from benchmark signal models are overlaid for comparisons, with $m(\chinoonepm) = 250, 200$~\GeV and $m(\ninoone) = 200, 170 (150)$~\GeV for the \ConepmNtwo (\ConepConem) channel.
}
\label{fig:c1c1_sr}
\end{figure}

The comparison between the observed and expected yields in all SRs is shown in Figure~\ref{fig:c1c1_sr} for the \ConepConem and the \ConepmNtwo channels. The results are shown after a background-only fit. In the \ConepConem regions the background predictions are found to be compatible with data, with the largest deviation having a local significance of $1.6 \sigma$ in the \RunThr LH SR\@. In the \ConepmNtwo channel, the largest downwards deviation from the predicted backgrounds is seen for the fourth bin of \RunThr LH SR, with a local significance of $2.4 \sigma$.

\begin{figure}[h!]
\centering
\subfloat[]{\includegraphics[width=0.48\textwidth]{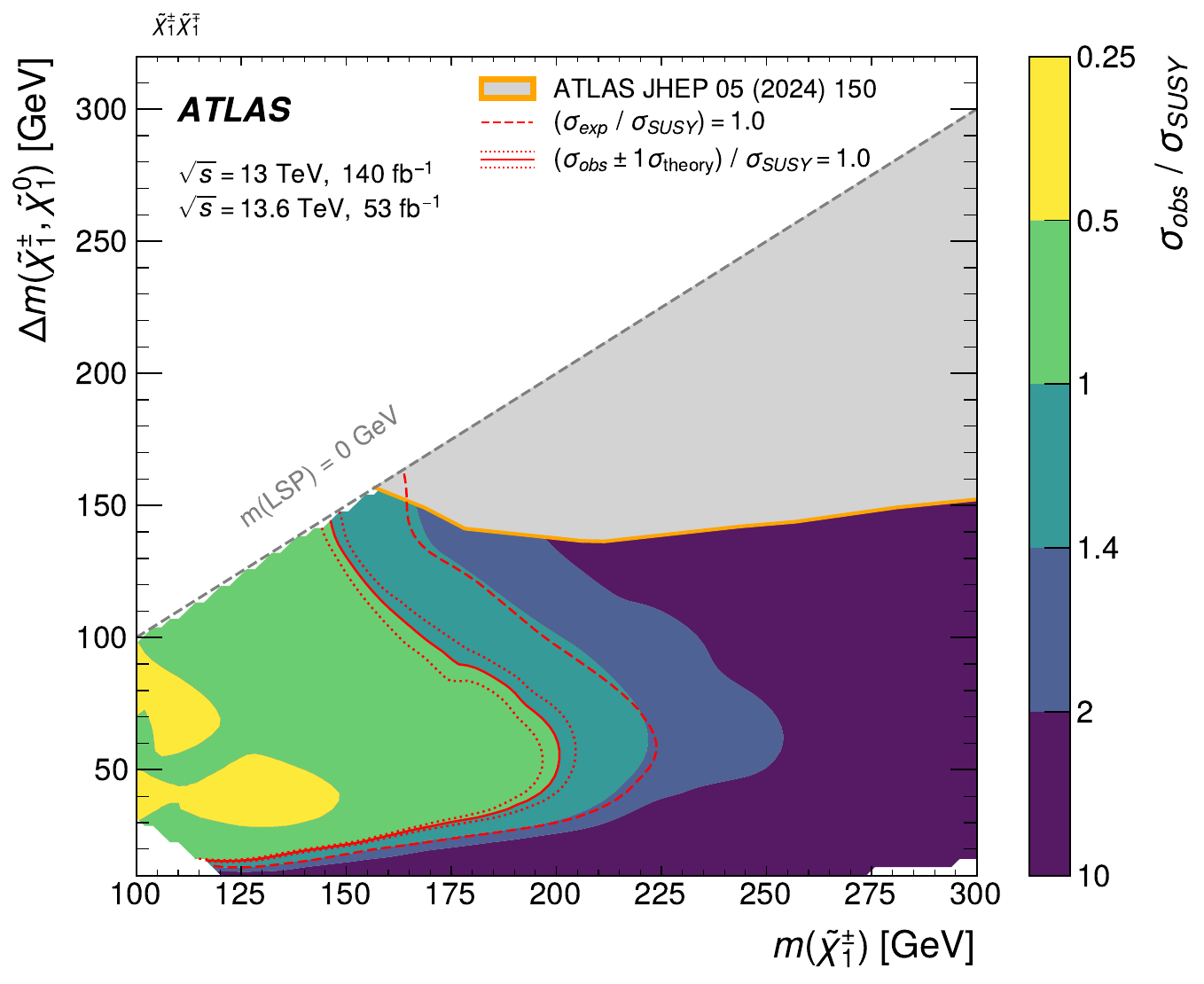}\label{fig:alimit-weak1}}
\subfloat[]{\includegraphics[width=0.48\textwidth]{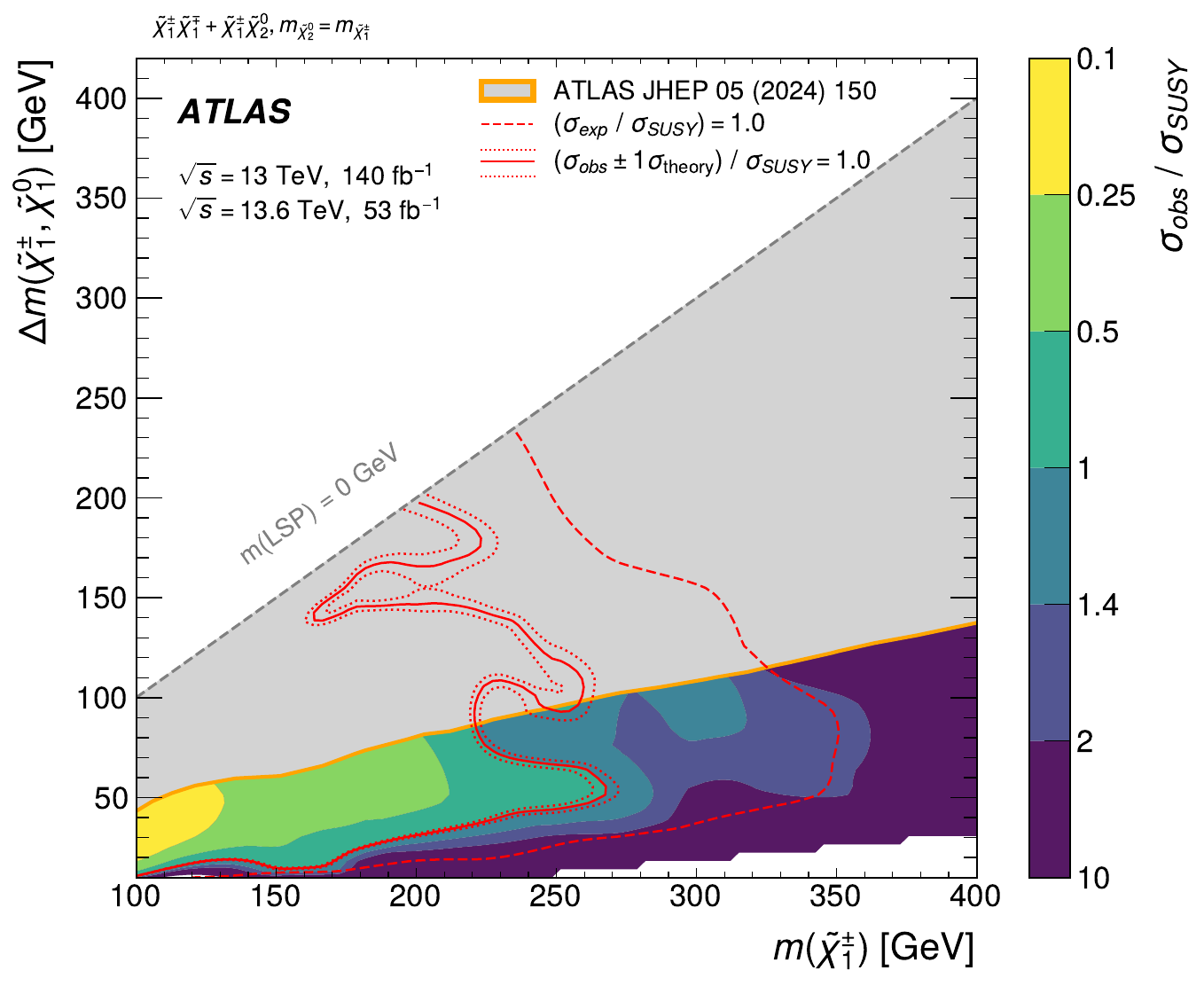}\label{fig:alimit-weak2}}
\caption{
Upper limit on the observed cross-section normalised to the theory-predicted SUSY cross-section, obtained from (a) the \ConepConem channel and (b) the \ConepmNtwo channel with both \ConepConem and \ConepmNtwo production assumed. The solid gray areas at higher \dmcn are excluded by the previous ATLAS searches. Dashed red line shows the expected exclusion limits at $95\%$ CL.
}
\label{fig:alimit-weak}
\end{figure}

In the absence of a significant excess over the expected SM background, the observed and expected numbers of events in the signal regions are used to
set exclusion limits at \percent{95} CL on the \ConepConem and \ConepmNtwo productions using the model-dependent fits. The LH \ConepConem SR and HH \ConepConem SR are statistically combined to set limits on \ConepConem production, shown in Figure~\ref{fig:alimit-weak1}. LH \ConepmNtwo SR and HH \ConepmNtwo SR are combined to derive limits on the combined production of \ConepConem and \ConepmNtwo processes, shown in Figure~\ref{fig:alimit-weak2}.

Chargino masses up to $175$~\GeV with $\dmcn = 30~\GeV$ are excluded.
For production of chargino pairs or mass-degenerate charginos and next-to-lightest neutralinos,
chargino masses up to around $200~\GeV$ with $\dmcn = 30~\GeV$ are excluded. The expected limits in this scenario reach chargino masses of $350$~\GeV with $\dmcn = 50$~\GeV, and completely cover the gap with the previous results, focused on the non-compressed final states, at higher $\dmcn$.
All limits apply to scenarios where the neutralinos and charginos decay solely via intermediate $\tau$-sleptons and $\tau$-sneutrinos.
These limits significantly extend previous results~\cite{SUSY-2023-03} in the compressed region with small differences between $\chinoonepm$/$\ninotwo$ and $\ninoone$ masses.
The most compatible signal mass point is $m(\ninotwo) = 200$~\GeV, $m(\ninoone) = 180$~\GeV, with the local $p$-value of $0.0013$ ($3.0 \sigma$ local significance) for the combination of the two signal production modes. No chargino pair production model is compatible with the observed data with the local $p$-value below $0.05$.

\FloatBarrier

\subsection{Model-independent limits}
For each of the channels and SRs considered in the analysis, a model-independent limit on the presence of new physics is set. The last bin of each SR is included in the fit (one at a time) with all the corresponding CRs, and the profile-likelihood-ratio test statistic is evaluated. The classifier scores are, by definition, tuned for a specific set of SUSY model. So for a valid reinterpretation, the new physics model has to have sufficiently similar kinematic properties to populate the high-classifier-score tail. The limits are set for \RunTwo and \RunThr separately, such that no assumptions about the dependence of the new physics cross-section on the $\sqrt{s}$ are made. The results are summarised in Table~\ref{tab:discSR}. Upper limits of $0.02{-}0.25$~fb are placed on the cross-section of a new physics process appearing in these regions. For the cut-and-count approach, the model-independent limits are set in every SR, as no clear \enquote{best} region exists. The results are summarised in Table~\ref{tab:discSR_cnc}.

\FloatBarrier

\begin{table}[t]
\caption{
Upper limits at $95\%$ CL on the visible cross-section \sigvis,
on the number of signal events ($S_\mathrm{obs}^{95}$),
and on the number of signal events given the expected number (and $\pm 1\sigma$ excursions of the expectation) of background events ($S_\mathrm{exp}^{95}$).
The total background numbers are provided after a fit in both CRs and the SR considered.
The last two columns indicate the \CLb value (i.e.\ the confidence level observed for the background-only hypothesis),
the discovery $p$-value ($p_0$),
and its associated significance $Z$. The regions are taken from all of the ML-based channels considered.
}
\label{tab:discSR}

\renewcommand{\arraystretch}{1.2}
\centering
\begin{adjustbox}{width=0.8\textwidth}
\footnotesize
\begin{tabular}{lccccccc}
\toprule{Signal region} &  Total bkg & Data & \sigvis[fb]  &  $S_\mathrm{exp}^{95}$  & $S_\mathrm{obs}^{95}$ & \CLb & $p_0$ ($Z$)  \\
\midrule
\RunTwo \\
\midrule
Stau NoISR \nojet SR  &  4.8 $\pm$ 1.9  &  5  &  0.068  &  9.1$^{+3.5}_{-2.2}$   &  9.5  &  0.6 &  0.39 (0.3)  \\
Stau NoISR \wjet SR  &  8.7 $\pm$ 1.2  &  15  &  0.11  &  8$^{+4}_{-2.4}$   &  15.8  &  1.0  &  0.01 (2.3) \\
SR-HH stau ISR  &  1.5 $\pm$ 0.6  &  0  &        0.018  &  4.0$^{+2.4}_{-1.4}$   &  2.5  &  0.1  &  0.50 (0.0)  \\
SR-LH stau ISR  &  5.0 $\pm$ 0.9  &  4  &        0.036  &  6.0$^{+3.1}_{-1.9}$   &  5.0  &  0.3  &  0.50 (0.0) \\
SR-HH \ConepmNtwo  &  4.8 $\pm$ 1.2  &  4  &     0.038  &  6.2$^{+3.1}_{-2.0}$   &  5.3  &  0.4  &  0.50 (0.0) \\
SR-LH \ConepmNtwo  &  6.9 $\pm$ 1.2  &  9  &     0.067  &  6.9$^{+2.6}_{-2.2}$   &  9.4  &  0.8  &  0.20 (0.9) \\
SR-HH \ConepConem  &  26.4 $\pm$ 2.9  &  30  &   0.13  &  13$^{+6}_{-4}$   &  17.5  &  0.8  &  0.20 (0.8) \\
SR-LH \ConepConem  &  31.1 $\pm$ 3.1  &  34  &   0.12  &  14$^{+5}_{-4}$   &  17.4  &  0.7  &  0.27 (0.6) \\
\midrule
\RunThr  \\
\midrule
Stau NoISR \nojet SR & 11.0 $\pm$ 2.7  &  11  &   0.24   &  13$^{+7}_{-4}$   &  12.5  &  0.5   &  0.50 (0.0) \\
Stau NoISR \wjet SR & 6.3 $\pm$ 1.6  &  4     &   0.095   &  7.4$^{+3.5}_{-2.3}$   &  4.9  &  0.1   &  0.50 (0.0) \\
SR-HH stau ISR  &  3.6 $\pm$ 1.0  &  2  &         0.073   &  5.4$^{+2.9}_{-1.8}$   &  3.9  &  0.2   &  0.50 (0.0) \\
SR-LH stau ISR  &  6.8 $\pm$ 1.2  &  7  &         0.13   &  6.9$^{+2.6}_{-2.2}$   &  7.1  &  0.5 &  0.47 (0.1)  \\
SR-HH \ConepmNtwo  &  4.5 $\pm$ 1.4  &  2  &      0.071   &  6.2$^{+3.1}_{-2.0}$   &  3.8  &  0.1  &  0.50 (0.0) \\
SR-LH \ConepmNtwo  &  6.9 $\pm$ 1.3  &  10  &     0.18   &  7.0$^{+2.5}_{-2.2}$   &  9.5  &  0.9   &  0.10 (1.3) \\
SR-HH \ConepConem  &  16.7 $\pm$ 2.4  &  18  &    0.24   &  11$^{+5}_{-3.2}$   &  12.6  &  0.6   &  0.34 (0.4) \\
SR-LH \ConepConem  &  14.4 $\pm$ 2.3  &  13  &    0.16   &  10$^{+4}_{-3.0}$   &  8.6  &  0.3  &  0.50 (0.0) \\
\bottomrule
\end{tabular}
\end{adjustbox}
\end{table}

\begin{table}[t]
\caption{
Upper limits at $95\%$ CL on the visible cross-section \sigvis,
on the number of signal events ($S_\mathrm{obs}^{95}$),
and on the number of signal events given the expected number (and $\pm 1\sigma$ excursions of the expectation) of background events ($S_\mathrm{exp}^{95}$).
The total background numbers are provided after a fit in both CRs and the SR considered.
The last two columns indicate the \CLb value (i.e.\ the confidence level observed for the background-only hypothesis),
the discovery $p$-value ($p_0$),
and its associated significance $Z$. The regions are taken from the cut-and-count channel.
}
\label{tab:discSR_cnc}
\renewcommand{\arraystretch}{1.2}

\centering
\begin{adjustbox}{width=0.8\textwidth}
\footnotesize
\begin{tabular}{lccccccc}
\toprule{Signal region} &  Total bkg & Data & \sigvis[fb]  &  $S_\mathrm{exp}^{95}$  & $S_\mathrm{obs}^{95}$ & \CLb & $p_0$ ($Z$)  \\
\midrule
\RunTwo \\
\midrule
HH CnC SR1   &  2.8 $\pm$ 0.9  &  3  &       0.037  &  5.0$^{+2.7}_{-1.6}$   &  5.2  &  0.5  &  0.45 (0.1)  \\
HH CnC SR2   &  3.4 $\pm$ 0.9  &  3  &       0.035  &  5.3$^{+2.8}_{-1.7}$   &  4.9  &  0.4  &  0.50 (0.0)  \\
HH CnC SR3   &  3.3 $\pm$ 0.9  &  4  &       0.043  &  5.2$^{+2.8}_{-1.7}$   &  6.1  &  0.6  &  0.34 (0.4)  \\
HH CnC SR4   &  1.9 $\pm$ 0.8  &  1  &       0.024  &  4.4$^{+2.5}_{-1.5}$   &  3.3  &  0.3  &  0.50 (0.0)  \\
LH CnC SR1 &  17.6 $\pm$ 2.7  &  19  &       0.094  &  11$^{+5}_{-3.3}$   &  13.2  &  0.7  &  0.33 (0.4)  \\
LH CnC SR2   &  37 $\pm$ 4  &  38  &         0.12  &  16$^{+7}_{-5}$   &  16.7  &  0.5  &  0.45 (0.1)  \\
LH CnC SR3   &  40 $\pm$ 4  &  38  &         0.11  &  18$^{+7}_{-5}$   &  15.6  &  0.4  &  0.50 (0.0)  \\
LH CnC SR4   &  49 $\pm$ 4  &  42  &         0.087  &  18$^{+8}_{-5}$   &  12.2  &  0.1  &  0.50 (0.0)  \\
LH Comp CnC SR1  & 12.9 $\pm$ 2.0  &  13  &  0.068  &  9$^{+4}_{-2.8}$   &  9.5  &  0.5  &  0.49 (0.0)  \\
LH Comp CnC SR2   &  5.2 $\pm$ 1.1  &  4  &  0.036  &  6.2$^{+3.2}_{-2.0}$   &  5.0  &  0.3  &  0.50 (0.0)  \\
\midrule
\RunThr  \\
\midrule
HH CnC SR1   &  1.5 $\pm$ 0.7  &  0  &    0.047    &  4.0$^{+2.4}_{-1.4}$   &  2.5  &  0.1  &  0.50 (0.0)  \\
HH CnC SR2   &  2.7 $\pm$ 0.8  &  4  &    0.12    &  4.9$^{+2.7}_{-1.6}$   &  6.6  &  0.8  &  0.20 (0.8)  \\
HH CnC SR3   &  1.3 $\pm$ 0.5  &  1  &    0.064    &  3.8$^{+2.3}_{-1.3}$   &  3.4  &  0.4  &  0.50 (0.0)  \\
HH CnC SR4   &  0.9 $\pm$ 0.4  &  1  &    0.067    &  3.3$^{+2.2}_{-1.2}$   &  3.5  &  0.5  &  0.43 (0.2)  \\
LH CnC SR1   &  6.7 $\pm$ 1.2  &  6  &    0.12    &  6.9$^{+3.4}_{-2.2}$   &  6.1  &  0.4  &  0.50 (0.0)  \\
LH CnC SR2   &  14.0 $\pm$ 1.9  &  18  &  0.27    &  10$^{+4}_{-2.9}$   &  14.2  &  0.9  &  0.11 (1.2)  \\
LH CnC SR3   &  14.4 $\pm$ 2.1  &  15  &  0.20    &  10$^{+4}_{-3.0}$   &  10.5  &  0.6  &  0.43 (0.2)  \\
LH CnC SR4   &  28 $\pm$ 5  &  31  &      0.48    &  20$^{+6}_{-4}$   &  25.2  &  0.8  &  0.08 (1.4)  \\
LH Comp CnC SR1 & 5.2 $\pm$ 1.2  &  7  &  0.16    &  6.3$^{+3.2}_{-2.0}$   &  8.6  &  0.8  &  0.18 (0.9)  \\
LH Comp CnC SR2 & 4.0 $\pm$ 1.2  &  3  &  0.089    &  5.8$^{+3.0}_{-1.8}$   &  4.7  &  0.3  &  0.50 (0.0)  \\
\bottomrule
\end{tabular}
\end{adjustbox}
\end{table}


%
\FloatBarrier
\section{Conclusion}
\label{sec:conclusion}

Searches for direct pair production of $\tau$-sleptons, for pair production of charginos, and for chargino-neutralino production are presented. The decays of charginos and neutralinos into the LSPs are assumed to occur exclusively via intermediate $\tau$-sleptons and $\tau$-sneutrinos. Final states with at least one hadronically decaying $\tau$-lepton and at least one other lepton, or at least two hadronically decaying $\tau$-leptons are considered. The searches use $\pp$ collision data at $\sqrt s = 13$~\TeV\ and $13.6$~\TeV\ delivered by the Large Hadron Collider to the ATLAS experiment in 2015--2018 and 2022--2023, respectively, corresponding to integrated luminosities of $140~\mathrm{fb}^{-1}$ and $53~\mathrm{fb}^{-1}$.
The improved calibration and identification algorithms, the larger data sample, the more sophisticated analysis techniques, and inclusion of production modes with ISR jets, improve the sensitivity to the compressed mass spectra compared with the previous searches. No significant deviations from the \SM predictions are observed and the results are interpreted in the context of simplified supersymmetric models. Given the observed data, no exclusion limits are set at $95\%$ CL for direct $\tau$-sleptons production when statistically combining both the \stauisr and \staunoisr channels. In the \stauisr channel alone, $\tau$-sleptons with masses below $115$~\GeV are excluded for lightest neutralinos with mass difference of $30$~\GeV. Chargino masses are excluded up to $175$~\GeV for mass splitting of $30$~\GeV. For mass-degenerate chargino pair and chargino-neutralino production, chargino masses are excluded up to $200$~\GeV for mass splitting of $30$~\GeV. These are the most stringent limits for the compressed scenarios of these models.


%

%

%
%
%
%
%
\section*{Acknowledgements}
%

%

%
%

%
%

We thank CERN for the very successful operation of the LHC and its injectors, as well as the support staff at
CERN and at our institutions worldwide without whom ATLAS could not be operated efficiently.

The crucial computing support from all WLCG partners is acknowledged gratefully, in particular from CERN, the ATLAS Tier-1 facilities at TRIUMF/SFU (Canada), NDGF (Denmark, Norway, Sweden), CC-IN2P3 (France), KIT/GridKA (Germany), INFN-CNAF (Italy), NL-T1 (Netherlands), PIC (Spain), RAL (UK) and BNL (USA), the Tier-2 facilities worldwide and large non-WLCG resource providers. Major contributors of computing resources are listed in Ref.~\cite{ATL-SOFT-PUB-2026-001}.

We gratefully acknowledge the support of ANPCyT, Argentina; YerPhI, Armenia; ARC, Australia; BMWFW and FWF, Austria; ANAS, Azerbaijan; CNPq and FAPESP, Brazil; NSERC, NRC and CFI, Canada; CERN; ANID, Chile; CAS, MOST, NSFC and FRFCU, China; Minciencias, Colombia; MEYS CR, Czech Republic; DNRF and DNSRC, Denmark; IN2P3-CNRS and CEA-DRF/IRFU, France; SRNSFG, Georgia; BMFTR, HGF and MPG, Germany; GSRI, Greece; RGC and Hong Kong SAR, China; ICHEP and Academy of Sciences and Humanities, Israel; INFN, Italy; MEXT and JSPS, Japan; CNRST, Morocco; NWO, Netherlands; RCN, Norway; MNiSW, Poland; FCT, Portugal; MNE/IFA, Romania; MSTDI, Serbia; MSSR, Slovakia; ARIS and MVZI, Slovenia; DSI/NRF, South Africa; MICIU/AEI, Spain; SRC and Wallenberg Foundation, Sweden; SERI, SNSF and Cantons of Bern and Geneva, Switzerland; NSTC, Taipei; TENMAK, T\"urkiye; STFC/UKRI, United Kingdom; DOE and NSF, United States of America.

Individual groups and members have received support from BCKDF, CANARIE, CRC and DRAC, Canada; CERN-CZ and PRIMUS, Czech Republic; COST, ERC, ERDF, Horizon 2020 and Marie Sk{\l}odowska-Curie Actions, European Union; Investissements d'Avenir Labex, Investissements d'Avenir Idex and ANR, France; DFG and AvH Foundation, Germany; Herakleitos, Thales and Aristeia programmes co-financed by EU-ESF and the Greek NSRF, Greece; BSF-NSF and MINERVA, Israel; NCN and NAWA, Poland; La Caixa Banking Foundation, CERCA and AGAUR programs from Generalitat de Catalunya and PROMETEO and GenT Programmes Generalitat Valenciana, Spain; G\"{o}ran Gustafssons Stiftelse, Sweden; The Royal Society and Leverhulme Trust, United Kingdom; Eric and Wendy Schmidt Fund for Strategic Innovation, United States of America.

In addition, individual members wish to acknowledge support from Chile: Agencia Nacional de Investigaci\'on y Desarrollo (ANID FONDECYT reg. 1230987, FONDECYT 1230812, FONDECYT 1240864, Fondecyt 3240661, Fondecyt Regular 1240721); China: Fundamental Research Funds for the Central Universities (010-63263105), Chinese Ministry of Science and Technology (MOST-2023YFA1605700, MOST-2023YFA1609300), National Natural Science Foundation of China (NSFC 12275265, NSFC-W2543005); Czech Republic: Czech Science Foundation (GACR - 24-11373S), Ministry of Education Youth and Sports (ERC-CZ-LL2327, FORTE CZ.02.01.01/00/22\_008/0004632); EU: H2020 European Research Council (ERC - 101002463); European Union: European Research Council (BARD No. 101116429, ERC - 101219398, ERC 101089007), European Regional Development Fund (HE COFUND GA No.101081355, ERDF), Marie Sklodowska-Curie Actions (GAP-101168829); France: Agence Nationale de la Recherche (ANR-22-EDIR-0002, ANR-24-CE31-0504-01); Germany: Deutsche Forschungsgemeinschaft (DFG - 469666862); China: Research Grants Council (GRF); Italy: Istituto Nazionale di Fisica Nucleare (LHC-MIUR - 28003/2025), Ministero dell'Università e della Ricerca (NextGenEU I53D23000820006 M4C2.1.1, SOE2024\_0000023); Japan: Japan Society for the Promotion of Science (JSPS KAKENHI  JP25H0063, JSPS KAKENHI JP22H04944, JSPS KAKENHI JP24K23939, JSPS KAKENHI JP24KK0251, JSPS KAKENHI JP25H00650, JSPS KAKENHI JP25H01291, JSPS KAKENHI JP25K01011, JSPS KAKENHI JP25K01023, JSPS KAKENHI JP25KK0047); Poland: Polish National Science Centre (NCN OPUS 2023/51/B/ST2/02507, NCN OPUS nr 2022/47/B/ST2/03059, UMO-2019/34/E/ST2/00393, UMO-2021/42/E/ST2/00350, UMO-2022/47/O/ST2/00148, UMO-2023/49/B/ST2/04085, UMO-2023/51/B/ST2/00920, UMO-2024/53/N/ST2/00869); Spain: Agència de Gestió d'Ajuts Universitaris i de Recerca. (AGAUR - 2023 BP 00141), Ministry of Science and Innovation (RYC2019-028510-I, RYC2020-030254-I, RYC2021-031273-I, RYC2022-038164-I), Ministerio de Ciencia, Innovación y Universidades/Agencia Estatal de Investigaci\'on (EU NextGenerationEU (PRTR-C17.I1), PID2022-142604OB-C22); Sweden: Carl Trygger Foundation (Carl Trygger Foundation CTS 22:2312), Swedish Research Council (Swedish Research Council 2023-04654, VR 2021-03651, VR 2022-03845, VR 2022-04683, VR 2023-03403, VR 2024-05451, VR 2025-05940), Knut and Alice Wallenberg Foundation (KAW 2023.0366); United Kingdom: The Binks Trust, Royal Society (NIF-R1-231091); United States of America: U.S. Department of Energy (ECA DE-AC02-76SF00515), John Templeton Foundation (John Templeton Foundation 63206), Neubauer Family Foundation.

%
%


%
\clearpage
\appendix
\part*{Appendix}
\addcontentsline{toc}{part}{Appendix}

\section{ML Technical Details}
\label{app:ml_app}


%

\subsection{Direct stau channels}
The input parameters of the DNNs used for the \staunoisr channel are gathered in Tables~\ref{tab:inputs_lep_had}-\ref{tab:inputs_lep_had3}. The hyperparameters used for training are shown in Table~\ref{tab:dnn_Description}. The input parameters of the DNNs used for the \stauisr channel are gathered in Tables~\ref{app:stauisr_input1}-\ref{app:stauisr_input4}. The hyperparameters used are gathered in Table~\ref{app:tab:model_parameters}.

\begin{table}[h]
\centering
\begin{tabular}{|c|c|c|}
\hline
Channel & \wjet & \nojet \\
\hline
Number of input variables & 91 & 84 \\
Number of hidden layers & 5 & 4\\
Number of nodes [per layer] & [64, 48, 32, 16, 16]&  [48, 32, 16, 16]\\
Dropout [per layer] & [0.1, 0.1, 0.1, 0.0, 0.0]& [0.125, 0.125, 0.0, 0.0]\\
Maximum training epochs & \multicolumn{2}{c|}{100} \\
Number of output variables & \multicolumn{2}{c|}{6}  \\
Activation function (hidden layers) & \multicolumn{2}{c|}{ReLU} \\
Predictive activation & \multicolumn{2}{c|}{Softmax} \\
Optimiser & \multicolumn{2}{c|}{Adam} \\
Learning rate & \multicolumn{2}{c|}{0.0005} \\
$\beta_1$, $\beta_2$, $\epsilon$ & \multicolumn{2}{c|}{0.9, 0.999, $1\times10^{-7}$ (\textit{default})} \\
Loss function & \multicolumn{2}{c|}{Categorical cross-entropy} \\
Batch size & \multicolumn{2}{c|}{2048} \\
\hline
\end{tabular}
\caption{\label{tab:dnn_Description} Summary of the hyperparameters for the \staunoisr channel DNN training.}
\end{table}

\begin{table}[h]
\caption{\label{tab:inputs_lep_had} Description of the input variables used by the deep neural networks in the \staunoisr channel. The ticks ($\checkmark$) correspond to variables used in the given jet selection. For rows referring to more than one variable, the variations for each variable are given inside the curly brackets ($\{\}$), where the ticks and dashes ($-$) inside correspond to the presence of each variable in the order that they are presented in the curly brackets under the \enquote{Variable} and \enquote{Description} columns.}
\centering
\begin{adjustbox}{width=1\textwidth}
\begin{tabular}{lcccc}
\toprule
Category & Variable & Description & \wjet & \nojet \\
\midrule
\multirow{4}{*}{\makecell{ hadronic \\ tau-lepton }} & \tauhadvis \pt & transverse momentum& \checkmark & \checkmark \\
& \tauhadvis $\eta$ & pseudorapidity&  & \checkmark \\
& \tauhadvis $E$ & energy & \checkmark & \checkmark \\
& ${\tauhadvis}^{\mathrm{NNDM}}$ & number of charged/neutral pions decay mode
& \checkmark & \checkmark \\
\midrule
\multirow{7}{*}{\makecell{lepton} }& $\ell$ \pt & transverse momentum& \checkmark & \checkmark \\
& $\ell$ $\eta$ & pseudorapidity& \checkmark & \checkmark \\
& $\ell$ $E$ & energy & \checkmark & \checkmark \\
& $\ell$ $(+,-)$ & charge& \checkmark & \checkmark \\
& $\ell$ $(e,\mu)$  & flavour& \checkmark & \checkmark \\
& $\ell$ $d_0/\sigma_{d_0}$  & transverse impact parameter significance& \checkmark & \checkmark \\
& $\ell$ $z_0\sin\theta$  & longitudinal impact parameter & \checkmark & \checkmark \\
\midrule
\multirow{7}{*}{\makecell{missing \\transverse \\momentum}} & $\met $ & missing transverse momentum magnitude& \checkmark & \checkmark \\
& $\met$ sig & $\met$ significance& \checkmark & \checkmark \\
& $\sum \pt$ & \makecell{Scalar sum of $\pt$ of objects used to\\ calculate $\met$}& \checkmark & \checkmark \\
&  ${\pt}^{\mathrm{hard}}_{,\{20, 30, 40, 50\}} $ & \makecell{
magnitude of vector sum of hard objects, \\ using jets with $\pt >\{20, 30, 40, 50\}~\GeV$ } & $\{\checkmark,\checkmark,\checkmark,\checkmark\}$ & $\{\checkmark,\checkmark,\checkmark,\checkmark\}$ \\
&  ${\pt}^{\mathrm{soft}}_{,\{20, 30, 40, 50\}} $ & $|(\pTmiss+\mathbf{\pt}^{\mathrm{hard}}_{,\{20, 30, 40, 50\}})\cdot -1 |$ & $\{\checkmark,\checkmark,\checkmark,- \}$ & $\{\checkmark,\checkmark,\checkmark,\checkmark\}$ \\
&  ${\pt}^{\mathrm{minus}}_{,\{20, 30, 40, 50\}} $ & $|\pTmiss-\mathbf{\pt}^{\mathrm{hard}}_{,\{20, 30, 40, 50\}} |$ & $\{\checkmark,\checkmark,-,- \}$ & $\{\checkmark,\checkmark,\checkmark,\checkmark\}$ \\
\midrule
\multirow{5}{*}{\makecell{leading \\ jet}}& jet \pt & transverse momentum& \checkmark & \\
& jet $\eta$ & pseudorapidity& \checkmark & \\
& jet $E$ & energy & \checkmark & \\
& jet $m$  & invariant mass& \checkmark & \\
& jet $b$-tag quantile & $b$-tagging pseudo-continuous working point  & \checkmark & \\
\midrule
\multirow{6}{*}{\makecell{azimuthal \\ differences}} & $\Delta\phi(\pTmiss, \ell )$& between $\pTmiss$ and lepton & \checkmark & \checkmark \\
& $\Delta\phi(\pTmiss, \tau )$& between $\pTmiss$ and tau-lepton & \checkmark & \checkmark \\
& $\Delta\phi(\ell, \tau )$& between lepton and tau-lepton & \checkmark & \checkmark \\
& $\Delta\phi(\pTmiss, j_1)$& between $\pTmiss$ and leading jet & \checkmark & \\
& $\Delta\phi(\pTmiss, \mathbf{\pt}^{\mathrm{hard}}_{,\{20, 50\}})$& between $\pTmiss$ and $\mathbf{\pt}^{\mathrm{hard}}_{,\{20, 50\}}$ & $\{\checkmark,- \}$ & $\{\checkmark,\checkmark\}$ \\
& $\Delta\phi(\pTmiss, \mathbf{\pt}^{\mathrm{soft}}_{,\{20, 50\}})$& between $\pTmiss$ and $\mathbf{\pt}^{\mathrm{soft}}_{,\{20, 50\}}$ & $\{\checkmark,\checkmark \}$ & $\{\checkmark,\checkmark\}$ \\
\bottomrule
\end{tabular}
\end{adjustbox}
\end{table}

\begin{table}[h]
\caption{\label{tab:inputs_lep_had2} Description of the input variables used by the deep neural networks in the \staunoisr channel. The ticks ($\checkmark$) correspond to variables used in the given jet selection. For rows referring to more than one variable, the variations for each variable are given inside the curly brackets ($\{\}$), where the ticks and dashes ($-$) inside correspond to the presence of each variable in the order that they are presented in the curly brackets under the \enquote{Variable} and \enquote{Description} columns.}
\centering
\begin{adjustbox}{width=1\textwidth}
\begin{tabular}{lcccc}
\toprule
Category & Variable & Description & \wjet & \nojet \\
\midrule
\multirow{13}{*}{\makecell{azimuthal \\ differences}} & $\Delta\phi(\mathbf{\pt}^{\mathrm{hard}}_{,\{20, 50\}}, \ell)$& between $\mathbf{\pt}^{\mathrm{hard}}_{,\{20, 50\}}$ and lepton & $\{\checkmark,- \}$ & $\{\checkmark,\checkmark\}$ \\
& $\Delta\phi(\mathbf{\pt}^{\mathrm{hard}}_{,\{20, 50\}}, \tau)$& between $\mathbf{\pt}^{\mathrm{hard}}_{,\{20, 50\}}$ and tau-lepton & $\{-,\checkmark\}$ & $\{\checkmark,\checkmark\}$ \\

& $\Delta\phi(\mathbf{\pt}^{\mathrm{soft}}_{,\{20, 50\}}, \ell)$& between $\mathbf{\pt}^{\mathrm{soft}}_{,\{20, 50\}}$ and lepton & $\{-,\checkmark \}$ & $\{\checkmark,\checkmark\}$ \\
& $\Delta\phi(\mathbf{\pt}^{\mathrm{soft}}_{,\{20, 50\}}, \tau)$& between $\mathbf{\pt}^{\mathrm{soft}}_{,\{20, 50\}}$ and tau-lepton & $\{\checkmark,- \}$ & $\{\checkmark,\checkmark\}$ \\

& $\Delta\phi(\mathbf{\pt}^{\mathrm{hard}}_{,\{20, 50\}}, \mathbf{\pt}^{\mathrm{soft}}_{,\{20, 50\}})$& between $\mathbf{\pt}^{\mathrm{hard}}_{,\{20, 50\}},$ and $\mathbf{\pt}^{\mathrm{soft}}_{,\{20, 50\}}$ & $\{-,\checkmark \}$ & $\{\checkmark,\checkmark\}$ \\

& $\Delta\phi(\pTmiss, \mathbf{\pt}^{\mathrm{minus}}_{,\{20, 50\}})$& between $\pTmiss$ and $\mathbf{\pt}^{\mathrm{minus}}_{,\{20, 50\}}$ & $\{-,- \}$ & $\{\checkmark,\checkmark \}$ \\

& $\Delta\phi(\mathbf{\pt}^{\mathrm{hard}}_{,\{20, 50\}}, \mathbf{\pt}^{\mathrm{minus}}_{,\{20, 50\}})$& between $\mathbf{\pt}^{\mathrm{hard}}_{,\{20, 50\}},$ and $\mathbf{\pt}^{\mathrm{minus}}_{,\{20, 50\}}$ & $\{\checkmark,- \}$ & $\{\checkmark,\checkmark\}$ \\

& $\Delta\phi(\mathbf{\pt}^{\mathrm{soft}}_{,\{20, 50\}}, \mathbf{\pt}^{\mathrm{minus}}_{,\{20, 50\}})$& between $\mathbf{\pt}^{\mathrm{soft}}_{,\{20, 50\}},$ and $\mathbf{\pt}^{\mathrm{minus}}_{,\{20, 50\}}$ & $\{\checkmark,\checkmark \}$ & $\{\checkmark,\checkmark\}$ \\

& $\Delta\phi(\mathbf{\pt}^{\mathrm{minus}}_{,\{20, 50\}}, \ell)$& between $\mathbf{\pt}^{\mathrm{minus}}_{,\{20, 50\}}$ and lepton & $\{\checkmark,- \}$ & $\{\checkmark,  \checkmark\}$ \\
& $\Delta\phi(\mathbf{\pt}^{\mathrm{minus}}_{,\{20, 50\}}, \tau)$& between $\mathbf{\pt}^{\mathrm{minus}}_{,\{20, 50\}}$ and tau-lepton & $\{-, \checkmark \}$ & $\{  \checkmark,\checkmark\}$ \\

& $\Delta\phi(\mathbf{\pt}^{\mathrm{hard}}_{,50}, j_1)$& between $\mathbf{\pt}^{\mathrm{hard}}_{,50}$ and leading jet & \checkmark & \\
& $\Delta\phi(\mathbf{\pt}^{\mathrm{soft}}_{,20}, j_1)$& between $\mathbf{\pt}^{\mathrm{soft}}_{,20}$ and leading jet & \checkmark & \\
& $\Delta\phi(\mathbf{\pt}^{\mathrm{minus}}_{,\{20, 50\}}, j_1)$& between $\mathbf{\pt}^{\mathrm{minus}}_{,\{20, 50\}}$ and leading jet & $\{\checkmark, \checkmark \}$ & $\{-,- \}$ \\
\midrule
\multirow{9}{*}{\makecell{other \\ angular \\ differences}}
& $\Delta\eta(\ell, \tau)$& $\eta$ difference between lepton and tau-lepton & \checkmark & \checkmark \\
& $\Delta\eta(\ell, j_1)$&  $\eta$ difference between lepton and leading jet & \checkmark &  \\
& $\Delta R(\ell, \tau)$& \makecell{angular distance between \\ lepton and tau-lepton} & \checkmark & \checkmark \\
& $\Delta R(\ell, j_1)$& \makecell{angular distance between \\ lepton and leading jet} & \checkmark &  \\
& $\Delta R(\tau, j_1)$& \makecell{angular distance between \\ tau-lepton and leading jet} & \checkmark & \\
& $\sum \cos \Delta \phi$& $\cos\Delta\phi(\pTmiss, \ell ) +  \cos\Delta\phi(\pTmiss, \tau )$ & \checkmark & \checkmark \\
& $\met$ centrality  & $\frac{A+B}{\sqrt{A^2+B^2}}$, where $\{A,B\} = \frac{\sin\Delta\phi(\pTmiss, \{\ell,\tau\} )}{\sin\Delta\phi(\tau, \ell )}$%
& \checkmark & \checkmark \\
& $\phi_{\{1,2\}}$ &  \makecell{$\phi_1 = \phi_{\mathrm{max}} - \phi_\text{MET}$ and $\phi_2 = \phi_\text{MET} - \phi_{\mathrm{min}}$, \\ where $\phi_{\mathrm{\{max, min\}}} = \mathrm{\{max, min\}(\phi_\ell, \phi_\tau)}$}   & $\{\checkmark,\checkmark\}$ & $\{\checkmark,\checkmark\}$ \\
& $\cos{\theta^*}$ &  $\left| \tanh \left(\frac{\Delta y (\tau,\ell)}{2} \right) \right|$, where $y$ is the rapidity& \checkmark & \checkmark \\
\midrule
\multirow{2}{*}{\makecell{balance}}
& $\tau-\ell$ balance & $\frac{\pt(\tau)-\pt(\ell)}{\pt(\tau)+\pt(\ell)}$ & \checkmark & \checkmark \\
& $\tau+\ell-\met$ balance & $\frac{\pt(\tau)+\pt(\ell)-\met }{\pt(\tau)+\pt(\ell)+\met }$ & \checkmark & \checkmark \\
\midrule
\multirow{6}{*}{\makecell{mass \\ variables}}
& $\mt(\ell,\met)$  & transverse mass of lepton with $\met$ & \checkmark & \checkmark \\
& $\mt(\tau,\met)$  & transverse mass of tau-lepton with $\met$ & \checkmark & \checkmark \\
& $\sum \mt$ & $\mt(\ell,\met)$ + $\mt(\tau,\met)$  & \checkmark & \checkmark \\
& $m_\mathrm{CT}(\ell,\tau) $
& \makecell{lepton and tau-lepton contransverse mass: \\ $\sqrt{2 \pt(\ell) \pt(\tau) [ 1 + \cos\Delta\phi(\ell,\tau) ]}$} & \checkmark & \checkmark \\
& $m(\tau,\ell) $ & invariant mass of tau-lepton and lepton & \checkmark & \checkmark \\
& \mttwo ($m_\chi=\{0,10,20,30,40,50,60\}$) & \makecell{symmetric stransverse mass with invisible \\ masses of $\{0,10,20,30,40,50,60\}~\GeV$ }
& $\{7 \checkmark\}$ & $\{7\checkmark\}$ \\
\bottomrule
\end{tabular}
\end{adjustbox}
\end{table}

\begin{table}[h]
\caption{\label{tab:inputs_lep_had3} Description of the input variables used by the deep neural networks in the \staunoisr channel. The ticks ($\checkmark$) correspond to variables used in the given jet selection. For rows referring to more than one variable, the variations for each variable are given inside the curly brackets ($\{\}$), where the ticks and dashes ($-$) inside correspond to the presence of each variable in the order that they are presented in the curly brackets under the \enquote{Variable} and \enquote{Description} columns.}
\centering
\begin{adjustbox}{width=1\textwidth}
\begin{tabular}{lcccc}
\toprule
Category & Variable & Description & \wjet & \nojet \\
\midrule
\multirow{3}{*}{\makecell{mass \\ variables}}
& $H_\mathrm{T}(40)$  & scalar sum of $\pt$ of $>40~\GeV$ jets  & \checkmark &  \\
& $m_\mathrm{eff}(40)$  & scalar sum of $H_\mathrm{T}(40)$, lepton $\pt$, and $\met$  & \checkmark & \checkmark \\
& $m_\mathrm{eff, \tau}$(40)  & scalar sum of $m_\mathrm{eff}(40)$ and tau-lepton $\pt$  & \checkmark & \checkmark \\
\midrule
\multirow{2}{*}{\makecell{jet \\ multiplicity}}
& $N_\text{jets}^{\{20,30,40,50\}}$  & number of jets with $\pt > \{20,30,40,50\}~\GeV$ & $\{\checkmark,\checkmark,\checkmark,\checkmark\}$ & $\{-,-,-,-\}$  \\
& $N_{b\text{-jet}}^{90\% \text{WP}}$  & \makecell{number of jets satisfying the 90\% \\ $b$-tagging working point} & \checkmark & \checkmark \\
\midrule
&  $\Delta\phi(\pTmiss, \tau_{\{1,2\}}^{\mathrm{reco}} )$  & \makecell{azimuthal angle between $\pTmiss$ and \\ reconstructed tau-lepton $\{1,2\}$} & $\{\checkmark,\checkmark\}$ & $\{\checkmark,\checkmark\}$  \\
&  $\Delta\phi(\tau_{1}^{\mathrm{reco}},\tau_{2}^{\mathrm{reco}} )$  & \makecell{azimuthal angle between reconstructed \\ tau-leptons} & \checkmark & \checkmark \\
\multirow{6}{*}{\makecell{collinear \\ approximation \\ tau variables}}&  $\Delta R(\tau_{1}^{\mathrm{reco}},\tau_{2}^{\mathrm{reco}} )$  & \makecell{angular distance between reconstructed \\ tau-leptons} & \checkmark & \checkmark \\

& $\tau_{1-2}^{\mathrm{r}}$ balance & $\frac{\pt(\tau_{1}^{\mathrm{r}})-\pt(\tau_{2}^{\mathrm{r}})}{\pt(\tau_{1}^{\mathrm{r}})+\pt(\tau_{2}^{\mathrm{r}})}$ & \checkmark & \checkmark \\
& $m_\mathrm{CT}(\tau_{1}^{\mathrm{r}},\tau_{2}^{\mathrm{r}} ) $ & \makecell{reconstructed tau contransverse mass: \\ $\sqrt{2 \pt(\tau_{1}^{\mathrm{r}}) \pt(\tau_{2}^{\mathrm{r}}) [ 1 + \cos\Delta\phi(\tau_{1}^{\mathrm{r}},\tau_{2}^{\mathrm{r}}) ]}$} & \checkmark & \checkmark \\
& $\tau^{\mathrm{r}}\cos{\theta^*}$  &  $\left| \tanh \left(\frac{\Delta y (\tau_{1}^{\mathrm{r}},\tau_{2}^{\mathrm{r}} )}{2} \right) \right|$, where $y$ is the rapidity& \checkmark & \checkmark \\
&  $\Delta\phi(\tau_{\{1,2\}}^{\mathrm{r}},j_1)$  & \makecell{azimuthal angle between reconstructed \\ tau-lepton $\{1,2\}$ and leading jet}  & $\{\checkmark,\checkmark\}$ & $\{-,-\}$ \\
&  $\Delta R(\tau_{\{1,2\}}^{\mathrm{r}},j_1)$  & \makecell{angular distance between reconstructed \\ tau-lepton $\{1,2\}$ and leading jet}  & $\{\checkmark,\checkmark\}$ & $\{-,-\}$ \\
\bottomrule
\end{tabular}
\end{adjustbox}
\end{table}

\FloatBarrier

\begin{table}[htbp]
\centering
\caption{Summary of hyperparameters for the \stauisr channel DNN training.}
\label{app:tab:model_parameters}
\begin{tabular}{@{}lll@{}}
\toprule
Hyperparameter & Details & Value \\
\midrule
\multirow{2}{*}{Data split ratio} & \makecell{Train + Validation : Test \\ (5-fold cross-validation) } & 0.9 : 0.1 \\
& \makecell{Train : Test \\ (prediction model)} & 0.9 : 0.1 \\
Importance factor & Top & 0.5 \\
& Z & 0.5 \\
Hidden layer neuron numbers &  & 256, 128, 128, 64, 32 \\
Optimiser &  & Adam \\
Initial learning rate &  & 0.001 \\
Activation function &  & Leaky ReLU \\
Batch size &  & 128 \\
Dropout rate &  & 0.3 \\
L2 regularisation ($\lambda$) &  & 0.001 \\
\multirow{3}{*}{\makecell{Reduce learning \\ rate on plateau}} & Monitor & \makecell[l]{Validation Loss (cross-validation) \\ Train Loss (prediction model)} \\
& Factor & 0.5 \\
& Patience & 20 \\
\multirow{2}{*}{\makecell{Early stopping \\ (cross-validation)}} & Monitor & Validation Loss \\
& Patience & 50 \\
\bottomrule
\end{tabular}
\end{table}

\begin{table}[h]
\caption{Description of the input variables used by the deep neural networks in the \stauisr channel.}
\label{app:stauisr_input1}
\centering
\begin{tabular}{lcc}
\toprule
Category & Variable & Description \\
\midrule
\multirow{8}{*}{\makecell{leading \\ hadronic \\ tau-lepton}}
& $\tau_1 \pt$  & transverse momentum  \\
& $\tau_1 \eta$ & pseudorapidity  \\
& $\tau_1 \phi$ & azimuthal angle  \\
& $\tau_1 E$    & energy  \\
& $\tau_1 m$    & mass  \\
& $\tau_1 \mt$  & transverse mass  \\
& $\mttauone$ & transverse mass with \met  \\
& $\tau_1 N_\text{prong}$ & number of charged tracks \\
\midrule

\multirow{7}{*}{\makecell[t]{subleading \\ hadronic \\ tau-lepton or \\ light lepton}}
& $\tau_2 \pt$   & transverse momentum  \\
& $\tau_2 \eta$  & pseudorapidity  \\
& $\tau_2 \phi$  & azimuthal angle  \\
& $\tau_2 E$     & energy  \\
& $\tau_2 m$     & mass  \\
& $\tau_2 \mt$   & transverse mass  \\
& $\mttautwo $ & transverse mass with \met  \\
& $\tau_2 N_\text{prong}$ & number of charged tracks (HH only)\\
\midrule

\multirow{5}{*}{\makecell[t]{light lepton\\ (LH only)}}
& $\ell = e/\mu$ & flavor \\
& $d_0$ & transverse impact parameter \\
& $d_0\ \text{sig}$ & transverse impact parameter significance \\
& $z_0 \sin\theta$ & longitudinal impact parameter \\
& $z_0 \sin\theta\ \text{sig}$ & longitudinal impact parameter significance \\
\midrule

\multirow{7}{*}{\makecell{leading \\ jet}}
& $j_1 \pt$  & transverse momentum  \\
& $j_1 \eta$ & pseudorapidity  \\
& $j_1 \phi$ & azimuthal angle  \\
& $j_1 E$    & energy  \\
& $j_1 m$    & mass  \\
& $j_1 \mt$  & transverse mass  \\
& $m_\text{T}^{j_1}$ & transverse mass with \met  \\
\midrule

\multirow{7}{*}{\makecell{subleading \\ jet}}
& $j_2 \pt$  & transverse momentum  \\
& $j_2 \eta$ & pseudorapidity  \\
& $j_2 \phi$ & azimuthal angle  \\
& $j_2 E$    & energy  \\
& $j_2 m$    & mass  \\
& $j_2 \mt$  & transverse mass  \\
& $m_\text{T}^{j_2}$ & transverse mass with \met  \\
\midrule

\multirow{8}{*}{multiplicity} & $N_{\text{jet}}$ & number of jets \\
& $N_{\text{jet}}(\pt > 50~\text{GeV})$ &  number of $\pt > 50~\text{GeV}$ jets \\
& $N_{\text{jet}}(\pt > 100~\text{GeV})$ & number of $\pt > 100~\text{GeV}$ jets \\
& $N_{\text{jet}}(\pt > 150~\text{GeV})$ & number of $\pt > 150~\text{GeV}$ jets \\
& $N_{\text{jet}}(\pt > 200~\text{GeV})$ & number of $\pt > 200~\text{GeV}$ jets \\
& $N_{\text{jet}}(\pt > 300~\text{GeV})$ & number of $\pt > 300~\text{GeV}$ jets \\
& $\max(b\text{-quantile})$ & maximum jet $b$-tagging quantile\\
& $N_{\text{PV}}$ & number of primary vertices \\
\bottomrule
\end{tabular}
\end{table}

\begin{table}[h]
\caption{Description of the input variables used by the deep neural networks in the \stauisr channel.}
\label{app:stauisr_input2}
\centering
\begin{tabular}{lcc}
\toprule
Category & Variable & Description \\
\midrule
\multirow{10}{*}{\makecell{vector\\ sum of\\ jets}}
& $\sum j \pt$ & transverse momentum of vector sum of jets  \\
& $\sum j \eta$ & pseudorapidity of vector sum of jets  \\
& $\sum j \phi$ & azimuthal angle of vector sum of jets  \\
& $\sum j E$ & energy of vector sum of jets  \\
& $\sum j m$ & mass of vector sum of jets  \\
& $\sum_{\pt>50 ~\text{GeV}} j \pt$ & transverse momentum of vector sum of $\pt > 50~\text{GeV}$ jets \\
& $\sum_{\pt>50 ~\text{GeV}} j \eta$ & pseudorapidity of vector sum of $\pt > 50~\text{GeV}$ jets \\
& $\sum_{\pt>50 ~\text{GeV}} j \phi$ & azimuthal angle of vector sum of $\pt > 50~\text{GeV}$ jets  \\
& $\sum_{\pt>50 ~\text{GeV}} j E$ & energy of vector sum of $\pt > 50~\text{GeV}$ jets \\
& $\sum_{\pt>50 ~\text{GeV}} j m$ & mass of vector sum of $\pt > 50~\text{GeV}$ jets \\
\midrule

\multirow{15}{*}{balance} & $\met,\HT$ balance        & $\dfrac{\pt(\met) - \pt(\HT)}{\pt(\met) + \pt(\HT)}$  \\
& $\met,H_{\text{T},50}$ balance    & $\dfrac{\pt(\met) - \pt(H_{\text{T},50})}{\pt(\met) + \pt(H_{\text{T},50})}$  \\
& $\tau_1,\tau_2$ balance          & $\dfrac{\pt(\tau_1) - \pt(\tau_2)}{\pt(\tau_1) + \pt(\tau_2)}$  \\
& $\met,(\tau_1+\tau_2)$ balance  & $\dfrac{\pt(\met) - (\pt(\tau_1)+\pt(\tau_2))}{\pt(\met) + (\pt(\tau_1)+\pt(\tau_2))}$  \\
& $\met,\tau_2$ balance      & $\dfrac{\pt(\met) - \pt(\tau_2)}{\pt(\met) + \pt(\tau_2)}$  \\
& $\met,\tau_1$ balance      & $\dfrac{\pt(\met) - \pt(\tau_1)}{\pt(\met) + \pt(\tau_1)}$  \\
& $j_1,(\tau_1+\tau_2)$ balance     & $\dfrac{\pt(j_1) - (\pt(\tau_1)+\pt(\tau_2))}{\pt(j_1) + (\pt(\tau_1)+\pt(\tau_2))}$  \\
& $j_1,\tau_2$ balance             & $\dfrac{\pt(j_1) - \pt(\tau_2)}{\pt(j_1) + \pt(\tau_2)}$  \\
& $j_1,\tau_1$ balance             & $\dfrac{\pt(j_1) - \pt(\tau_1)}{\pt(j_1) + \pt(\tau_1)}$  \\
& $\HT,(\tau_1+\tau_2)$ balance     & $\dfrac{\pt(\HT) - (\pt(\tau_1)+\pt(\tau_2))}{\pt(\HT) + (\pt(\tau_1)+\pt(\tau_2))}$  \\
& $\HT,\tau_2$ balance             & $\dfrac{\pt(\HT) - \pt(\tau_2)}{\pt(\HT) + \pt(\tau_2)}$  \\
& $\HT,\tau_1$ balance             & $\dfrac{\pt(\HT) - \pt(\tau_1)}{\pt(\HT) + \pt(\tau_1)}$  \\
& $H_{\text{T},50},(\tau_1+\tau_2)$ balance & $\dfrac{\pt(H_{\text{T},50}) - (\pt(\tau_1)+\pt(\tau_2))}{\pt(H_{\text{T},50}) + (\pt(\tau_1)+\pt(\tau_2))}$  \\
& $H_{\text{T},50},\tau_2$ balance        & $\dfrac{\pt(H_{\text{T},50}) - \pt(\tau_2)}{\pt(H_{\text{T},50}) + \pt(\tau_2)}$  \\
& $H_{\text{T},50},\tau_1$ balance        & $\dfrac{\pt(H_{\text{T},50}) - \pt(\tau_1)}{\pt(H_{\text{T},50}) + \pt(\tau_1)}$  \\
\midrule
event geometry& $A$ & aplanarity  \\
& $S$ & sphericity  \\
& $S_\text{T}$ & transverse sphericity~\cite{ALICE:2012cor} \\
\bottomrule
\end{tabular}
\end{table}

\begin{table}[h]
\caption{Description of the input variables used by the deep neural networks in the \stauisr channel.}
\centering
\label{app:stauisr_input3}
\begin{adjustbox}{width=1\textwidth}
\begin{tabular}{lcc}
\toprule
Category & Variable & Description \\
\midrule

\multirow{10}{*}{\met variables}
& $E_\text{T}^{\text{miss}}$ & missing transverse energy  \\
& $E_\text{T}^{\text{miss}} \phi$ & azimuthal angle of missing transverse energy  \\
& $E_\text{T}^{\text{miss, jet}}$ & jet component of missing transverse energy  \\
& $E_\text{T}^{\text{miss, trk}}$ & track (soft) component of missing transverse energy  \\
& $E_\text{T}^{\text{miss, jet}} \phi$ & azimuthal angle of jet component of \met  \\
& $E_\text{T}^{\text{miss, trk}} \phi$ & azimuthal angle of track component of \met  \\
& \met sig & \met significance  \\
& $\sum p_\text{T}$ & scalar sum of transverse momenta (sumet)  \\
& $\sum p_\text{T}^{\text{jet}}$ & jet component of $\sum \pt$  \\
& $\sum p_\text{T}^{\text{trk}}$ & track (soft) component of $\sum \pt$  \\
\midrule

\multirow{7}{*}{\makecell{momentum \\ variables}}
& di-$\tau \pt$ & transverse momentum of di-tau system  \\
& di-$\tau \mt$ & transverse mass of di-tau system  \\
& $\text{boost}~ \pt$ & \makecell{transverse momentum of the boost system\\ ($\mathbf{p}_\text{T}(\tau_1)+\mathbf{p}_\text{T}(\tau_2)+\mathbf{p}_\text{T}^\text{miss} - \sum\mathbf{p}_\text{T}(j)$)}  \\
& $\text{boost}~ \mt$ & transverse mass of the boost system  \\
& $\text{soft}~ \pt$ & transverse momentum of the soft system ($\mathbf{p}_T^\text{miss} + \sum\mathbf{p}_T(j)$)   \\
& $\text{soft}~ \mt$ & transverse mass of the soft system  \\
& $\met \pt^{\text{proj}}$ & projection of \met onto bisector of two $\tau$  \\
& $(\tau_1,\tau_2) \pt^{\text{proj}}$ & projection of $\pt$ of $\tau_1, \tau_2$ onto bisector of two $\tau$  \\
& $j_1 \pt^{\text{proj}}$ & projection of $\pt$ of leading jet onto bisector of two $\tau$  \\
& $\sum j \pt^{\text{proj}}$ & projection of $\pt$ of jets onto bisector of two $\tau$  \\
& $\sum_{p_T>50 \text{GeV}} j \pt^{\text{proj}}$ & projection of $\pt$ of $\pt>50~\text{GeV}$ jets onto bisector of two $\tau$  \\
\midrule
\multirow{18}{*}{mass variables}
& $m(\tau_1, \tau_2)$ & invariant mass of $\tau_1$ and $\tau_2$  \\
& $\mt(\tau_1, \tau_2, \met)$ & transverse mass of $\tau_1$, $\tau_2$, and \met  \\
& $\HT$ & scalar sum of $\pt$ of jets  \\
& $\HT(50)$ & scalar sum of $\pt$ of $\pt > 50~\text{GeV}$ jets \\
& $m_{\text{CT}}$ & contransverse mass  \\
& $\mttwo(m_\chi=0)$ & stransverse mass with   $0~\text{GeV}$   invisible mass \\
& $\mttwo(m_\chi=50)$ & stransverse mass with  $50~\text{GeV}$  invisible mass  \\
& $\mttwo(m_\chi=70)$ & stransverse mass with  $70~\text{GeV}$  invisible mass  \\
& $\mttwo(m_\chi=100)$ & stransverse mass with $100~\text{GeV}$ invisible mass   \\
& $\mttwo(m_\chi=120)$ & stransverse mass with $120~\text{GeV}$ invisible mass   \\
& $\mttwo(m_\chi=150)$ & stransverse mass with $150~\text{GeV}$ invisible mass   \\
& $m_{\text{eff}}(\tau\tau)$ & scalar sum of \pt of $\tau_1$ and $\tau_2$  \\
& $m_{\text{eff}}(\met)$ &  scalar sum of \pt of $\tau_1$, $\tau_2$ and \met  \\
& $m_{\text{eff}}(j_1)$ &   scalar sum of \pt of $\tau_1$, $\tau_2$, \met, and leading jet  \\
& $m_{\text{eff}}(\HT)$ &  scalar sum of \pt of $\tau_1$, $\tau_2$, \met, and all jets  \\
& $m_{\text{eff}}(\HT(50))$ &  scalar sum of \pt of $\tau_1$, $\tau_2$, \met, and $\pt > 50~\text{GeV}$ jets  \\
& $\mt(\tau_1) + \mt(\tau_2)$ & sum of transverse masses of $\tau_1$ and $\tau_2$  \\
& $\mt(\tau_1,\met) + \mt(\tau_2,\met)$ & sum of transverse masses of $\tau_1$ and $\tau_2$ with \met  \\

\bottomrule
\end{tabular}
\end{adjustbox}
\end{table}

\begin{table}[h]
\caption{Description of the input variables used by the deep neural networks in the \stauisr channel.}
\label{app:stauisr_input4}
\centering
\begin{adjustbox}{width=1\textwidth}
\begin{tabular}{lcc}
\toprule
Category & Variable & Description \\
\midrule

\multirow{30}{*}{\makecell{angular \\ differences}}
& $\Delta \phi(\tau_1, j_1)$ & $\phi$ difference between $\tau_1$ and leading jet  \\
& $\Delta \phi(\tau_1, \met)$ & $\phi$ difference between $\tau_1$ and \met  \\
& $\Delta \phi(\tau_1, \sum j)$ & $\phi$ difference between $\tau_1$ and total jet system  \\
& $\Delta \phi(\tau_1, \text{soft})$ & \makecell{$\phi$ difference between $\tau_1$ and soft system \\ $\mathbf{p}_T^\text{miss} + \sum\mathbf{p}_T(j)$ }  \\
& $\Delta \phi(\tau_1, \text{boost})$ & \makecell{$\phi$ difference between $\tau_1$ and boost system \\ $\mathbf{p}_\text{T}(\tau_1)+\mathbf{p}_\text{T}(\tau_2)+\mathbf{p}_\text{T}^\text{miss} - \sum\mathbf{p}_\text{T}(j)$} \\
& $\Delta \phi(\tau_2, j_1)$ & $\phi$ difference between $\tau_2$ and leading jet  \\
& $\Delta \phi(\tau_2, \met)$ & $\phi$ difference between $\tau_2$ and \met  \\
& $\Delta \phi(\tau_2, \sum j)$ & $\phi$ difference between $\tau_2$ and total jet system  \\
& $\Delta \phi(\tau_2, \text{soft})$ & $\phi$ difference between $\tau_2$ and soft system  \\
& $\Delta \phi(\tau_2, \text{boost})$ & $\phi$ difference between $\tau_2$ and boost system  \\
& $\Delta \phi(j_1, \met)$ & $\phi$ difference between the leading jet and \met  \\
& $\Delta \phi(j_1, \sum j)$ & $\phi$ difference between the leading jet and total jet system  \\
& $\Delta \phi(j_1, \text{soft})$ & $\phi$ difference between the leading jet and soft system  \\
& $\Delta \phi(j_1, \text{boost})$ & $\phi$ difference between the leading jet and boost system  \\
& $\Delta \phi(\met, \sum j)$ & $\phi$ difference between \met and total jet system  \\
& $\Delta \phi(\met, \text{soft})$ & $\phi$ difference between \met and soft system  \\
& $\Delta \phi(\met, \text{boost})$ & $\phi$ difference between \met and boost system  \\
& $\Delta \phi(\sum j, \text{soft})$ & $\phi$ difference between total jet system and soft system  \\
& $\Delta \phi(\sum j, \text{boost})$ & $\phi$ difference between total jet system and boost system  \\
& $\Delta \phi(\text{soft}, \text{boost})$ & $\phi$ difference between soft system and boost system  \\
& $\Delta \eta(\tau_1, \tau_2)$ & $\eta$ difference between $\tau_1$ and $\tau_2$  \\
& $\Delta \eta(\tau_1, j_1)$ & $\eta$ difference between $\tau_1$ and leading jet  \\
& $\Delta \eta(\tau_2, j_1)$ & $\eta$ difference between $\tau_2$ and leading jet  \\
& $\Delta \eta(\tau_1, \sum j)$ & $\eta$ difference between $\tau_1$ and total jet system  \\
& $\Delta R(\tau_1, \tau_2)$ & angular distance between $\tau_1$ and $\tau_2$  \\
& $\Delta R(\tau_1, j_1)$ & angular distance between $\tau_1$ and leading jet  \\
& $\Delta R(\tau_2, j_1)$ & angular distance between $\tau_2$ and leading jet  \\
& $\Delta R(\tau_2, \sum j)$ & angular distance between $\tau_2$ and total jet system  \\
& $\cos\Delta\phi(\tau_1, \met) + \cos\Delta\phi(\tau_2, \met)$ & sum of cosine of $\phi$ separations with \met  \\
\midrule

\multirow{5}{*}{\makecell{other\\ angular\\ differences}}
& $\phi_{\max}$ & $\max(\phi(\tau_1),\phi(\tau_2))$  \\
& $\phi_{\min}$ & $\min(\phi(\tau_1),\phi(\tau_2))$  \\
& $\phi_{1}$ & $\phi_{\max}-\phi_\text{MET}$  \\
& $\phi_{2}$ & $\phi_\text{MET}-\phi_{\min}$  \\
& $\phi_{1}$+$\phi_{2}$ & $\phi_1+\phi_2=\phi_{\max}-\phi_{\min}$  \\
\midrule
\multirow{5}{*}{\makecell{collinear \\ approximation \\ tau variables}}
& $\chi_1$ & scale factor of $\tau_1$ 4-momenta  \\
& $\chi_2$ & scale factor of $\tau_2$ 4-momenta  \\
& $\pt(\tau_1^r)$ & corrected $\tau_1$ transverse momentum  \\
& $\pt(\tau_2^r)$ & corrected $\tau_2$ transverse momentum  \\
& $\mreco$ & invariant mass of corrected taus  \\
& balance of corrected taus &  $\dfrac{\pt(\tau_1^r) - \pt(\tau_2^r)}{\pt(\tau_1^r) + \pt(\tau_2^r)}$   \\

\bottomrule
\end{tabular}
\end{adjustbox}
\end{table}

\FloatBarrier

\subsection{Intermediate stau channels}

The hyperparameters used to train BDTs are gathered in Table~\ref{tab:hyper_c1c1} for the \ConepConem channel, and in Table~\ref{tab:hyper_c1n2} for the \ConepmNtwo channel. The input variables used for the \ConepConem channel are gathered in Tables~\ref{tab:BDTvar_LH}-\ref{tab:BDTvar_HH}. The input variables for the \ConepmNtwo channel are gathered in Table~\ref{tab:bdt_variables_C1N2}. An additional set of frames is used here, defined in the following way. Take the direction with maximum momentum projection of all particles in the final state as the thrust axis. The thrust axis direction without \met is the ISR-frame. Then the S-frame is defined as the thrust axis direction with $\met$, and V-frame is defined as the S-frame minus \met.

\begin{table}[htbp]
\centering
\caption{Hyperparameters used for BDT training, \ConepConem.}
\label{tab:hyper_c1c1}
\begin{tabular}{c|cc}
\hline
Hyperparameter   & \lephad & \hadhad   \\
\hline
Shrinkage         & 0.05    & 0.1       \\
Number of trees   & 300     & 400       \\
Max depth         & 6       & 8         \\
Minimum node size & 1\%     & 1\%       \\
\hline
\end{tabular}
\label{tab:hp_c1c1}
\end{table}

\begin{table}[h]
\centering
\caption{Hyperparameters used for BDT training, \ConepmNtwo.}
\label{tab:hyper_c1n2}
\begin{tabular}{c|cc}
\hline
Hyperparameter   & \lephad & \hadhad   \\
\hline
Shrinkage         & 0.03    & 0.03       \\
Number of trees   & 300     & 400       \\
Max depth         & 6       & 8         \\
Minimum node size & 1\%     & 1\%       \\
\hline
\end{tabular}
\end{table}

\begin{table}[htbp]
\centering
\caption{Kinematic variables used in the BDT training for the \lephad channel, \ConepConem.}
\begin{tabular}{lcc}
\toprule
Category & Variable & Description \\
\midrule
\multirow{3}{*}{\makecell{angular \\ correlations}}
& $\Delta R(\tauhadvis,\met)$ & angular distance between $\tauhadvis$ and $\met$ \\
& $\Delta R(\tauhadvis,\ell)$ & angular distance between $\tauhadvis$ and lepton \\
& $\Delta\phi(\tauhadvis,\ell)$ & azimuthal separation between $\tauhadvis$ and lepton \\
\midrule
\multirow{26}{*}{\makecell{kinematic \\ variables}}
& $E_{\ell}$ & lepton energy \\
& $\mt^{\ell}$ & transverse mass of lepton \\
& $\tau \pt$ & leading $\tauhadvis$ transverse momentum \\
& \met significance & missing transverse momentum significance \\
& N$_j$ & number of jets \\
& $\min(\Delta\phi(j,\met))$ & minimum $\Delta\phi$ between jets and $\met$ \\
& $\max(\Delta\phi(j,\ell))$ & maximum $\Delta\phi$ between jets and lepton \\
& $\mttau$ & transverse mass of $\tau$ and $\met$ system \\
& $\mtlep$ & transverse mass of lepton and $\met$ system \\
& $\min(\mt(\tau,\met),\mt(\ell,\met))$ & minimum transverse mass \\
& $\sqrt{\mt(\tau,\met)^2+\mt(\ell,\met)^2}$ & quadratic sum of transverse masses \\
& $\mt(\tau,\met)+\mt(\ell,\met)$ & sum of transverse masses \\
& $m_\text{inv}(\tauhadvis,\ell)$ & invariant mass of $\tau$ and lepton \\
& $m_\text{inv}(j_1,\ell)$ & invariant mass of leading jet and lepton \\
& \mttwo ($m_\chi=110$) & stransverse mass with invisible mass of $110$~\GeV \\
& $m_\mathrm{CT}(\tau,\ell)$ & contransverse mass of $\tau$ and lepton \\
& $\pt^{\text{proj}}(\tau_1,\tau_2)$ & projection of $\tau$ pair \pt on bisector \\
& $\pt^{\text{proj}}(j_1)$ & projection of leading jet \pt on bisector \\
& $\met/\pt^{\tauhadvis}$ & $\met$ to $\tau$ \pt ratio \\
& $\met/\pt^{lep}$ & $\met$ to lepton \pt ratio \\
& $\met/(\pt^{\tauhadvis}+\pt^{lep})$ & $\met$ to visible \pt ratio \\
& $\met/\sum \pt^{j_1}$ & $\met$ to jet \pt sum ratio \\
& $\pt^{j_1}/\pt^{\tauhadvis}$ & jet to $\tau$ \pt ratio \\
& $\pt^{j_1}/\pt^{lep}$ & jet to lepton \pt ratio \\
& $\pt^{j_1}/(\pt^{\tauhadvis}+\pt^{lep})$ & jet to visible \pt ratio \\
& \pt in V-frame &  transverse momentum in the V-frame  \\
\bottomrule
\end{tabular}
\label{tab:BDTvar_LH}
\end{table}

\begin{table}[htbp]
\centering
\caption{Kinematic variables used in the BDT training for the \hadhad channel, \ConepConem.}
\begin{tabular}{lcc}
\toprule
Category & Variable & Description \\
\midrule
\multirow{5}{*}{\makecell{angular \\ correlations}}
& $\Delta\phi(\text{V-frame},\text{ISR-frame})$ & azimuthal separation between frames \\
& $\Delta\phi(\text{V-frame},\met)$ & azimuthal separation between V-frame and \met \\
& $\Delta\phi(\tau_1,\met)$ & azimuthal separation between leading tau and \met \\
& $\max(\Delta\phi(j_1,\tau_1))$ & maximum $\Delta\phi$ between jet and $\tau_1$ \\
& $\max(\Delta\phi(j_1,\tau_2))$ & maximum $\Delta\phi$ between jet and $\tau_2$ \\
\midrule
\multirow{24}{*}{\makecell{kinematic \\ variables}}
& \met significance & missing transverse momentum significance \\
& \mttwo ($m_\chi=110$) & stransverse mass with invisible mass of $110$~\GeV \\
& $\pt^{\text{proj}}(j_1)$ & projection of leading jet \pt on bisector \\
& $\met/\sum \pt(j)$ & $\met$ to jet \pt sum ratio \\
& $\pt^{\text{proj}}(\met)$ & projection of \met onto the bisector of the two $\tau$ \\
& $m_\mathrm{CT}(\tau_1,\tau_2)$ & contransverse mass of the two tau-leptons \\
& jet $E$ & jet energy \\
& $N_{S_{\tau}}$ & number of taus along thrust axis \\
& $\met/\pt^{\tau_2}$ & \met to subleading tau \pt ratio \\
& $\met/\sum \pt^\tau$ & \met to tau \pt sum ratio \\
& $\pt(j_1)/\pt(\tau_1)$ & jet to leading tau \pt ratio \\
& $m_\text{inv}(j_1,\met)$ & invariant mass of jet and \met system \\
& $m_\text{inv}(\tau_2,\met)$ & invariant mass of $\tau_2$ and \met \\
& $\mt(\tau)$ & transverse mass of $\tau_2$ \\
& $\mt(\tau_1+\tau_2+\met)$ & transverse mass of full system \\
& $\mt(\tau_2+\met)$ & transverse mass of $\tau_2$ and \met \\
& $\min(\mt(\tau_1,\met),\mt(\tau_2,\met))$ & minimum tau transverse mass \\
& $\mt(\tau_1,\met)+\mt(\tau_2,\met)$ & sum of tau transverse masses \\
& $\mt(\tau_1)+\mt(\tau_2)$ & sum of tau transverse masses \\
& $\mt(\tau_1)+\mt(\tau_2)+\mt(j_1)$ & sum of tau and jet transverse masses \\
& $m_\text{inv}(\tau_1,\tau_2)$ & invariant mass of tau pair \\
& $\mttwo$ & stransverse mass \\
& $\met$ & missing transverse momentum \\
& \pt in V-frame &  transverse momentum in the V-frame \\
\bottomrule
\end{tabular}
\label{tab:BDTvar_HH}
\end{table}

\FloatBarrier

\begin{table}[htbp]
\centering
\caption{Variables used in BDT training for the \hadhad and \lephad channels, \ConepmNtwo.}
\label{tab:bdt_variables_C1N2}
\begin{adjustbox}{width=1\textwidth}
\footnotesize
\begin{tabular}{lcc}
\toprule
Category & Variable & Description \\
\midrule
\multirow{13}{*}{\makecell{angular \\ correlations}}
& $\Delta\phi(\tau_1, \met)$ & azimuthal separation between the leading tau-lepton and missing transverse momentum \\
& $\Delta\eta(\tau_1, \tau_2)$ & pseudorapidity difference between the two tau-leptons \\
& $\max[\Delta\phi(\met, \tau)]$ & maximum azimuthal separation between missing transverse momentum and the tau-leptons \\
& $\Delta\phi(\tau_1+\tau_2, \text{bisector})$ & azimuthal separation between the tau-pair system and the bisector\\
& $\Delta\phi(\tau_1, \tau_2)$ & azimuthal separation between the two tau-leptons \\
& $\Delta\phi_z(\text{bisector}, \met)$ & azimuthal separation between the taus bisector and missing transverse momentum \\
& $\min[\Delta\phi(\met, \tau)]$ & minimum azimuthal separation between missing transverse momentum and the tau-leptons \\
& $\Delta\phi(\tau_2, \met)$ & azimuthal separation between the subleading tau-lepton and missing transverse momentum \\
& $\min[\Delta\phi(\tau, jet)]$ & minimum azimuthal separation between the tau-leptons and jets \\
& $\Delta R(\tau_2, \met)$ & angular distance between the subleading tau-lepton and missing transverse momentum \\
& $\max[\Delta R(\met, \tau)]$ & maximum angular distance between missing transverse momentum and the tau-leptons \\
& $\min[\Delta R(\tau, jet)]$ & minimum angular distance between the tau-leptons and jets \\
& $\Delta R(\tau_1, \tau_2)$ & angular distance between the two tau-leptons \\
\midrule
\multirow{19}{*}{\makecell{kinematic \\ variables}}
& missing transverse impact angle & missing transverse impact angle \\
& $\pt(\tau_1) + \pt(\tau_2)$ & scalar sum of the transverse momenta of the two tau-leptons \\
& \mttwo ($m_\chi=150$) & stransverse mass with invisible mass of $150$~\GeV \\
& projection of $\tau\tau$ system & projection of the tau-pair transverse momentum onto the bisector \\
& $\mreco$ & reconstructed invariant mass of the tau-pair system \\
& reconstructed stau mass  & stau mass after recursive jigsaw reconstruction~\cite{Jackson:2017gcy} \\
& $m_\mathrm{CT}(\tau_1,\tau_2)$ & contransverse mass \\
& projection of $\tau_1$ & projection of the leading tau-lepton transverse momentum onto the bisector \\
& \met significance & missing transverse momentum significance \\
& $\met/(\pt(\tau_1)+\pt(\tau_2))$ & ratio of missing transverse momentum to the scalar sum of the two tau-lepton transverse momenta \\
& $\met/\pt(\tau_1)$ & ratio of missing transverse momentum to the leading tau-lepton transverse momentum \\
& $\met/\HT^\text{inc}$ (50~\GeV cut) & ratio of missing transverse momentum to inclusive $\HT$ with a 50~\GeV jet threshold \\
& $\met/m_{\text{eff}}^{1j}$ & ratio of missing transverse momentum to the effective mass in the one-jet category \\
& $\Delta\phi(\tau_1, \met) + \Delta\phi(\tau_2, \met)$ & sum of the cosines of the azimuthal angle differences \\
& sum of tau \pt & scalar sum of the tau-lepton transverse momenta \\
& $\met/\pt(\tau_2)$ & ratio of missing transverse momentum to the subleading tau-lepton transverse momentum \\
& $\eta_{\tau_2}$ & pseudorapidity of the subleading tau-lepton \\
& $\sum \mathbf{p}_{\text{T}}(j)$ &  vector sum of \pt of jets \\
& \pt in V-frame & transverse momentum in the V-frame \\
\bottomrule
\end{tabular}
\end{adjustbox}
\end{table}

\FloatBarrier


%
%
%
%

%
%

%
%
\printbibliography
\clearpage
\input{atlas_authlist}
%

%
%

%

%
%
%
%

\end{document}

%% file: atlas_authlist.tex
 
\begin{flushleft}
\hypersetup{urlcolor=black}
{\Large The ATLAS Collaboration}

\bigskip

\AtlasOrcid[0000-0002-6665-4934]{G.~Aad}$^\textrm{\scriptsize 102}$,
\AtlasOrcid[0000-0001-7616-1554]{E.~Aakvaag}$^\textrm{\scriptsize 17}$,
\AtlasOrcid[0000-0002-5888-2734]{B.~Abbott}$^\textrm{\scriptsize 121}$,
\AtlasOrcid[0000-0002-0287-5869]{S.~Abdelhameed}$^\textrm{\scriptsize 83b}$,
\AtlasOrcid[0000-0002-1002-1652]{K.~Abeling}$^\textrm{\scriptsize 54}$,
\AtlasOrcid[0000-0001-5763-2760]{N.J.~Abicht}$^\textrm{\scriptsize 48}$,
\AtlasOrcid[0000-0002-8496-9294]{S.H.~Abidi}$^\textrm{\scriptsize 30}$,
\AtlasOrcid[0009-0003-6578-220X]{M.~Aboelela}$^\textrm{\scriptsize 44}$,
\AtlasOrcid[0000-0002-9987-2292]{A.~Aboulhorma}$^\textrm{\scriptsize 36e}$,
\AtlasOrcid[0000-0001-5329-6640]{H.~Abramowicz}$^\textrm{\scriptsize 154}$
% ...
\bigskip
\\

% ...
$^{17}$Department for Physics and Technology, University of Bergen, Bergen; Norway.\\
% ...
$^{30}$Physics Department, Brookhaven National Laboratory, Upton NY; United States of America.\\
% ...
$^{36}$$^{(a)}$Facult\'e des Sciences Ain Chock, Universit\'e Hassan II de Casablanca;$^{(b)}$Facult\'{e} des Sciences, Universit\'{e} Ibn-Tofail, K\'{e}nitra;$^{(c)}$Facult\'e des Sciences Semlalia, Universit\'e Cadi Ayyad, LPHEA-Marrakech;$^{(d)}$LPMR, Facult\'e des Sciences, Universit\'e Mohamed Premier, Oujda;$^{(e)}$Facult\'e des sciences, Universit\'e Mohammed V, Rabat;$^{(f)}$Institute of Applied Physics, Mohammed VI Polytechnic University, Ben Guerir; Morocco.\\
% ...
$^{44}$Physics Department, Southern Methodist University, Dallas TX; United States of America.\\
% ...
$^{48}$Fakult\"{a}t Physik, Technische Universit{\"a}t Dortmund, Dortmund; Germany.\\
% ...
$^{54}$II. Physikalisches Institut, Georg-August-Universit\"{a}t G\"ottingen, G\"ottingen; Germany.\\
% ...
$^{83}$$^{(a)}$Khalifa University of Science and Technology, Abu Dhabi;$^{(b)}$New York University Abu Dhabi, Abu Dhabi;$^{(c)}$United Arab Emirates University, Al Ain;$^{(d)}$University of Sharjah, Sharjah; United Arab Emirates.\\
% ...
$^{102}$CPPM, Aix-Marseille Universit\'e, CNRS/IN2P3, Marseille; France.\\
% ...
$^{121}$Homer L. Dodge Department of Physics and Astronomy, University of Oklahoma, Norman OK; United States of America.\\
% ...
$^{154}$Raymond and Beverly Sackler School of Physics and Astronomy, Tel Aviv University, Tel Aviv; Israel.\\
% ...
$^{a}$ Also at Affiliated with an institute formerly covered by a cooperation agreement with CERN.\\
$^{b}$ Also at An-Najah National University, Nablus; Palestine.\\
$^{c}$ Also at Borough of Manhattan Community College, City University of New York, New York NY; United States of America.\\
$^{d}$ Also at Center for Interdisciplinary Research and Innovation (CIRI-AUTH), Thessaloniki; Greece.\\
$^{e}$ Also at Centre of Physics of the Universities of Minho and Porto (CF-UM-UP); Portugal.\\
$^{f}$ Also at CERN, Geneva; Switzerland.\\
% ...

\end{flushleft}
